\documentclass{emulateapj}  
\slugcomment{Submitted to ApJ Supplements}
\DeclareGraphicsExtensions{.pdf,.png}

\newcommand{\kms}{\ensuremath{{\rm km~s}^{-1}}}

\newcommand{\delm}{$\Delta\rm{m}_{15}$}

\newcommand{\He}{\ion{He}{1}}

\newcommand{\synNi}{$^{56}$\rm{Ni}}

\newcommand{\ninlit}{37}
\newcommand{\nnoninlit}{27}
\newcommand{\nphotnoninlit}{6}
\newcommand{\nsntot}{64}
\newcommand{\nsnir}{25}
\newcommand{\nsnopt}{61} 

\newcommand{\nsnboth}{22} 
\newcommand{\nsnspec}{54}                                              

\newcommand{\totoptphot}{4543}
\newcommand{\totirphot}{2142}

\newcommand{\ubvri}{\protect\hbox{$U\!BV\!RI$} }

\newcommand{\jhk}{\protect\hbox{$JHK_s$} }
\newcommand{\sloanubvri}{\protect\hbox{$U\!BV\!r'i'$} }

\newcommand{\sloanupbvri}{\protect\hbox{$u'\!BV\!r'i'$} }

\newcommand{\secref}[1]{Section~\ref{#1}}
\newcommand{\tabref}[1]{Table~\ref{#1}}
\newcommand{\figref}[1]{Figure~\ref{#1}}
\newcommand{\maxep}{$\mathrm{JD}_\mathrm{Vmax}$} 
\newcommand{\cfaweb}{\url{http://www.cfa.harvard.edu/oir/Research/supernova/}}
\newcommand{\nyuweb}{\url{http://www.cosmo.nyu.edu/SNYU}}
\usepackage{hyperref}

\begin{document}

\title{Multi-color Optical and NIR  Light Curves  of \nsntot\ Stripped-Envelope Core-Collapse Supernovae }


\author{F.~B.~Bianco \altaffilmark{1},
M.~Modjaz\altaffilmark{1}, 
M.~Hicken\altaffilmark{2},
A.~Friedman\altaffilmark{2,3},
R.~P.~Kirshner\altaffilmark{2},
J.~S.~Bloom\altaffilmark{4,5},
P.~Challis\altaffilmark{2},
G.H.~Marion\altaffilmark{2,6},
W.~M.~Wood-Vasey\altaffilmark{7} 
}

\altaffiltext{1}{Center for Cosmology and Particle Physics, New York University, 4 Washington Place, New York, NY 10003, USA}
\altaffiltext{2}{Harvard-Smithsonian Center for Astrophysics, 60 Garden Street, Cambridge, MA, 02138.}
\altaffiltext{3}{Center for Theoretical Physics and Department of Physics, Massachusetts Institute of Technology, Cambridge, MA 02139}
\altaffiltext{4}{Department of Astronomy, University of California, Berkeley, CA 94720-3411, USA}
\altaffiltext{5}{Physic Division, Lawrence Berkeley National Laboratory, 1 Cyclotron Road, Berkeley, CA, 94720, USA}
\altaffiltext{6}{Astronomy Department, University of Texas at Austin, Austin, TX 78712, USA}
\altaffiltext{7}{PITT PACC, Department of Physics \& Astronomy, 3941 O'Hara St, University of Pittsburgh, Pittsburgh, PA 15260.}

\begin{abstract}
We present a densely-sampled, homogeneous set of light curves of $\nsntot$\ low redshift ($z\la0.05$) stripped-envelope supernovae (SN of type IIb, Ib, Ic and Ic-bl). These data were obtained between 2001 and 2009 at the Fred L. Whipple Observatory (FLWO) on Mt. Hopkins in Arizona, with the optical FLWO 1.2-m 
and the near-infrared PAIRITEL 1.3-m telescopes.
Our dataset consists of $\totoptphot$\ optical photometric measurements on $\nsnopt$\ SN, including a combination of $\ubvri$, $\sloanubvri$, and $\sloanupbvri$, and $\totirphot$\  
$JHK_s$ near-infrared measurements
on $\nsnir$\ SN. 

This sample constitutes the most extensive \emph{multi-color} data set of stripped-envelope SN to date.  Our photometry is based on template-subtracted images to eliminate any potential host galaxy light contamination. This work presents these photometric data, compares them with data in the literature, and estimates basic statistical quantities: date of maximum, color, and photometric properties. 
We identify promising color trends that may permit the identification of  stripped-envelope SN subtypes from their photometry alone.
Many of these SN were observed spectroscopically by the CfA SN group, and the spectra are
presented in a companion paper~\citep{Modjaz14}. 
A thorough exploration that combines the CfA photometry and spectroscopy of stripped-envelope core-collapse SN will be presented in a follow-up paper.

\end{abstract}

\bibliographystyle{apj}

\section{Introduction}\label{sec:intro_sec}
Stripped-envelope core-collapse supernovae (stripped SN) arise from the spectacular death of massive stars
that have been stripped of their outer layers of hydrogen and helium. In this paper we present photometric data in optical and near infra-red (NIR) wavelengths for 64 stripped SN, data that we collected between 2001 and 2009 at the Fred L. Whipple Observatory (FLWO) on Mt. Hopkins in Arizona. 

Stripped SN include SN of
types Ib, Ic, and IIb. Type Ib (SN Ib) and Type Ic SN (SN Ic) are SN that do not show hydrogen lines (thus Type I), but do not exhibit the strong \ion{Si}{2} absorption lines characteristic of SN Ia~(\citealt{Uomoto_Kirshner_1986},~\citealt{Clocchiatti97}).
SN Ib show  conspicuous lines of \He, while SN Ic do not.
SN IIb change as they age: initially they show strong hydrogen features (hence the Type II classification), but over time the Balmer series decreases in strength, while the series of \He\ lines characteristic of SN Ib grows stronger~(e.g., \citealt{Filippenko_Matheson_Ho_1993}).
Finally, broad-lined SN Ic (SN Ic-bl) exhibit broad and blended lines in a SN Ic-like spectrum, indicative of very high
 expansion velocities
\citep{onzalez_Leibundgut_et_al__1998,r_Gonzalez_Hainaut_et_al__2001,y_Kouveliotou_Deng_et_al__2006,Diamond-Stanic_Hao_et_al__2006,2012ApJ...756..184S}.
SN Ic-bl are the only type of SN that have been observed in conjunction with long-duration GRBs (e.g.,
\citealt{onzalez_Leibundgut_et_al__1998,Schild_Krisciunas_et_al__2003,rsen_Castro-Tirado_et_al__2003,Diamond-Stanic_Hao_et_al__2006}). See
\citet{2006ARA&A..44..507W}, \citet{2011AN....332..434M}, and \citet{2012grbu.book..169H} for reviews of GRB-SN connections.
For a review of SN spectroscopic classification, see  \citet{Filippenko_1997}.

Stripped SN have been studied less than SN Ia. 
These SN, however, are intrinsically almost as common per volume as SN Ia \citep{2011MNRAS.412.1441L}, and they hold vital clues about the death and explosion properties of very massive stars
\citep{Uomoto_Kirshner_1986}, and their nucleosynthesis products that contribute
to the Universe's chemical enrichment \citep{dge_Burbidge_Fowler_Hoyle_1957,aga_Umeda_Kobayashi_Maeda_2006}. The characteristics of the progenitor channels, and their link to each SN class and subclass are not yet well understood. Nor do we know which is the dominant process responsible for stripping these massive stars of their outer layers:  models propose stripping may occur through strong winds
\citep{Woosley_Langer_Weaver_1993}, or binary interaction
\citep{Nomoto_Iwamoto_Suzuki_1995,nds_Rappaport_Heger_Pfahl_2004}.

Several stripped SN have been studied in detail individually, beginning with the SN Ic ~1994I in the nearby galaxy M~51 (e.g.,
\citealt{Treffers_Filippenko_Paik_1996,1995ApJ...450L..11F}). Because of its proximity, it was well observed over  many wavelengths and it is
commonly referred to as the ``prototypical'' normal SN Ic (e.g., \citealt{Leibundgut_Baron_Kirshner_2006} and \citealt{2006MNRAS.369.1939S}).

However, in order to assess the peculiarities of these explosions and to understand the characteristics of stripped SN, well observed SN must be evaluated in the context of a sample large enough to be studied with a statistical approach. For example, SN~1994I appears to be \emph{non-typical}:
\citet{Richardson_Branch_Baron_2006} and \citet{rd_Sand_Moon_Arcavi_Green_2011} showed that SN~1994I had a faster light curve than any other SN Ic in the literature and is a 2$\sigma$ outlier of the overall distribution of light curves of SN Ib and SN Ic.

\citet{Richardson_Branch_Baron_2006} compiled light
curves of 27 stripped-envelope SN from the literature, of which one-third had been found
or observed with photographic plates. However, photographic plate surveys are strongly biased against dim SN or SN near the nucleus of the host galaxy. Modern CCD surveys, analyzed with image subtraction techniques (\citealt{gensen_Uomoto_Gunn_et_al__2002}) should instead be nearly complete, barring large amounts of host
galaxy dust extinction.

A sample of stripped SN was presented in \citet{rd_Sand_Moon_Arcavi_Green_2011} -- D11 henceforth:  25 SN Ib, Ic and Ic-bl, observed in 2 bands. Eighteen of these objects were also observed within our program. D11 concluded that SN Ib and Ic are indistinguishable photometrically. Furthermore, from the peak luminosity D11 sets constraints to the \synNi~  mass generated in the explosion, and assuming that SN Ib and SN Ic have the same photospheric velocities, D11 derives constraints on the ejecta mass from the light curve shape.
This pioneering study of stripped SN, however, presents data in just two bands and does
not employ galaxy subtraction.  As we show in \secref{sec:litdata}, galaxy
subtraction can be important for producing accurate light curves.

 \citet{2011MNRAS.412.1441L} presented unfiltered light curves of SN that were discovered as part of the Lick Observatory SN Search (LOSS,  \citealt{2001LOSS}), including about 30 stripped SN (5 of which are included in this study). Those unfiltered light
curves were crucial for calculating the SN luminosity function and the LOSS SN rates,
however they are in a single, non-standard band.

A collection of UV light curves of core-collapse SN from \textit{SWIFT}~\citep{y_Burrows_Cominsky_et_al__2004}, including 15 stripped SN (6 of which are in our sample), is presented in ~\citet{Pritchard13}.

Understanding the full range of massive star explosion properties
requires the study of a large and comprehensive SN sample with
homogeneous and densely-sampled data.
Moreover, the current SN classification scheme, outlined above,
is based on spectroscopy. As we enter the era of all-sky optical transient
searches, with hundreds, even thousands of SN to be discovered each night \citep{LSST}, we will simply be unable to obtain systematic spectroscopic follow-up data of most objects. Devising photometric criteria for classifying SN without spectra is important \citep{2014arXiv1401.3317S}.  The first step in this process is to obtain well-sampled light curves of SN Ib, SN Ic and SN IIb.

This work presents a densely sampled, multi-color, homogeneous data set of stripped SN, supported and complemented by spectroscopic data (\citealt{Modjaz14}, henceforth M14).  Since 1993, spectroscopic \emph{and} photometric monitoring of nearby and
newly-discovered SN at the Fred L. Whipple Observatory (FLWO) on
Mt. Hopkins in Arizona has been undertaken by the  Harvard-Smithsonian Center for Astrophysics
(CfA)\footnote{\cfaweb}.  Furthermore, the CfA conducted a parallel near infrared (NIR) photometric campaign with PAIRITEL at FLWO starting in 2004. While, due to their cosmological relevance, SN Ia were prioritized targets throughout the campaign
\citep{er_Grashius_Schild_et_al__1999,2006AJ....131..527J, 2009ApJ...700..331H,ind_Brown_Caldwell_et_al__2012},
an intense
follow-up program of stripped SN began in 2004, in
addition to the SN Ia follow-up. Here we present
photometric data of nearby ($z \la 0.047$)
stripped SN collected between 2001 and 2009.
In a second paper (Bianco et al. in preparation) we will present a deeper analysis of the sample, integrate it with data from the literature, discuss statistical differences in the photometry and colors of different stripped SN subtypes, and derive constraints on their progenitors.

The data set presented in this paper includes $\totoptphot$\ optical photometric
observations of $\nsnopt$\ SN (\secref{sec:optphot_sec}), and $\totirphot$\ NIR observations of $\nsnir$\ SN (\secref{sec:irphot_sec}). All photometry presented here is available in the online version of the journal, and at the CfA\footnote{\cfaweb} and NYU\footnote{\nyuweb} supernova group Web sites.
The CfA spectroscopic observations of $\nsnspec$\ of our SN are presented in M14.

\section{Discovery}\label{sec:disc_sec}

The nearby SN we monitored  at the CfA were discovered by a variety of
professional SN searches, as well as amateurs using modern CCD
technology. Systematic SN searches include LOSS, the Texas SN
Search\footnote{\url{http://www.grad40.as.utexas.edu/\~quimby/tss}}
\citep{Quimby06phd}, The Chilean Automatic Supernova Search \citep{2012MmSAI..83..388H}, and the Nearby SN Factory \citep{Aldering02}.  SN~2008D was discovered in the X-Ray with SWIFT (\citealt{Ofek_Cucchiara_Rau_et_al__2008}, in X-Ray observations of SN~2007uy, an unrelated stripped SN discovered in the same galaxy). The LOSS survey, and many amateur SN searches, observe in a relatively small field-of-view (FOV, $8.7\arcmin \times 8.7\arcmin$~for LOSS) and
recursively monitor the same galaxies. Typically, these surveys concentrate on well-known luminous galaxies (e.g.,
\citealt{nko_Treffers_Riess_Hu_Qiu_2001,lind_Challis_Jha_Kirshner_2005,aiolino_Petrosian_Turatto_2005}). Conversely, the Texas SN Search and the
Nearby SN Factory are rolling searches with a large FOV (2 and 3
square degrees, respectively) with thousands of galaxies searched impartially.

We list the objects in our SN sample and their basic discovery data in
\tabref{tab:discoverytable}. Our decision to monitor a particular
newly-discovered SN Ib, SN Ic, or SN IIb was broadly informed by three
considerations:  \emph{accessibility}, \mbox{declination $\ga-20\arcdeg$}, \emph{brightness}, $m <$ 18 mag for spectroscopic
observations, and $m <$ 20 mag for optical photometry, and 
\emph{age}, SN whose spectra indicated a young age were given higher
priority.
Of course, the latter two criteria are correlated, since older SN are
dimmer. 

FLWO undergoes a shutdown during the month of August every
year, due to Arizona monsoon season, thus, we have no monitoring
data for one month each year.

The \ninlit\ SN in our sample that were studied in the literature prior to this work (and to M14) are noted in \tabref{tab:discoverytable}. 
Twenty-two of these were previously studied \emph{individually} (i.e., not just as part of a survey) in the literature.
However the optical and/or NIR  light curves are published for only 18 of these 22 SN. 
In some cases (e.g., SN~2007ke -- \secref{sec:07ke}) the only photometric data published are in a single band, while our data always provides multi-band coverage, in a minimum of three photometric bands. The photometry for an additional 17 stripped SN that are part of  our sample appeared in D11 in $V$ and $R$ bands.

The host galaxy characteristics for all of our objects are listed in \tabref{tab:hostgal_table}.

\section{Photometry Data and Reduction}\label{sec:data_sec}

Our optical and NIR photometric campaigns are described in detail below, and elsewhere \citep{2009ApJ...700..331H, ind_Brown_Caldwell_et_al__2012, Falco_Szentgyorgyi_et_al__2008, Friedman14}. We paid particular attention 
to removing galaxy contamination, as contaminating host galaxy light may affect both the estimates of the peak brightness of the SN and its decline rate \citep{1991AJ....101.1281B}. The optical sample is produced from template-subtracted images in all but 6 cases, where the SN is well removed from the host galaxy (\secref{sec:optphot_sec}). Thus $\sim90\%$ of our optical sample of \nsntot\ stripped SN have photometry based on template-subtracted images.
 Similarly, for 80\% of our objects with NIR coverage, NIR photometry is derived from template-subtracted images: 
all but 5 objects out of \nsnir. 

Photometry
in both the natural and standard system is available
in the supplementary material of this paper, as well as through
the CfA Web site.\footnote{\cfaweb}

Below we describe the photometry acquisition for both optical and NIR  photometry, as well as the image and photometric reduction pipelines.

\subsection{Optical Photometric Observations and Reductions}\label{sec:optphot_sec}

All optical photometry presented in this paper was obtained  with the FLWO 1.2m telescope during the CfA3 \citep{2009ApJ...700..331H} and CfA4 campaigns \citep{ind_Brown_Caldwell_et_al__2012}.
Three different cameras were used to acquire the photometry: the 4Shooter 2$\times$2 CCD
mosaic (for data before 2004 September), the Minicam CCD mosaic camera
(2004 September until 2005 July), and the Keplercam CCD mosaic Camera
(after 2005 August). All cameras are thinned,
back-illuminated CCDs, mounted at the f/8 Cassegrain focus of the
1.2m telescope. All $UBV$ photometry is obtained in Johnson $UBV$, with $B$ and $V$ Harris filters. At redder wavelengths, observations were conducted with the 4Shooter 2$\times$2 CCD
mosaic in Johnson $RI$  band-passes, with the Harris filter set, and after 2004 September with 
Sloan Digital Sky Survey (SDSS) $r'i'$ filters \citep{ek_Ichikawa_Lupton_et_al__2007,gensen_Uomoto_Gunn_et_al__2002}. In addition in 2009 January the Johnson $U$ filter broke and was replaced by an SDSS $u'$ filter. Two objects in our survey, SN~2009iz and SN~2009jf, have $u'$ data. The
typical FWHM in our data falls between 1.5\arcsec~and 3\arcsec~, with the larger values typically found in the CfA4 survey. 
To provide prompt and dense sampling, the SN were observed by observers at the telescopes for other programs, and supplemented by photometric observations on our scheduled nights.

The optical photometry presented here was produced at the same time and in the same way as the CfA3 and CfA4
SN Ia samples. The
detailed operations of the optical photometric pipeline are discussed in~\citet{2009ApJ...700..331H}, and~\citet{ind_Brown_Caldwell_et_al__2012} respectively.
In brief: we employed differential photometry by measuring the
brightness of the SN with respect to a set of comparison stars (ranging from a few to dozens) in the SN field.
We employed the photometry pipeline of the SuperMACHO and ESSENCE collaborations (see \citealt{rest05} and
\citealt{bbs_Suntzeff_Foley_et_al__2007} for details), adapted for the 1.2m FLWO.

\begin{figure}[tb]
\begin{center}
\includegraphics[width=0.7\columnwidth]{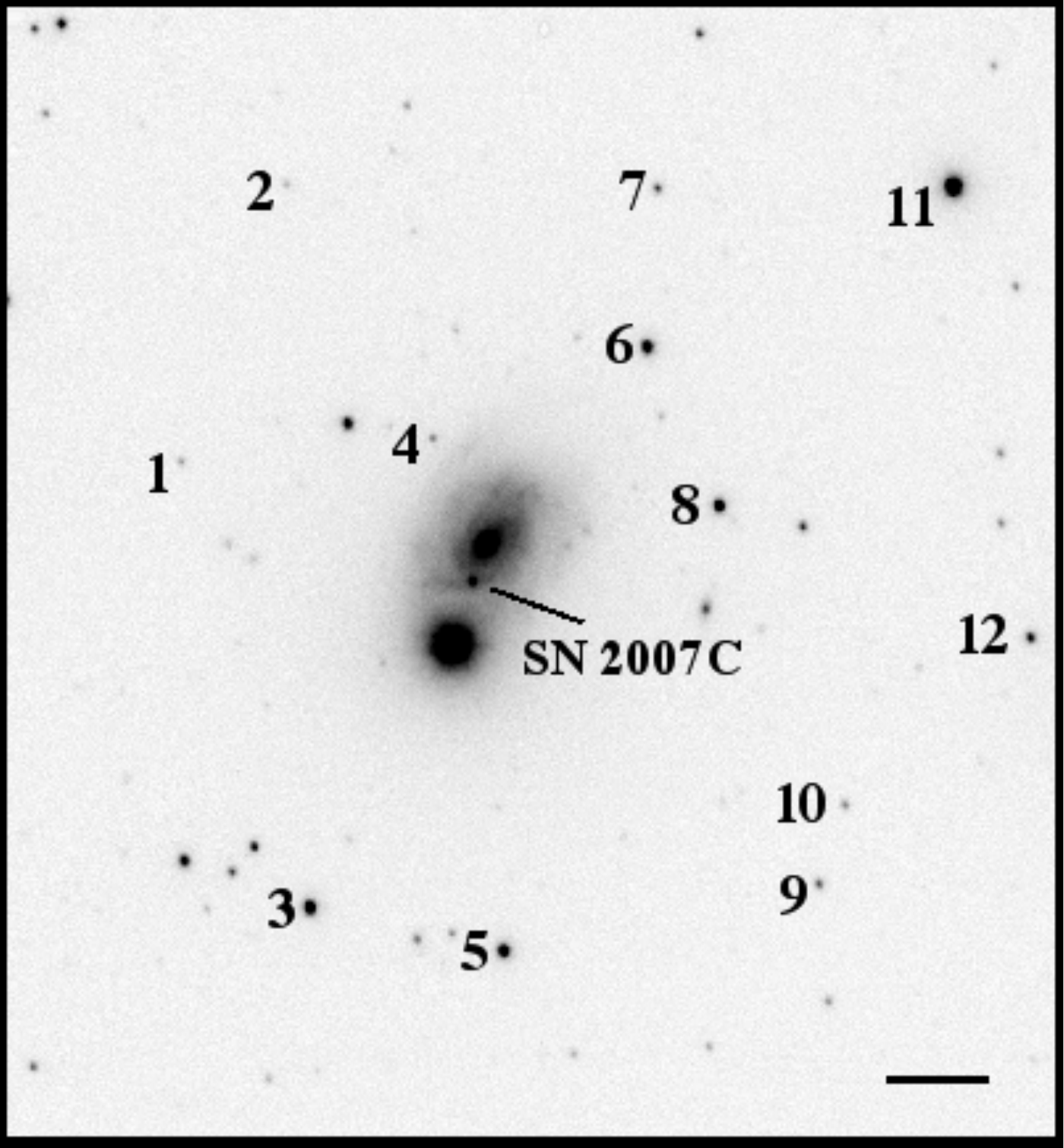}
\caption{Example supernova finding chart.  This image of SN~2007C was collected with Keplercam on the FLWO 1.2-m telescope (\url{http://linmax.sao.arizona.edu/FLWO/48/kep.primer.html})
 on 2007 January 21. The comparison stars are indicated with numbers from 1 to 12. 
North is top and East is to the left. The bar in the lower right corner indicates one arcminute.}
\label{fig:fchart}
\end{center}
\end{figure}

A finding chart is shown in \figref{fig:fchart}, with the field comparison stars marked. Comparison stars for each SN are available on the web\footnote{\cfaweb}, and the photometry for the comparison stars used to produce the optical light curve of SN~2005hg is shown, as an example, in \tabref{tab:compstar_2005hg}. The comparison stars were calibrated on photometric nights by observing standard stars from Landolt~\citep{Landolt_1992} and
\citet{gensen_Uomoto_Gunn_et_al__2002}. Aperture photometry in IRAF\footnote{IRAF (Image Reduction and Analysis Facility) is
distributed by the National Optical Astronomy Observatories, which are operated by the Association of Universities for Research in Astronomy, Inc., under cooperative agreement with the National Science
Foundation.} was used for this calibration. 

Color terms are obtained from the standard stars. The implicit color term equations have the following form: for Keplercam-chip2/Sloan, for example, \mbox{$(v-r)=1.0458(V-r')~+~$constant}. 
For the \ubvri\ filters, the lowercase/uppercase
letters in the color terms refer to instrumental/standard
magnitudes. For the $u'r'i'$ filters, the  lowercase letters refer to the instrumental magnitudes, whereas the primed lower case letters refer to the standard
magnitudes. Average color terms for each
setup used in our optical sample, along with the internal uncertainties
in the mean, are available in the supplementary material, as well as online.\footnote{\cfaweb}

Throughout the survey, five different sets of color terms were used, corresponding to four different camera/filter setups (4Shooter 2$\times$2 -- $\ubvri$, Minicam -- $\sloanubvri$, Keplercam -- $\sloanubvri$, and Keplercam --  $\sloanupbvri$), slight modifications to the photometric pipeline between CfA3 and CfA4 (all
4Shooter, Minicam and much of the Keplercam data before 2009 was processed during
CfA3 while some of the data before 2009 and all of it afterwards was processed during CfA4), and lastly, changes in the instrument transmission observed in mid 2009. 
For each light curve made available online, the instrument setup and pipeline used for the reduction are indicated in the file header as 4sh/$\ubvri$, mini/$\sloanubvri$, CfA3kep/$\sloanubvri$, CfA4kep1/$\sloanubvri$, and CfA4kep2/$\sloanupbvri$, respectively. Since no data in Johnson $U$ band was collected in 2009, all our $\sloanupbvri$ is to be processed with the CfA4kep2/$\sloanupbvri$ color-terms.
Anyone wishing to use the natural system
passbands must ensure that the proper passband is used to correct the photometry.  

 With the exception of 6 objects that are well removed from the host galaxy (SN~2002ap, SN~2004aw, SN~2006gi, SN~2007ce, SN~2007ru, and SN~2008aq), we derive photometric measurements from template-subtracted images (see \citealt{gensen_Uomoto_Gunn_et_al__2002}) using the robust algorithm of Alard \& Lupton
\citep{Alard_Lupton_1998,Alard_2000}. The template images
of SN host galaxies were obtained  under optimal seeing conditions, after the SN
had faded sufficiently, usually 6 months to 1 year after the end of the SN
observing campaign.  In \tabref{tab:hostgal_table} we report the characteristics of the host galaxies for all objects in our sample.

DoPHOT PSF photometry~\citep{Schechter_Mateo_Saha_1993} was used to measure the flux of the SN and its comparison stars.
  The majority of the stripped SN photometry in this work was produced during the CfA3 campaign and used only one host-galaxy image for host subtraction.  However, the CfA4 campaign used multiple host-galaxy images where possible, and for stripped SN produced during CfA4 we use the median photometry pipeline uncertainty as the uncertainty for each light curve point.  The CfA4 SN~Ia uncertainties \citep{ind_Brown_Caldwell_et_al__2012} also added, in quadrature, the standard deviation of the photometry values from the multiple host-image subtractions for a given point to produce the total uncertainty.  However, this overestimates the uncertainty \citep{2013arXiv1310.3824S}. In order to maintain consistency with the CfA3-era stripped SN,  we present the CfA4 data without adding the standard deviation to the CfA4-era uncertainties.   
The optical photometry of the \nsnopt\ SN is available for download\footnote{\cfaweb}
 and in the supplementary material for this paper. 

Optical CfA photometry of some of the SN listed in this paper has been previously published: SN~2005bf~\citep{jaz_Hicken_Challis_et_al__2005},  SN~2006aj/GRB060218~\citep{Diamond-Stanic_Hao_et_al__2006}, and SN~2008D~\citep{Kirshner_Kocevski_et_al__2009}. The optical data previously published for  SN~2005bf  were not based on template-subtracted images. Although  SN~2005bf  is well removed from its host, so host contamination was not significant, here we present the template-subtracted photometry, produced with the standard CfA photometric pipeline. Thus the light curves presented here for SN~2005bf supersede those previously published. 

\begin{figure}[tb]
\begin{center}
\centerline{
\includegraphics[width=0.52\columnwidth]{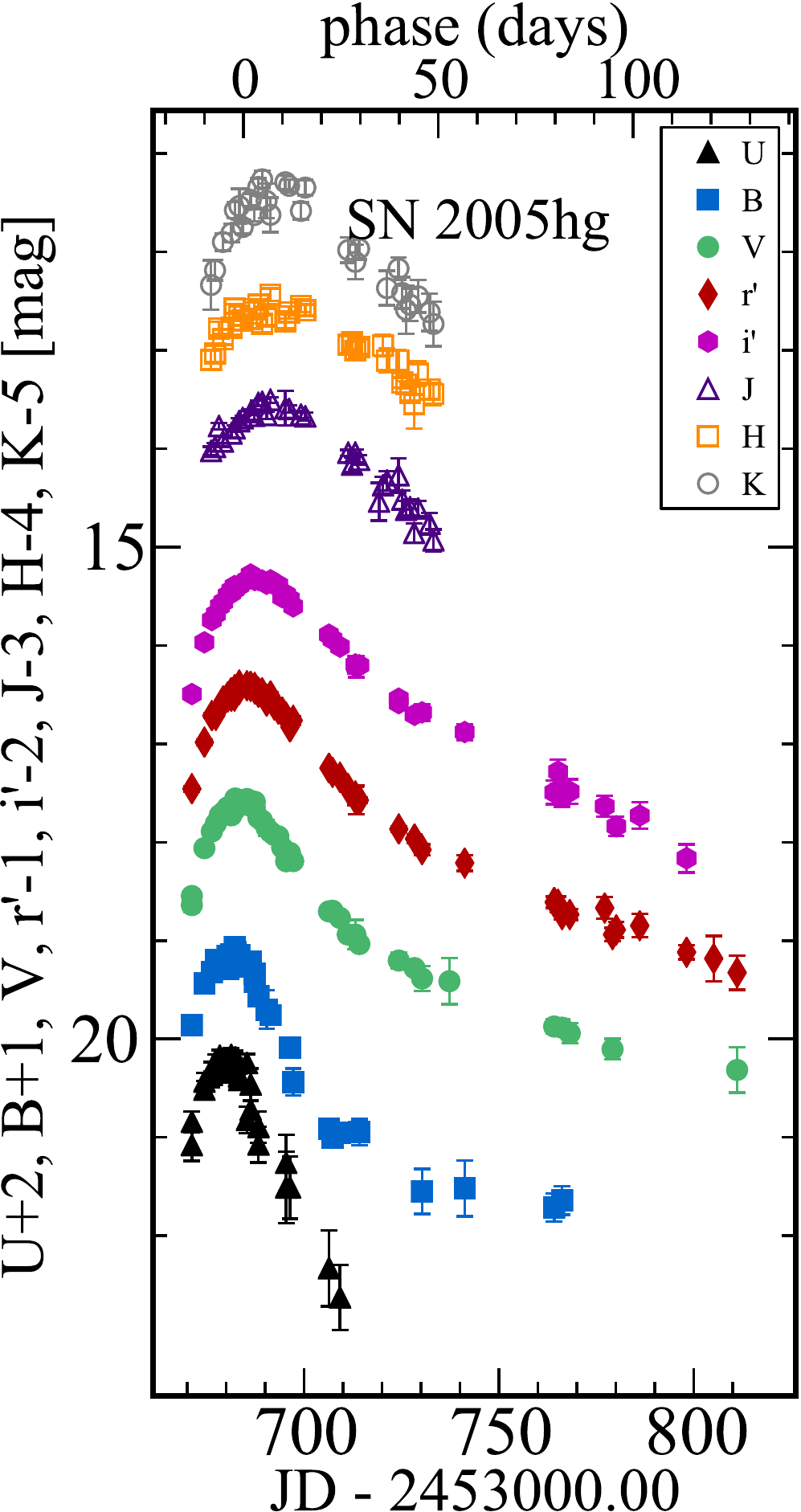}
\includegraphics[width=0.482\columnwidth]{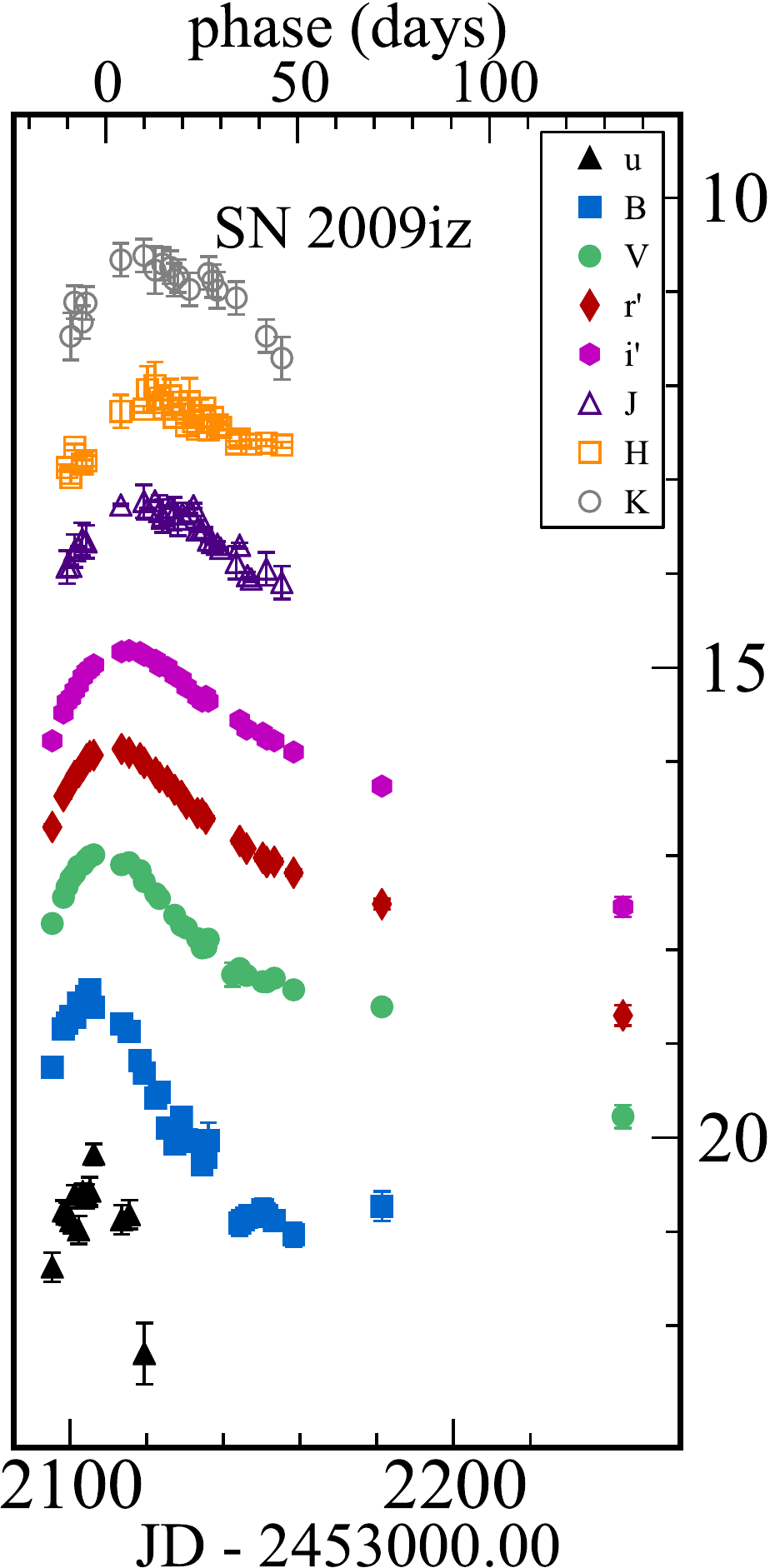}
}
\centerline{
\includegraphics[width=0.52\columnwidth]{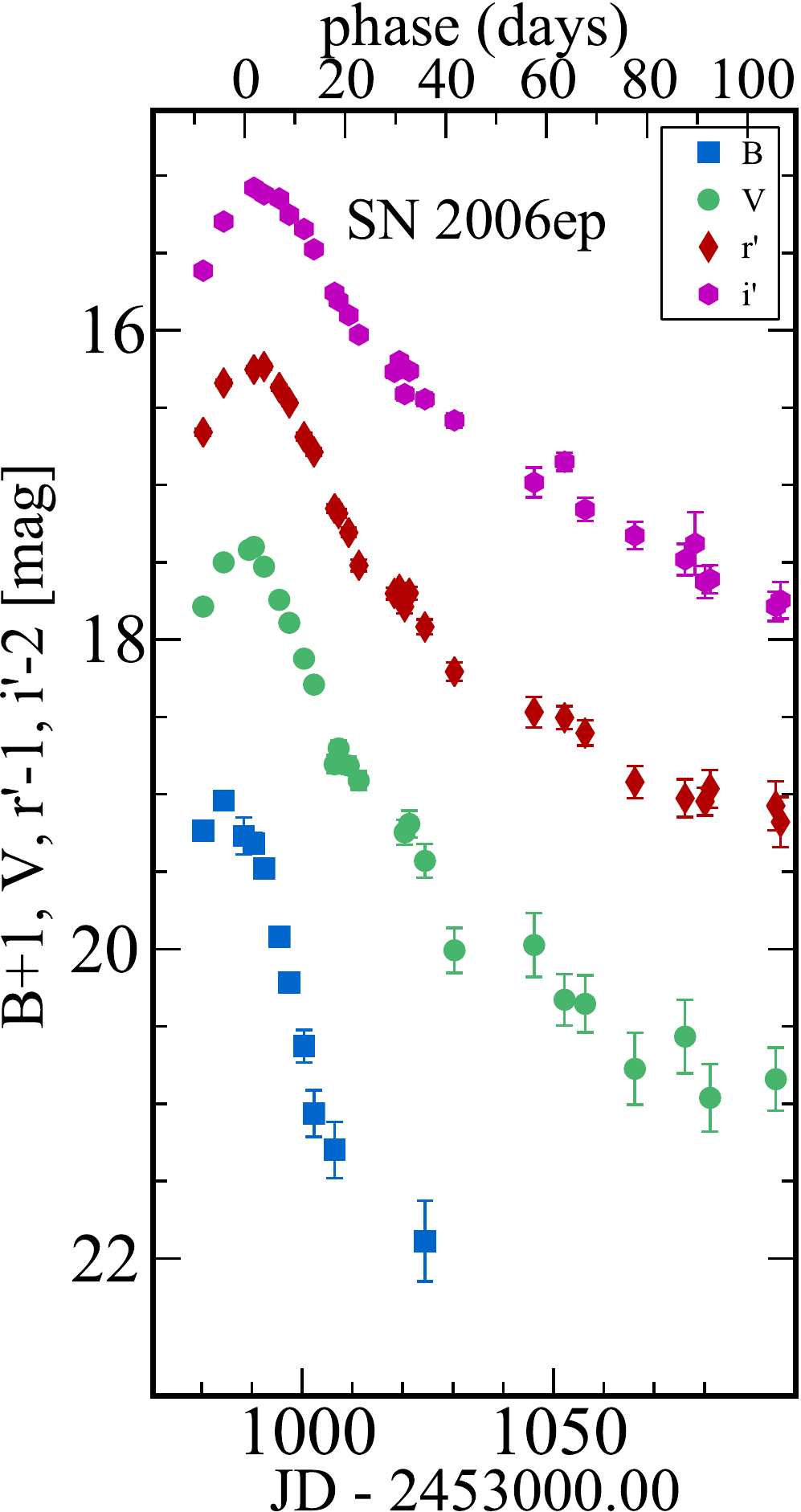}
\includegraphics[width=0.478\columnwidth]{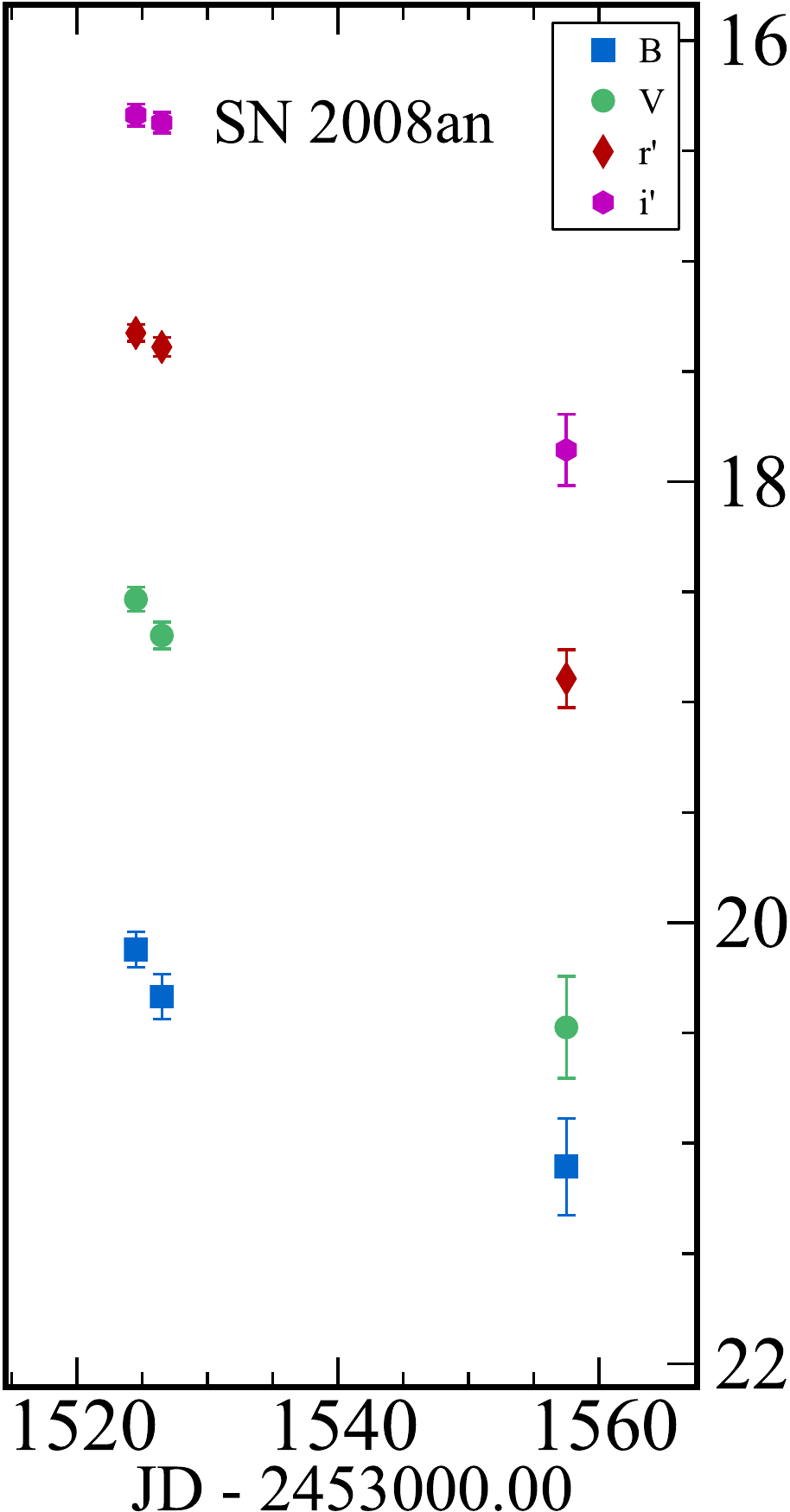}
}
\caption{Optical and NIR photometry of four SN chosen to represent the range of wavelength and cadence coverage in our sample.  The multi-color light curves in $U$ (black triangles), $B$ (blue squares)\
, $V$ (green circles),
 $r'$ (red diamonds), $i'$ (magenta hexagons), $J$ (purple empty triangles), $H$ (orange empty squares), and $K_s$ (gray empty circles) light curves are shown with offsets as indicated on the $y$-axis. }\label{fig:optphot}
\end{center}
\end{figure}

A sample CfA SN light curve is shown in \tabref{tab:snoptphot}, and the photometry for four objects, spanning the best and worst sampling quality, is shown in \figref{fig:optphot}. Note that for SN~2005hg, SN~2009iz, and SN~2006ep, where the epoch of maximum $V$-band brightness is known (see \secref{sec:sample_stat}), the epochs are expressed both as JD (bottom $x$-axis)  and as days since/to $V$-band peak (top $x$-axis). However, the epoch of maximum $V$-band brightness is \emph{not} known for SN~2008an.  Plots for all SN are available online.\footnote{\nyuweb}

\subsection{Near-infrared Photometry}\label{sec:irphot_sec}

For 25 SN in our sample, we obtained near infrared (NIR)
photometry with the fully automated 1.3-m Peters
Automated Infrared Telescope (PAIRITEL)\footnote{\url{http://www.pairitel.org/}} 
located at FLWO. 
PAIRITEL is a refurbishment of the 2MASS North telescope outfitted with the 2MASS South camera \citep{apps_Chester_Elias_et_al__2006} 
and is the first fully robotic and dedicated IR imaging system for the follow-up of transients \citep{Bloom06}.
The automation of PAIRITEL has enabled NIR SN with simultaneous $J$-, $H$- and $K_s$-band observations and nearly nightly cadence allows for densely
sampled PAIRITEL NIR SN light curves, from as many as $\sim$ 10 days before $V$-band maximum
brightness to $\sim$ 150 days past maximum.
PAIRITEL SN Ia data are published in \citet{Falco_Szentgyorgyi_et_al__2008}, \citet{Friedman12phd}, and \citet{Friedman14} will present the CfAIR2 sample of $\sim 100$ $JHK_s$ SN Ia light curves.  

PAIRITEL data for individual stripped-envelope SN have been published in \citet{jaz_Hicken_Challis_et_al__2005}, \citet{er_Skrutskie_Szentgyorgyi_2007}, \citet{Kirshner_Kocevski_et_al__2009}, \citet{Marion13}, and \cite{ock_Foley_Kirshner_et_al__2013}. 

PAIRITEL $J$-, $H$- and $K_s$-band images were acquired
simultaneously with the three NICMOS3 arrays in double correlated
reads with individual exposure times of 7.8 seconds. Individual images were dithered every fourth
exposure in order to remove bad pixels and aid subtraction of the bright NIR
sky. Each image consists of a 256$\times$256 array with a
plate scale of \mbox{$2\arcsec$~pixel$^{-1}$}, yielding an individual FOV of
\mbox{$8.5 \arcmin \times 8.5 \arcmin $}. 

Sky subtraction is a crucial step in NIR image processing. 
The PAIRITEL image reduction pipeline software \citep{Bloom06,Falco_Szentgyorgyi_et_al__2008,Friedman12phd} performed sky subtraction before cross-correlating, stacking and sub-sampling the processed images in order to produce the final, Nyquist-sampled image, with an effective pixel scale of \mbox{1$\arcsec$ pixel$^{-1}$}. 

The PAIRITEL imager does not have a shutter, thus independent determination of the dark current is impossible. 
For all SN, the sky+dark values for a given raw image were determined using a star-masked, pixel-by-pixel robust average through a temporal stack of unregistered raw images, which included removing the highest and lowest pixel values in the stack.  The temporal range of the raw image stack was set to $\pm 5$ minutes around the raw science image, which implicitly assumes that sky+dark values are approximately constant on $\sim 10$ minute time scales.
This reconstructed sky+dark image was then subtracted from the corresponding raw science image. For some SN fields with large host galaxies filling a fraction of the final FOV greater than $\sim 1/5$, a pixel-by-pixel robust average through the image series can lead to biased sky+dark values due to excess galaxy light falling in those pixels. However, overall systematic effects are negligible, biasing photometry to be fainter by only $\sim 1-2$ hundredths of a magnitude. The same sky+dark procedure was applied to all SN fields, including those with large host galaxies.  The sky+dark subtracted dithered science images are then registered and combined into final mosaiced images with SWarp \citep{Bertin02}, with a FOV of $12 \arcmin \times 12 \arcmin$.

Collection time ranged between 1800-second and 5400-seconds including overhead; the effective exposure times for the final mosaiced images ranged between 10 and 20 minutes. The effective seeing generally fell between 2{\arcsec} and 2.5{\arcsec} FWHM. The typical 30-minute signal-to-noise-ratio (SNR=10)
sensitivity limits are $\sim$18, 17.5, and 17 mag for $J$, $H$, and $K_s$ respectively. For fainter
sources, 10$\sigma$ point source sensitivities of 19.4, 18.5, and 18 mag are achievable
with 1.5 hours of dithered imaging \citep{Bloom03}.  

Photometric data points are calculated using forced DoPHOT photometry at the best fit SN centroid position. Photometry is generated for each SN from both  un-subtracted and  template-subtracted mosaiced images, the latter produced using the ESSENCE pipeline~\citep{rest05}. Typically, a minimum of 3 template images were obtained for each SN, after the SN had faded below our detection
limit, 6 to 12 months after discovery. The template-subtracted light curves are created as a nightly weighted average of the photometry produced using different templates. The most reliable photometry is ultimately chosen by visual inspection of the un-subtracted and subtracted mosaiced images, as well as considering the scatter in the photometric measurement obtained by each method. For template-subtracted light curves, a combination of automated and visual inspection also allowed removal of individual bad subtractions and outlier data points arising from poor quality science or template images.   

For SN not embedded in the host galaxy nucleus, or with little host galaxy light at the SN position, forced DoPHOT photometry on the un-subtracted mosaics was sometimes of higher quality than the galaxy subtracted light curves. We include in our sample forced photometry NIR light curves from \emph{un-subtracted} images for the following objects: SN~2004gq,
SN~2005ek, 
SN~2007ce, 
SN~2007uy, and 
SN~2008hh.
All other NIR SN light curves included here used photometry on the template-subtracted images, including  SN~2006aj, and  SN~2008D, for which  PAIRITEL photometry is already published  in \citet{er_Skrutskie_Szentgyorgyi_2007} and  \citet{Modjaz07phd}, and \citet{Kirshner_Kocevski_et_al__2009}, respectively. A new light curve, generated from template-subtracted images, is presented here for SN~2005bf, and supersedes previously published PAIRITEL data in \citealt{jaz_Hicken_Challis_et_al__2005}. SN~2005ek is well separated from the host galaxy; the light curve presented here is not based on host subtracted images, however we present additional PAIRITEL data points, together with those already included in \citealt{ock_Foley_Kirshner_et_al__2013}, and originally published in~\citealt{Modjaz07phd}.

For each SN field, the SN brightness was
determined using differential photometry against reference field stars
in the 2MASS point source catalog \citep{Cutri03}. 
Each field had $\sim$10--90 2MASS stars (which achieved 10$\sigma$ point
source sensitivities of $J$=15.8~mag, $H$=15.1~mag, $K_s$=14.3~mag; \citealt{apps_Chester_Elias_et_al__2006}). 
No color-term corrections were
required since our natural system photometry is already on the 2MASS system. 
We
extensively tested the accuracy and precision of both the PAIRITEL
reduction and our photometry pipeline by comparing our photometry of
2MASS stars in the SN observations to that in the 2MASS
catalog. 
The difference between the two photometry values is consistent with
zero everywhere in the magnitude range $J$ = 12-18 mag. Thus, we conclude our photometry is well anchored in the 2MASS system. 
Note however, that the difference uncertainties are expected
to be correlated, since the 2MASS photometry values were used to
compute the zeropoint of each image in the first place. 
More details of the PAIRITEL image processing and photometric pipelines are presented in \citet{Modjaz07phd,Falco_Szentgyorgyi_et_al__2008,Friedman12phd}, and \citet{Friedman14}.

A sample CfA NIR SN light curve, for SN~2005hg, is shown in \tabref{tab:snnirphot}, and NIR photometry is shown in  \figref{fig:optphot} for SN~2005hg and SN~2009iz. 

\section{CfA stripped SN sample statistics}\label{sec:sample_stat}
\begin{figure}[tb]
\begin{center}
\includegraphics[width=0.9\columnwidth]{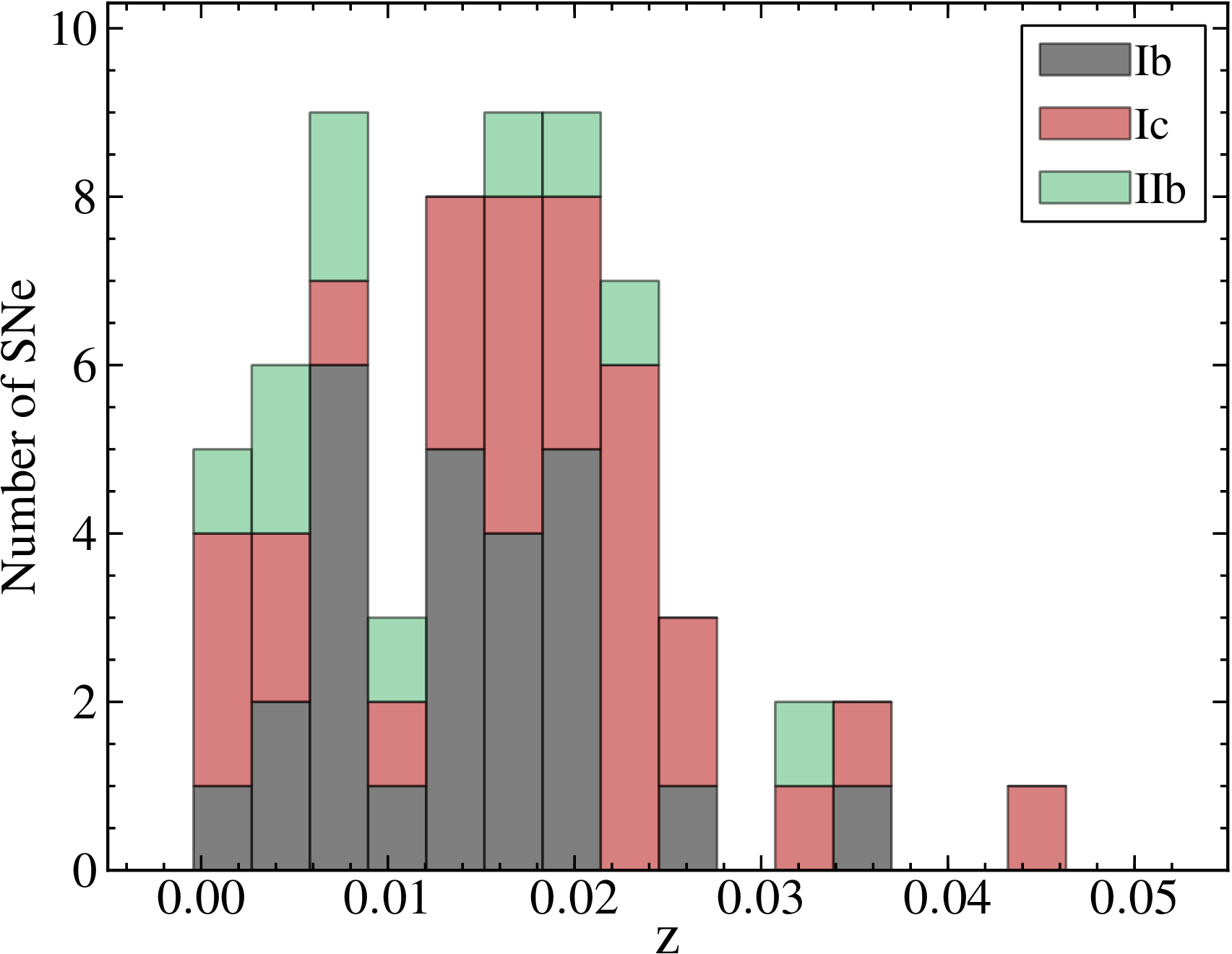}
\caption{Redshift distribution of SN in our sample.  Each color represents a SN type: SN Ib, Ic/Ic-bl, and IIb. The histograms are stacked: for each bin the total height of the bar indicates the total number of objects and the color segments represent the contribution of types within that redshift bin.}\label{fig:histz}
\end{center}
\end{figure}

\begin{figure}[tb]
\begin{center}
\includegraphics[width=0.9\columnwidth]{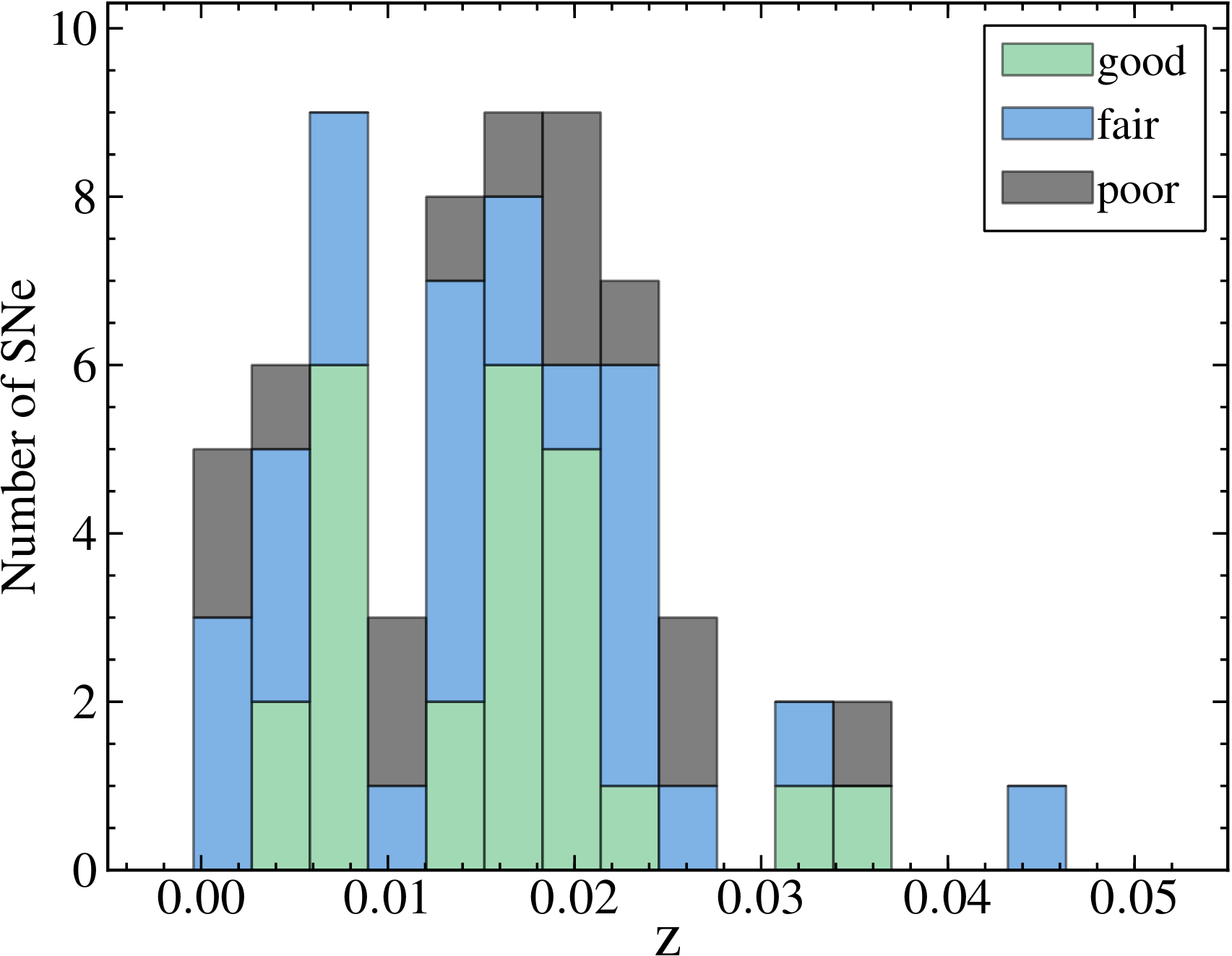}
\caption{Same as \figref{fig:histz}, but with the sample split by photometric quality, as described in \secref{sec:sample_stat} (good, fair, and poor light curve quality).}\label{fig:histz2}
\end{center}
\end{figure}

The sample of stripped SN we present contains a total of \nsntot\ SN observed between 2001 and 2009. The quality varies. The list below details our objects grouped based on their photometric quality. Several objects are then discussed in the later sections of this paper.
\begin{itemize}
\item
 Our best quality subset contains light curves in at least 4 bands, with data before and after the $V$ photometric peak. In this subset are multi-band light curves of 24 objects, 11 SN Ib, 5 SN Ic, 3 SN IIb, 3 SN Ic-bl, and two peculiar SN Ib (SN~2007uy, and SN~2009er). 
\item
An intermediate quality subsample contains multi-band light curves of 26 SN: 6 SN Ib, 8 SN Ic, 4 SN IIb, 4 SN Ic-bl, 1 SN Ic/Ic-bl (SN~2007iq), 2 SN Ib-n/IIb-n (Ib with narrow emission lines of H and He: SN~2005la and SN~2006jc), one Ca-rich Ib (SN~2007ke).
\item
A subset of 11 SN light curves for which we could not set good constraints on the date of maximum in any band, or which contains only a few epochs, or less than four photometric bands.
\item
Finally three objects (SN~2005ek, SN~2008ax, and SN~2008hh) have only NIR photometry.
\end{itemize}
The distribution of our objects in redshift is shown in \figref{fig:histz}, where we identify different SN types with different colors, and \figref{fig:histz2}, where the color indicates the quality of our photometry.

Most of the optical photometric measurements are in $B$, $V$, $r'$, and $i'$: 854, 1120, 1115, 1123 in each band respectively; 183 measurements were collected in $U$ band. In addition, the earliest objects observed within our program (SN~2001ej, SN~2001gd, SN~2002ap, SN~2003jd, SN~2004ao, and SN~2004aw) were imaged with $R$ and $I$ filter, in place of $r'$ and $i'$, with 68 and 60 points in each band for the 6 objects. In 2009 the Johnson $U$ filter was replaced by an SDSS $u'$ filter; two objects, SN~2009iz and SN~2009jf, have $u'$ band data, a total of 20 data points, 13 for SN~2009iz and 7 for SN~2009jf.
In the NIR we collected: 774 measurements in $J$, 738 in $H$, and 630 in $K_s$.
The photometry for each of our SN is made available as machine-readable tables in the supplementary material, on our Web site\footnote{\nyuweb} as plots, and in tabular form.\footnote{\cfaweb} In \tabref{tab:maxsUBV}, ~\ref{tab:maxsRI}, and  ~\ref{tab:maxsHJK} we present observational photometric characteristic for all objects in our sample: the epoch of maximum brightness, the peak  magnitude, and the decline rate in each filter, whenever it is possible to derive them.
Similarly to what is done for SN Ia, we measure the decline rate as \delm: the difference in magnitude between peak and 15 days after peak. We simply rely on a second-degree polynomial fit near the light curve peak to obtain these quantities.
Notice that these are presented as observational quantities: no $S$ or $K$-corrections are applied to compensate for the reddening effects of redshift, nor do we correct for dust extinction at this time. A more complete analysis of our photometric data will be presented in the companion paper (Bianco et al., in preparation), and such corrections will be discussed there. 
The maximum brightness, and the epoch of maximum, are measured as follows:
\begin{itemize}
\item
For each single-band light curve we select by eye a region around peak large enough to allow a quadratic fit (at least four points, typically several more) but small enough to follow a simple parabolic evolution. 
\item
A suite of $N$ Monte Carlo realizations is generated by drawing each data point from a Gaussian distribution centered on the  photometric data point, and with a standard deviation corresponding to the photometric error-bars. In each realization the boundaries of the region that is fit, particularly after peak where in most cases more photometric data points are available, are allowed to oscillate by including or removing up to three data points. The number $N$ of realizations depends on the number of data points $N_d$ used for that particular object: $N$ is the integer nearest to $N_d\log(N_d)^2$, but no smaller than 200:
$$N~=~\mathrm{argmin}\left({\mathrm{int}(N_d\log(N_d)^2),200}\right)$$
\item 
Each realization is fit with a second-degree polynomial. The epochs of maximum brightness and the corresponding magnitudes we report are the mean of the maximum epoch and magnitude distributions in the fit 
over the family of Monte Carlo realizations, and the errors are the corresponding standard deviations.
\end{itemize}

\begin{figure}[tb]
\begin{center}
\includegraphics[width=0.9\columnwidth]{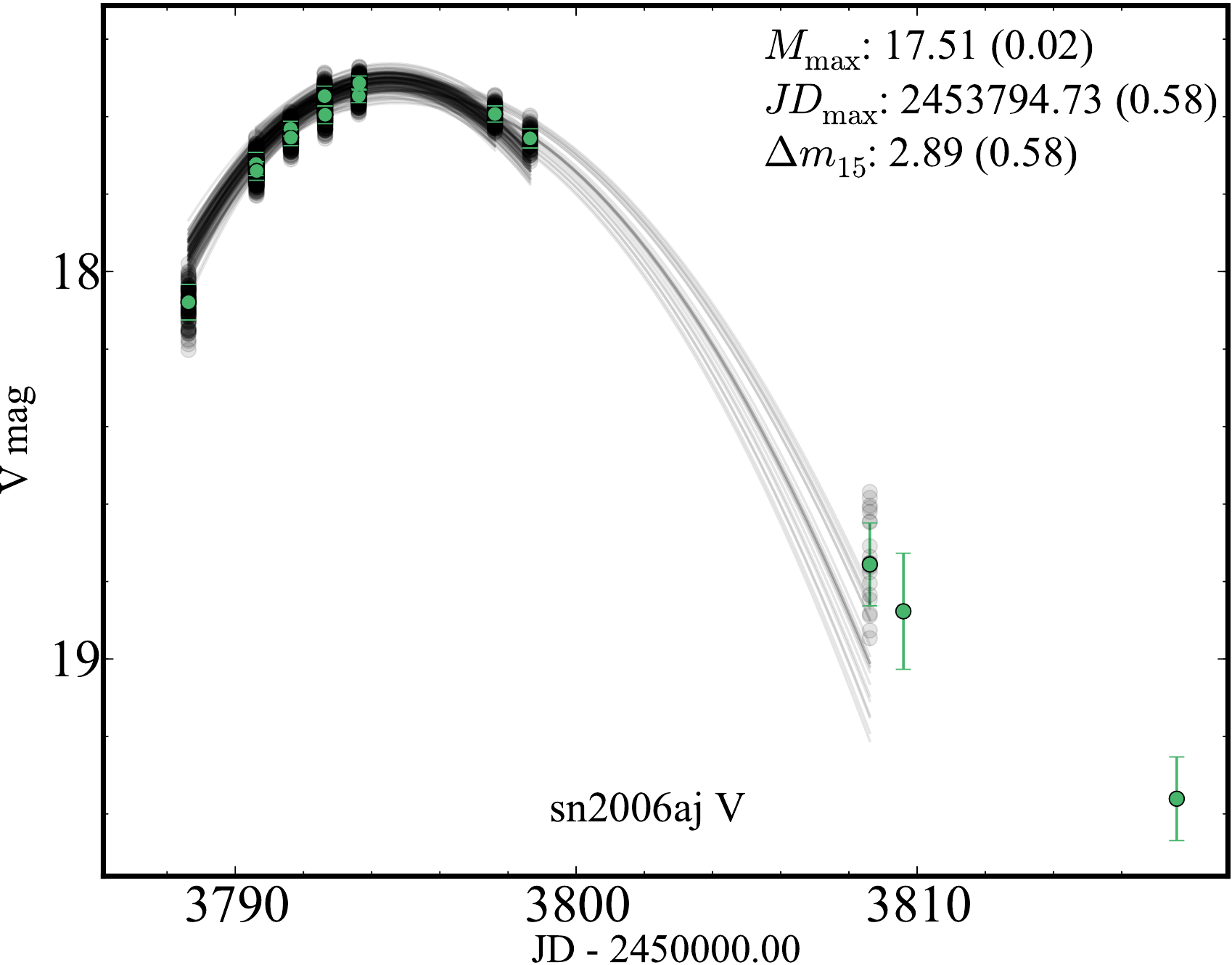}
\caption{Suite of Monte Carlo realization generated for SN~2006aj $V$-band near peak to determine the peak date and magnitude and associated uncertainties. In each realization a subset of data points is selected where one, two, or three data-points at the edges of the set may or may not be included. A synthetic photometric data set is then generated by drawing a data point (gray circle) for each epoch within a Gaussian distribution centered on the photometric datum (green circle), and with standard deviation equal to its error-bar. The synthetic photometry is fit with a parabola (gray line). The date of peak, the magnitude at peak, the \delm\ reported, and their related uncertainties, are calculated as the mean and standard deviation of the corresponding values across all realizations.}\label{fig:Vmax_mc}
\end{center}
\end{figure}

\figref{fig:Vmax_mc} shows the suite of Monte Carlo realization generated for SN~2006aj $V$-band near peak to determine the peak date and magnitude, and their uncertainties. Notice that, with a $\sim9$ day gap in the coverage starting about 4 days after peak, the determination of \maxep\ is indeed affected by the choice of boundaries to the region used for the parabola fit. By means of the Monte Carlo simulations this is reflected in a 0.61 days uncertainty. The \delm\ can be estimated as an extrapolation of the polynomial to 15 days. However, we only report this metric when it is sensible to do so: when data covers epochs near 15 days after maximum in the band considered, and the quadratic fit is consistent with these data. In the case of SN~2006aj, for example (\figref{fig:Vmax_mc}), these criteria are not fulfilled, and the \delm\ obtained through polynomial fitting, shown in the figure, is not reported in \tabref{tab:maxsUBV}. 

 In \tabref{tab:stats1} and  \tabref{tab:stats2}  we report the statistical differences we find across our sample in the date of maximum, and peak magnitude, compared to $V$ (helpful to estimate \maxep, which is usually the reference for spectral phases, even in absence of adequate $V$ coverage around peak).

\begin{deluxetable}{cccc}
\tablecaption{Peak epoch by band\tablenotemark{a}.}
\centering 
\tablehead{\colhead{Band} & \colhead{Weighted average} & \colhead{Median} & \colhead{Standard deviation} }
\startdata
 $U $  &  -1.2 &   -3.3 &2.1\\
 $B $  &  -2.3 &   -2.3 &1.3\\
 $R/r' $  & 1.8 &  1.5 &1.3\\
 $I/i' $  & 3.5 &  3.1 &1.5\\
 $J $  & 8.5 &  6.9 &3.3\\
 $H $  & 10.1 &  9.8 &4.3\\
 $K $  & 10.5 & 10.9 &5.0\\
\enddata 
\tablenotetext{a}{Difference in days between  V maximum epoch \maxep and peak epoch in each band. Negative values indicate the peak happens before \maxep}
\label{tab:stats1} 
\end{deluxetable}

\begin{deluxetable}{cccc}
\tablecaption{Magnitude at maximum brightness compared to $V_\mathrm{max}$}
\centering 
\tablehead{\colhead{Band} & \colhead{Weighted average} & \colhead{Median} & \colhead{Standard deviation} }
\startdata
 $U $  &  -0.13 &   -0.20  &0.27\\
 $B $  &  -0.70 &   -0.62 &0.16\\
 $R/r' $  & 0.21 &  0.19 &0.16\\
 $I/i' $  & 0.25 &  0.17 &0.32\\
 $J $  & 0.91 &  0.73 &0.71\\
 $H $  & 1.04 &  0.81 &0.81\\
 $K $  & 1.35 & 1.19 &0.87\\
\enddata 
\label{tab:stats2} 
\end{deluxetable}
\begin{figure*}[tb]
\begin{center}
\includegraphics[width=0.8\textwidth]{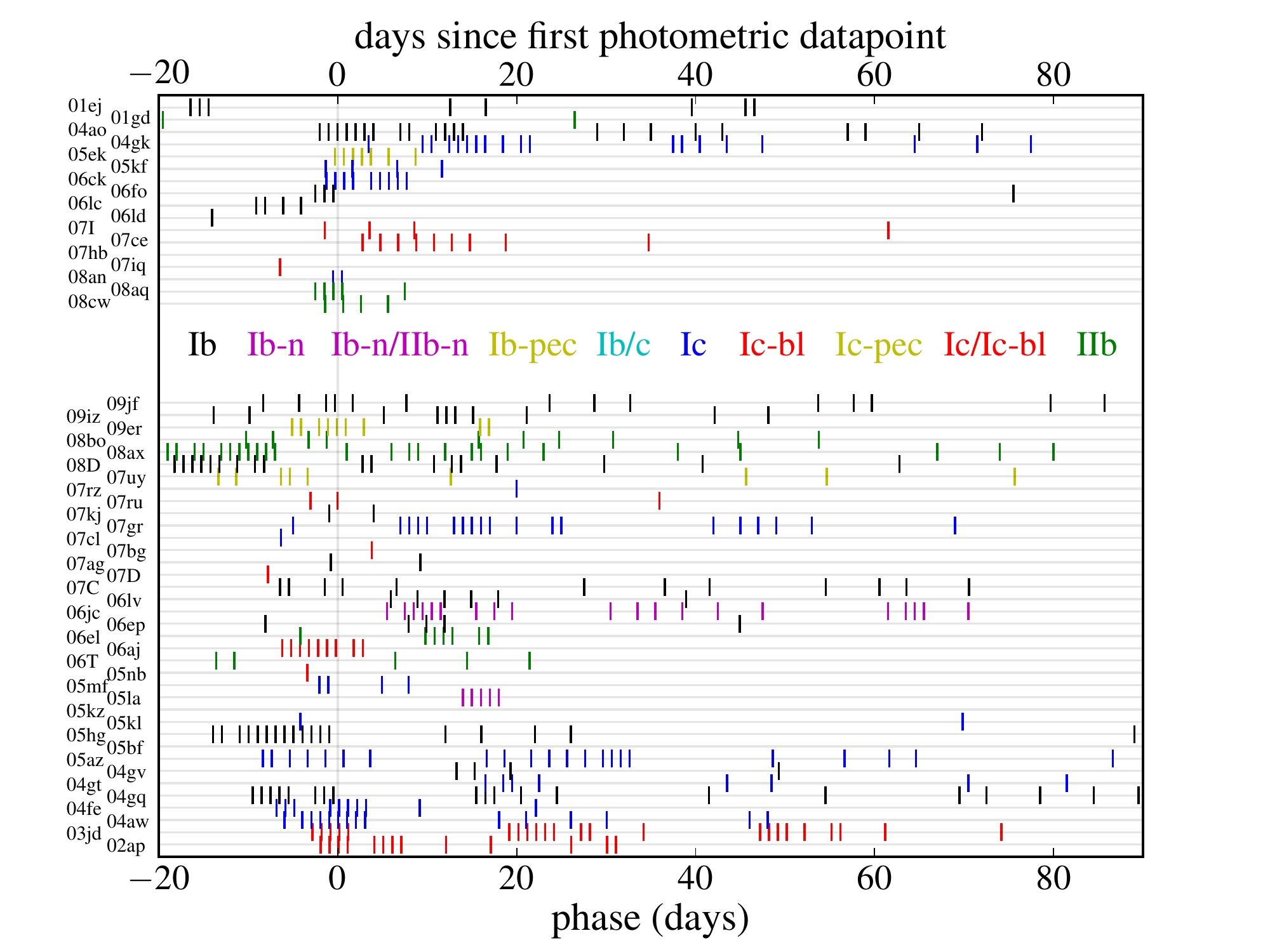}
\caption{Phases of spectroscopic observations for all SN in our sample for which CfA FLWO 1.5-m FAST spectra are available as published in M14. 
The SN classification is indicated by the color of the marks as defined by the inset list. 
For the top set of 17 SN the \maxep\ is not known and thus the phases of the spectra are defined as days since our first photometric measurement.
The phases for the bottom set of 37 SN are plotted with respect to the \maxep calculated as in \secref{sec:sample_stat}.}\label{fig:specep}
\end{center}
\end{figure*}

In \tabref{tab:hostgal_table} we report the characteristics of the host galaxies for all objects in our sample. We report the distance to the host galaxy (as heliocentric recession velocity), the absolute and apparent $B$ magnitude, when available, and distance modulus. These quantities are extracted from the HyperLEDA catalog\footnote{\url{http://www.leda.univ-lyon1.fr}} \citep{2003A&A...412...57P}. When not available in HyperLEDA, the NED catalog is used, and the cosmological parameters used, for consistency with HyperLEDA, are: $H_o = 70$~${\rm km~s^{\scriptscriptstyle -1}~Mpc^{\scriptscriptstyle -1}}$, $\Omega_m = 0.27$, and $\Omega_{\Lambda} = 0.73$

We report Galactic extinction ($E_{B-V}$) for each of our objects based on its sky coordinates, and on the most recent dust maps produced by \citet{2011ApJ...737..103S}. Note that these extinction maps deviate by about 10\% in high-extinction regions from the older, and commonly used \citet{1998ApJ...500..525S} maps. The photometry we provided is \emph{not} corrected for Galaxy, host, or cosmological extinction or reddening. 

\begin{figure}[th!]
\begin{center}
\centerline{
\includegraphics[width=0.85\columnwidth]{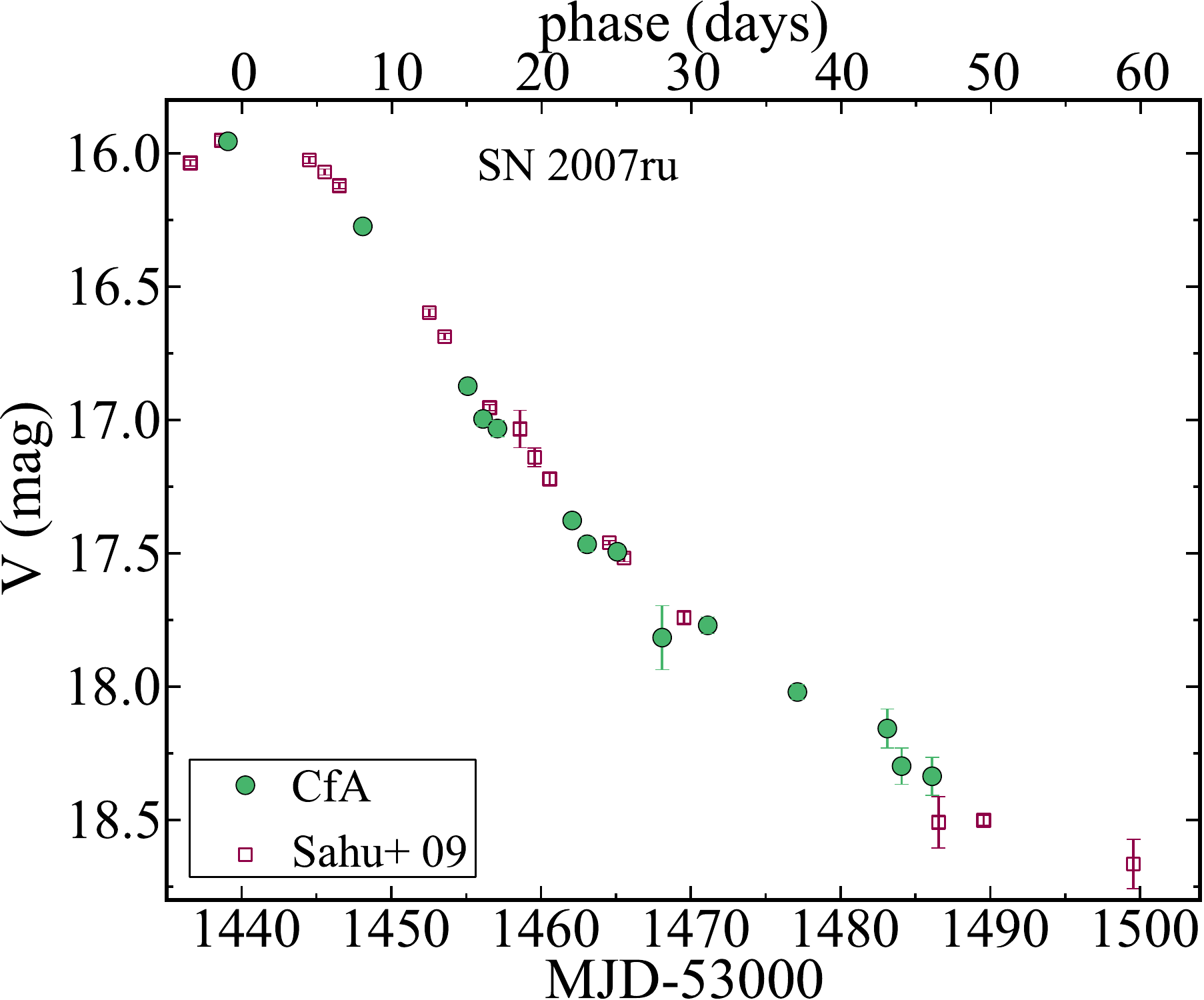}
}\centerline{\includegraphics[width=0.85\columnwidth]{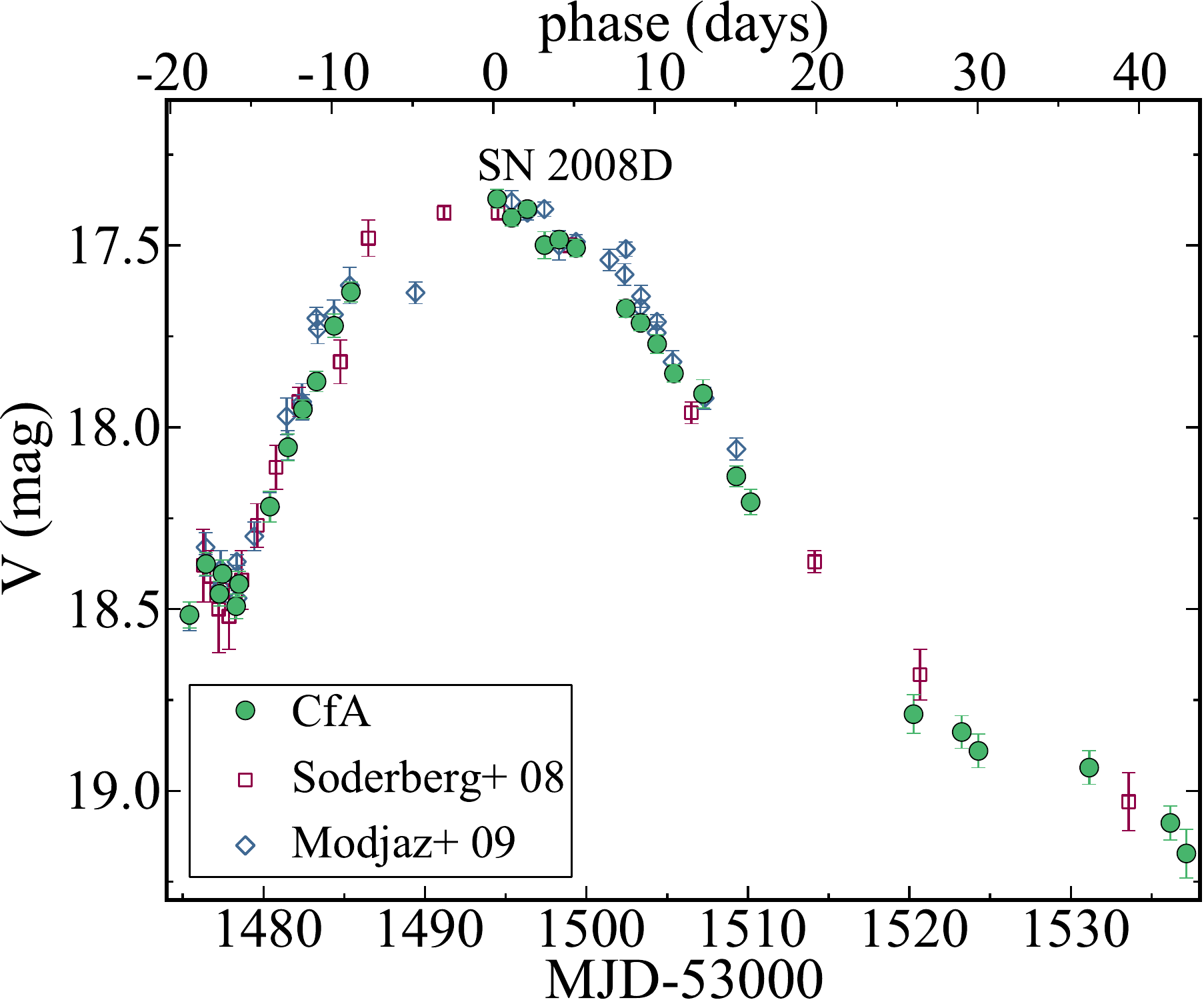}
}\centerline{\includegraphics[width=0.85\columnwidth]{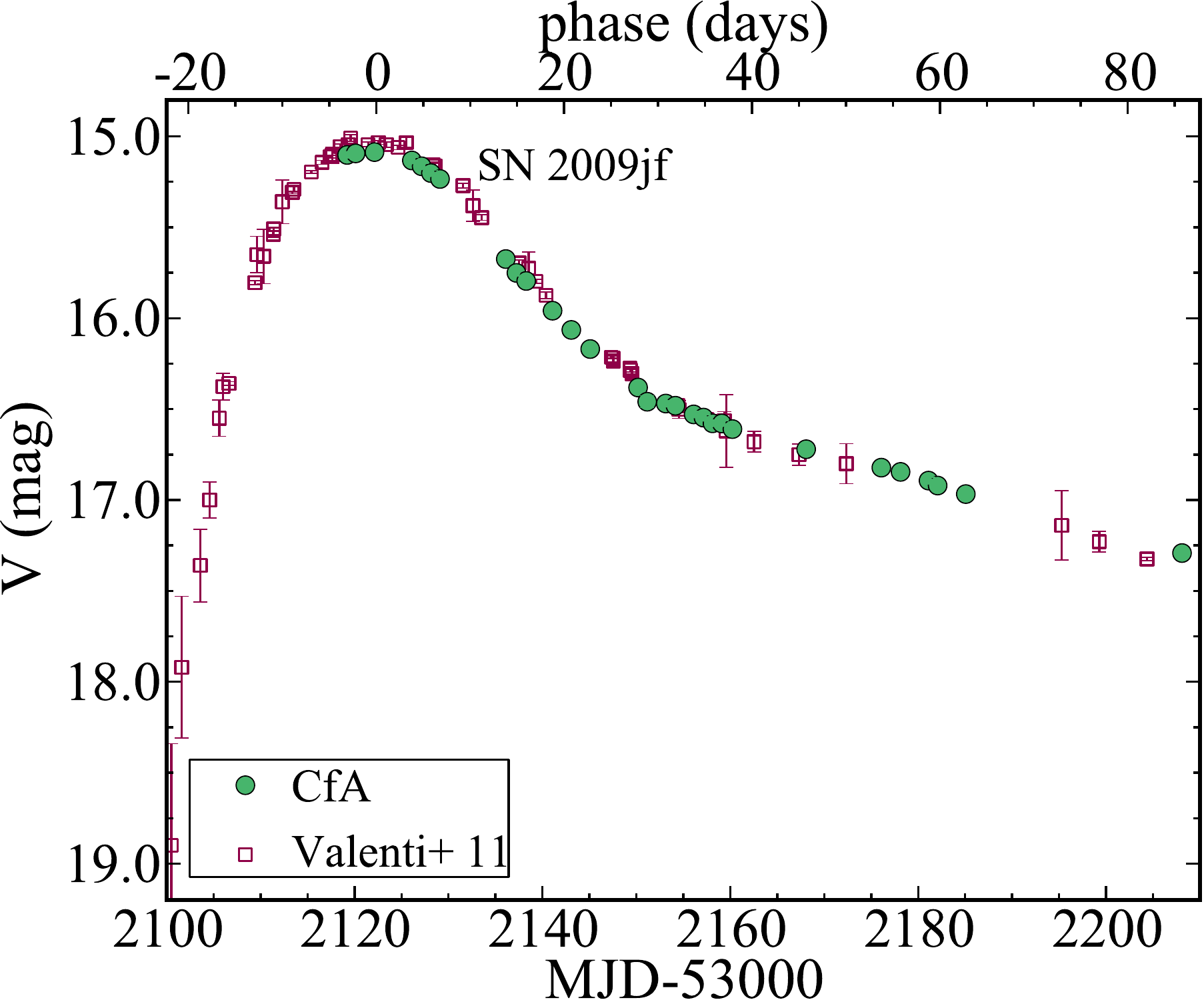}
}
\caption{CfA stripped SN $V$-band light curves compared with $V$-band light curves from the literature.  
 (Top) SN~2007ru, with data from \citet{2009ApJ...697..676S};
 (Middle) SN~2008D, including data from \citet{Ofek_Cucchiara_Rau_et_al__2008} and KAIT data from \citet{Kirshner_Kocevski_et_al__2009}; and 
 (Bottom) SN~2009jf with data from \citet{rello_Smartt_Ergon_et_al__2011}.  
Our photometry agrees with the literature photometry for these objects within the quoted errors. 
We have shown only $V$-band data here for clarity, but 
the photometry is similarly consistent in other bands.
}
\label{fig:litcomp}
\end{center}
\end{figure}

\begin{figure}[tb]
\begin{center}
\centerline{
\includegraphics[width=0.85\columnwidth]{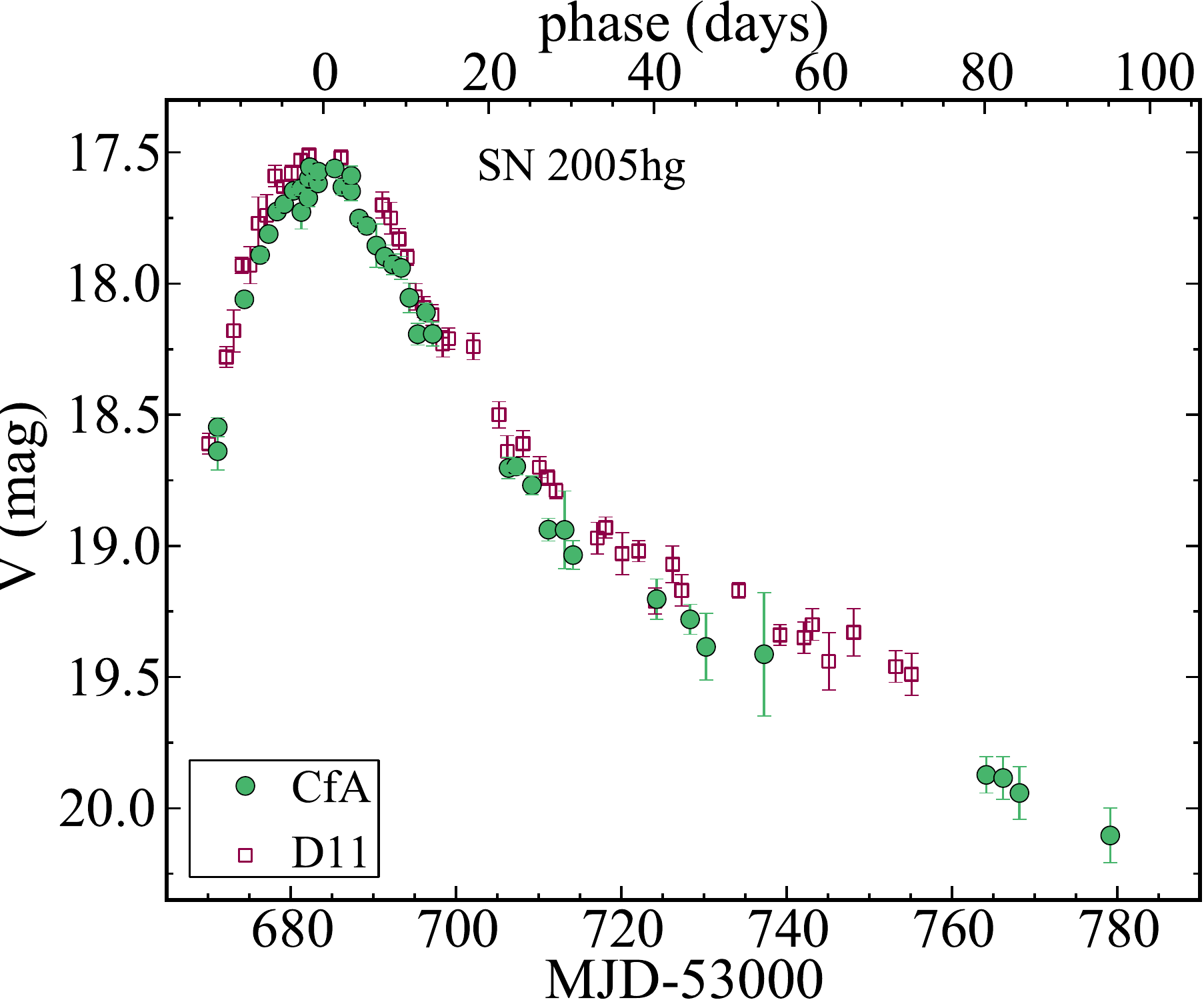}
}
\centerline{
\includegraphics[width=0.85\columnwidth]{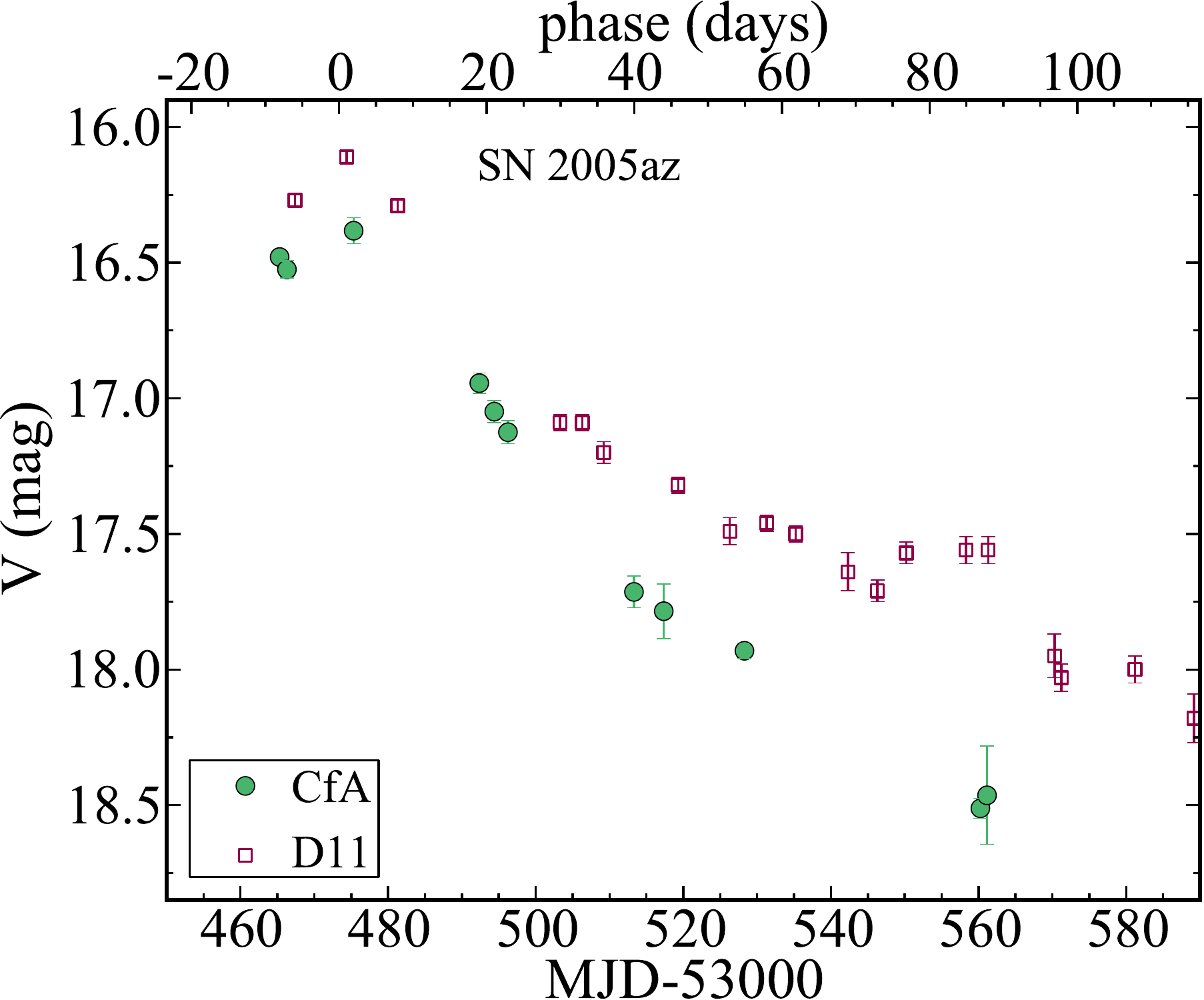}
}
\centerline{
\includegraphics[width=0.85\columnwidth]{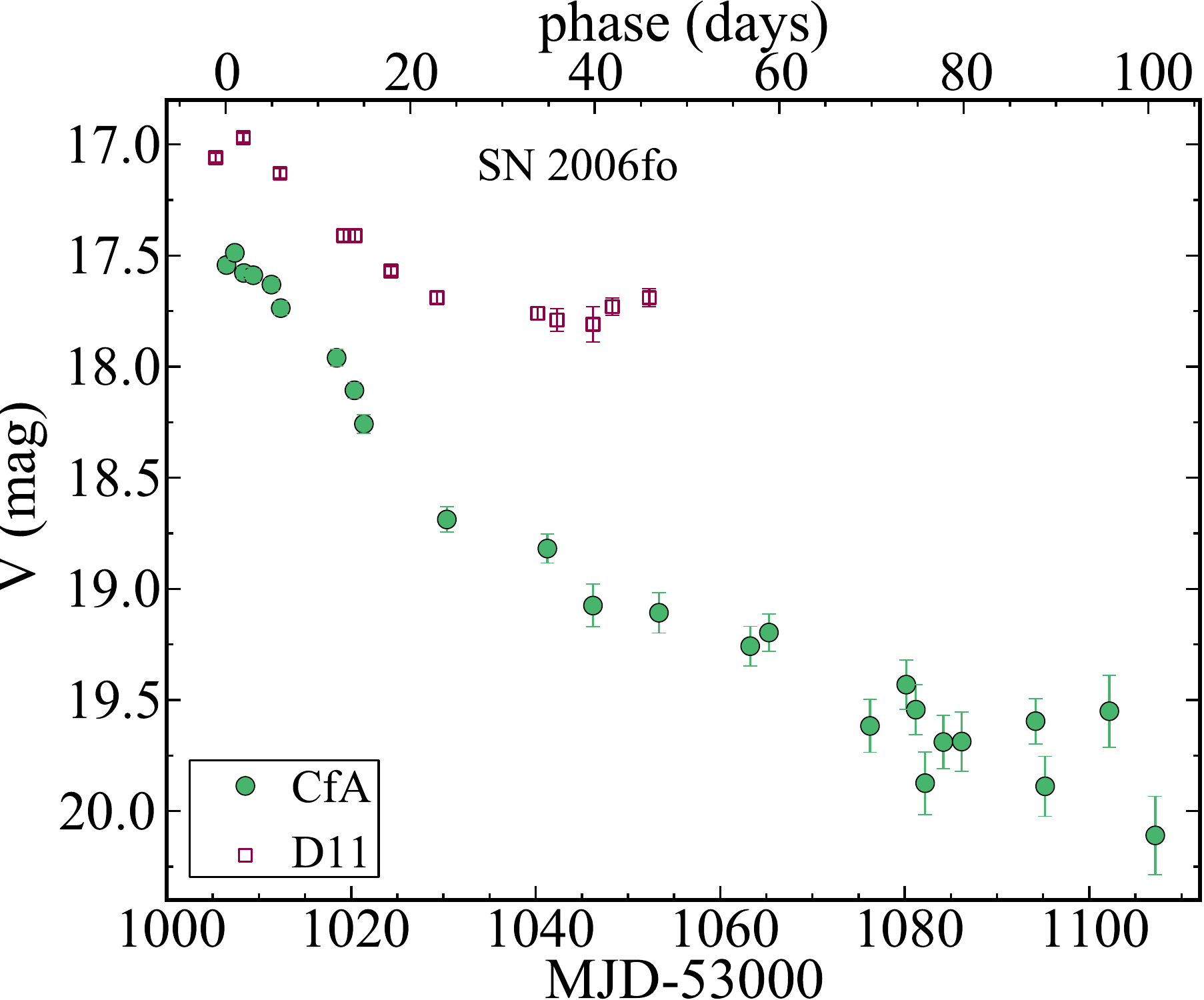}
}
\caption{$V$-band light curves from our survey and D11 for SN~2005hg (top), SN~2005az (middle), and SN~2006fo (bottom), respectively from the \emph{gold}, \emph{silver} and \emph{bronze} D11 subsamples (see text). SN~2005hg shows excellent agreement in $V$ (although an offset is present in $R$), while both SN~2005az and SN~2006fo show a significant photometric offset, with D11 brighter at all epochs, a behavior observed in the majority of the objects in common between the two samples. 
The discrepancy grows at later epochs, when the SN is fainter, which is consistent with the D11 photometry being brighter due to host galaxy contamination in the photometric measurements.  
Only $V$-band data is shown, as the $V$ standardized magnitudes are reported by both surveys in the same photometric system, but similar discrepancies are seen in $R$ after photometric transformations are applied.}
\label{fig:d11phot}
\end{center}
\end{figure}

Among the supernovae in this sample over 80\% of the objects have spectra collected within our SN program (M14): only SN~2005kz, SN~2006F, SN~2006ba, SN~2006bf, SN~2006cb, SN~2006gi, SN~2006ir, SN~2007aw, SN~2007ke, and SN~2009K 
do not have any spectral coverage obtained within our group. Spectral information for SN~2007ke exists in the literature, and it indicates that SN~2007ke is an unusual Ca-rich SN Ib \citep{Kasliwal13}. SN~2008hh does not have spectral coverage within the CfA sample and it only has NIR CfA photometric coverage. \figref{fig:specep} shows the epoch of all spectra obtained at FLWO for the objects in our sample. In the bottom portion of the plot all objects for which the epoch of maximum $V$ brightness (\maxep) is available are plotted against the bottom $x$-axis: the epoch of the spectra is expressed as days to (since) \maxep. In the top portion of the plot the 17 objects for which we have CfA spectra, but \maxep\  is not known (neither through our data nor in the literature) are plotted against the top $x$-axis, with the epoch expressed as day to (since) our first photometric measurement. Omitted from the plot are all spectra collected at epoch $\ga 90$ days, however our spectroscopic sample contains many nebular phase spectra (M14).

\section{Comparison with literature data}\label{sec:litdata}
\begin{figure}[tb]
\begin{center}
\includegraphics[width=1.0\columnwidth]{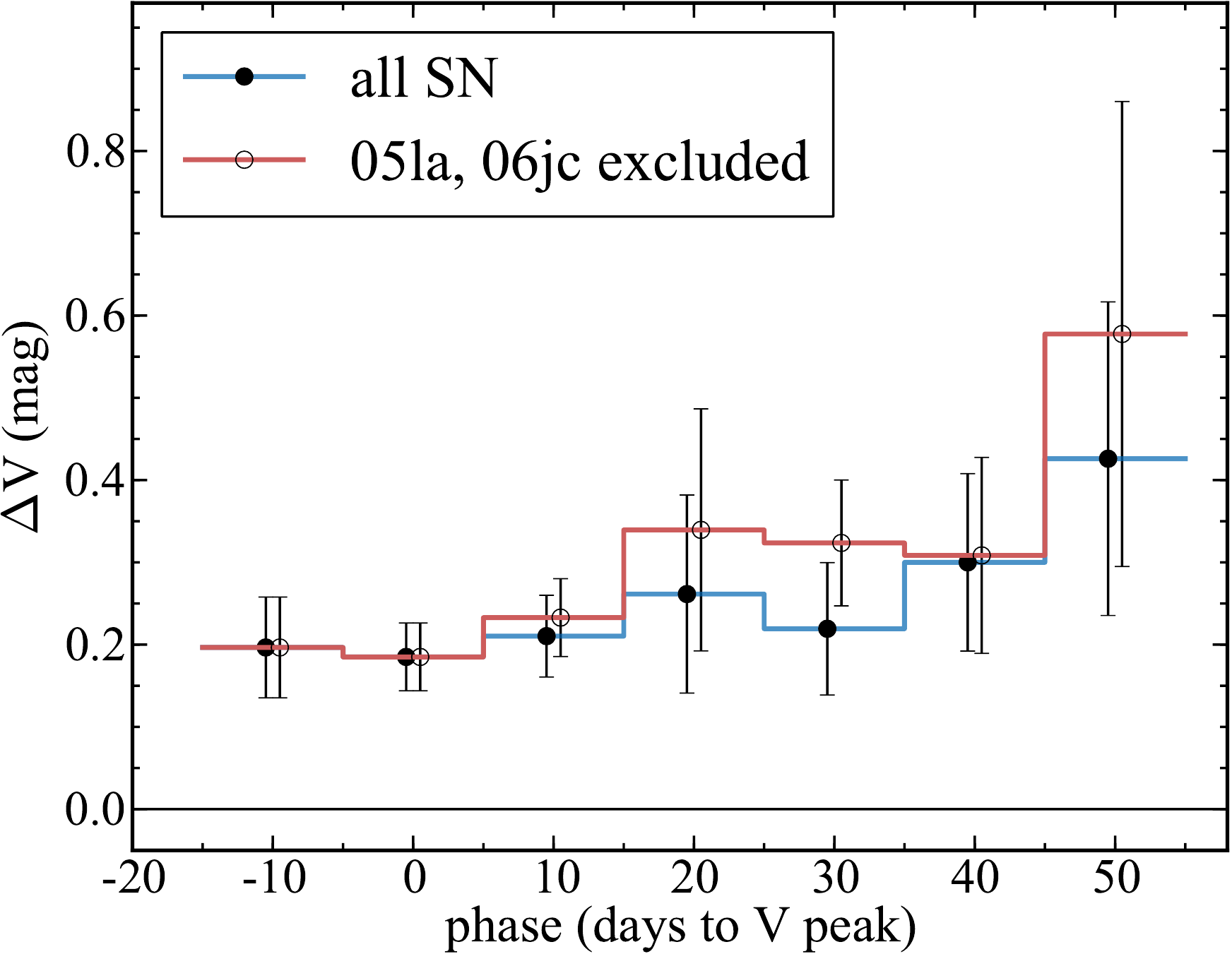}
\caption{$V$ magnitude difference between D11 and our (CfA) data ($\Delta{V}=V_{\rm CfA}-V_{\rm D11}$).  
For each light curve the data with the closest dates of observations between the surveys are used, and data are only included when the observations are separated by less than 5 days (10 days for epochs later than 45 days to peak). 
Error bars represent the error in the mean offset for each epoch ($\sigma/\sqrt{N}$).  The blue line (filled circles) includes all objects in common, while red line data (empty circles)  exclude SN~2005la and SN~2006jc (see text). 
The offset is minimum near peak magnitude, and increases as the SN get dimmer. 
Note that all values are positive, indicating that the D11 photometry is consistently, and significantly, brighter than our photometry at all epochs.}
\label{fig:d11comp}
\end{center}
\end{figure}

 \begin{figure}[t!]
\begin{center}
\centerline{
\includegraphics[width=0.97\columnwidth]{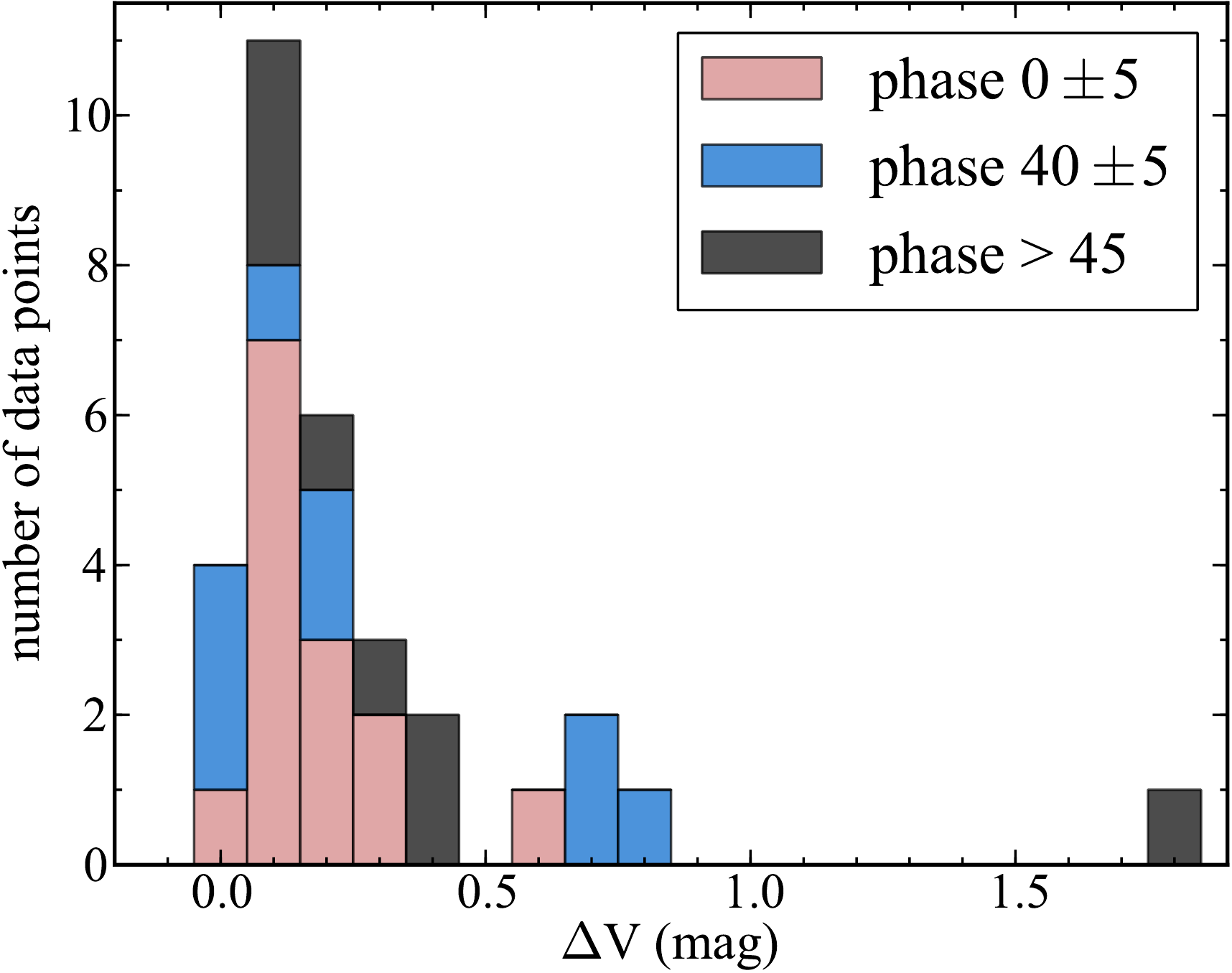}
}
\caption{
Distribution of the photometric discrepancy between the D11 data and our data (see \figref{fig:d11comp}) shown as a stacked histogram.
Different colors represent different epochs (blue: $0\pm5$ days; red: $40\pm5$ days; black: $>45$ days). All values are positive, indicating that the D11 photometry is consistently brighter than our photometry. The difference between D11 and our data is more prominently skewed toward high $\Delta{V}$ at late epochs than near peak.
The D11 light curves plateau at these late phases, an effect which we attribute to residual galaxy contamination in the D11 photometric analysis. }

\label{fig:d11comp2}
\end{center}
\end{figure}

\begin{figure}[t!]
\begin{center}
\centerline{
\includegraphics[width=0.97\columnwidth]{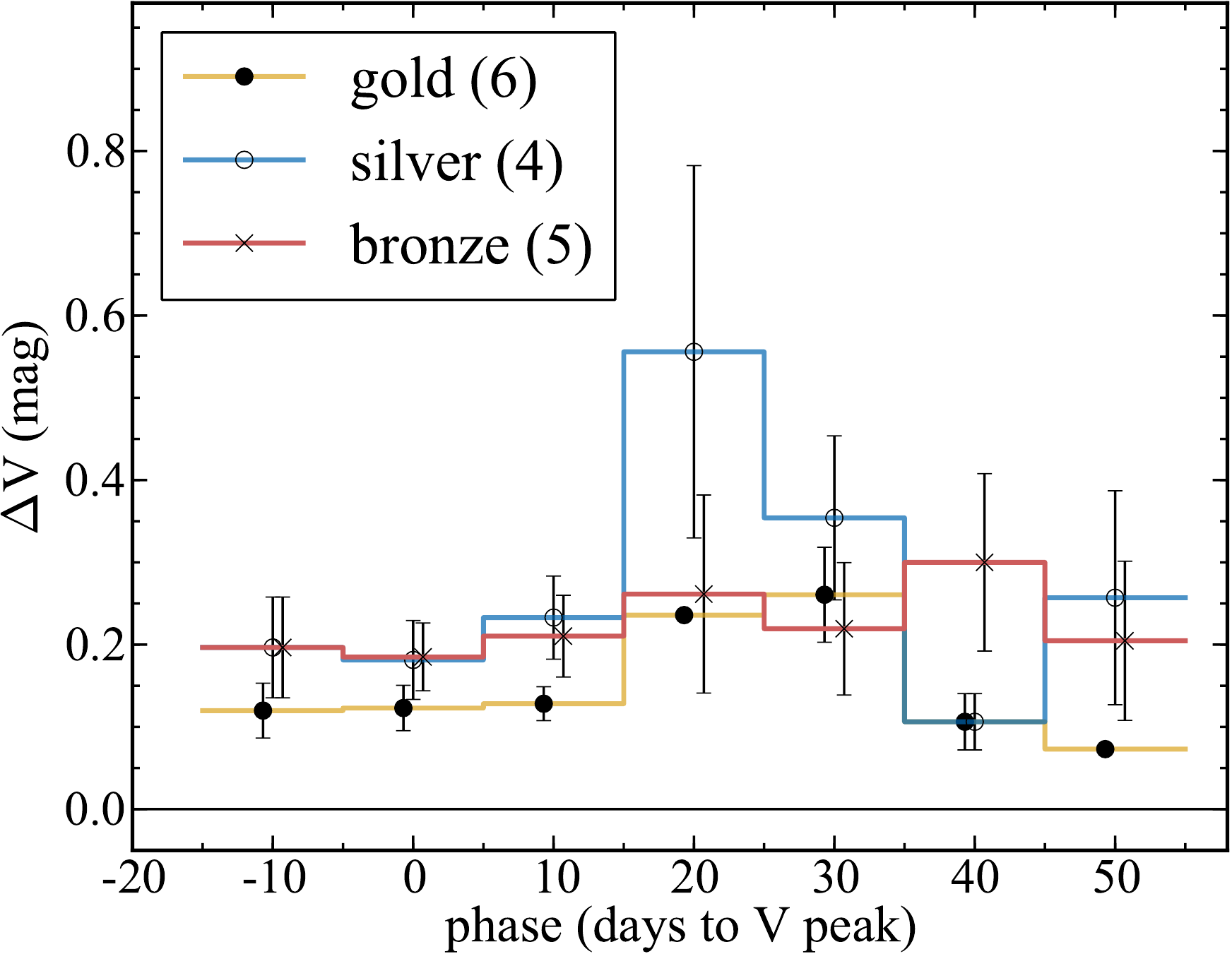}
}
\caption{$V$ magnitude difference between D11 and our (CfA) data averaged in 10-day bins of phase, as in \figref{fig:d11comp}.
The D11 data are shown separately for each of the different quality samples
in the D11 paper.  These samples were 
ranked from best to worst as \emph{gold} (yellow line, solid circles), \emph{silver} (blue line, empty circles), and \emph{bronze} (red line, exes). 
The number of objects in each subsample is shown in parenthesis.
D11 used only their gold and silver samples in their analysis. }
\label{fig:d11comp3}
\end{center}
\end{figure}
Out of the \nsntot\ objects that comprise our sample, \ninlit\  objects have  published photometry. The objects for which data is available in the literature are marked in \tabref{tab:discoverytable}. 
When photometric measurements exist for an object in the same photometric system of our monitoring program, or the photometric conversion is trivial (when the photometric system for the data available in the literature is well defined) we compare our data with the published photometry. In addition, when more data around maximum are available, our data are combined with the literature data to derive a more accurate date of maximum in $V$, following the procedure described in \secref{sec:sample_stat} and Figure \ref{fig:Vmax_mc}.

 We find that our photometry is generally consistent within the errors  with published photometry for the objects in our sample (\figref{fig:litcomp}), with one notable exception: D11 offered the most complete study of stripped SN light curves to date, and our samples share 17 stripped SN. D11 published photometry in $V$ and $R$. When compared with our photometry, only two objects appear to be in excellent agreement in both bands: SN~2004gk, and SN~2006jc. 
SN~2005hg and SN~2007C agree well in $V$, but shows an offset in $R$.
In addition for SN~2005la and SN~2006jc an independent confirmation of the magnitude is also available. The D11 photometry of
 SN~2006jc includes data from ~\citealt{am_Li_Chornock_Filippenko_2007}, and \citealt{2007Natur.447..829P}, and the photometry agrees well with ours. The D11 photometry of  SN~2005la includes measurements from 
\citealt{2008MNRAS.389..113P}, and although the light curves from both the D11 and our surveys are noisy, they are in reasonable agreement. 

For the remaining objects, where the coverage overlaps allowing a comparison, we notice that the D11 photometry is  brighter, with up to a magnitude difference in $R$ at peak and over 0.5 mag in $V$ (e.g., SN~2006fo, \figref{fig:d11phot}). The discrepancy typically increases as the SN evolve, growing as large as $\sim1.5$ in $V$ and $\sim2$ mag in $R$ at later epochs ($\ga~50~\mathrm{days}$, e.g., SN~2007D, SN~2006fo). \figref{fig:d11comp} shows the evolution of the mean discrepancy in $V$ band between the D11 and CfA surveys, as a function of phase.  Photometric measurements are included if the separation in time between the D11 and CfA photometry is less than 5 days for epochs earlier than 45 days after \maxep, and less than 10 days for later epochs. A single SN can contribute to each bin with one or more data points. Error bars represent the error in the mean offset for each epoch (standard deviation over square root of the number of data points that generates the mean). The mean evolution is shown for all objects, as well as after excluding SN~2005la and SN~2006jc. 

\figref{fig:d11comp2} shows a histogram of the distribution of $V$ band magnitude offsets for data points within 5 days of \maxep, within 5 days of phase=40 days after \maxep, and for any epoch later than 45 days after \maxep. Notice that all offsets are positive, or zero at best, indicating that the D11 is always brighter than the CfA photometry, and at later epochs the discrepancy increases, as the D11 light curves reach a plateau.

For all of these objects, our measurements are generated as PSF fitting photometry on host-subtracted images, while D11 performed PSF fitting photometry on the original images, without host subtraction. Thus we attribute this discrepancy to galaxy contamination in the D11 sample. The discrepancy is consistent with the well-known effect of host galaxy contamination on SN light curves, as the SN fades and
becomes less bright in comparison to the host-galaxy, as described in  \citep{1991AJ....101.1281B}. Additionally our photometry has proven to be consistent at the level of a few hundredths of a magnitude on average for a large sample of SN Ia \citep{ind_Brown_Caldwell_et_al__2012, 2009ApJ...700..331H}. Visual inspection of the SN images shows that several objects for which the difference is largest are in fact close to the core of the host galaxy (e.g., SN~2006fo) or in bright regions of the host galaxy arms (e.g., SN~2004gt, SN~2004fe).

D11 classified their sample by quality in three groups: a \emph{gold}, a \emph{silver}, and a \emph{bronze} subset. The latter is judged too poor to be used for the analysis and all inference in D11 is based on the gold and silver objects. We notice that, although still typically brighter, the gold sample is in best agreement with our data, while the silver and bronze sample shows the largest offsets (see \figref{fig:d11phot} and \figref{fig:d11comp3}). The host contamination in the D11 sample may affect  the time
evolution of the SN, and particularly the  \delm\ estimates, and the use of the $V-R$ color evolution in correcting the host galaxy extinction, since, in addition to
giving rise to an offset in magnitude, the contamination is worse at later epochs and is different in different bands. 

Finally, in addition to the 17 stripped SN common to our samples, D11 presents photometry or SN~2005eo, which was also monitored at the CfA. SN~2005eo is included in the D11 silver sample, and was originally classified as a SN Ic. SN~2005eo was however removed from our stripped SN sample, as we reclassified this object as a SN Ia (M14). This classification is further discussed in \secref{sec:05kl}.

\section{Colors and Color Evolution}\label{colorcolor}
Our multi-wavelength photometric coverage allows us to discuss the color characteristics and color evolution of the supernovae in our sample. 
While an in-depth discussion is beyond the scope of this paper, and it will be presented in (Bianco et al., in preparation), here we present the basic color evolution and color-color behavior of our stripped SN sample.

When we discuss colors, color evolution, and for all plots in color space, we correct the magnitude of all objects in our sample for Galactic extinction only. 
The Galactic extinction $E_{B-V}$ is obtained adopting the~\citet{2011ApJ...737..103S} recalibration of the~\citet{1998ApJ...500..525S} extinction maps, with the $E_{B-V}$ reported in \tabref{tab:hostgal_table}. For each photometric band we use the extinction coefficients $A_\lambda$ normalized to the photoelectric measurements of $E_{B-V}$ as reported in Table 6 of ~\citet{1998ApJ...500..525S}, which assume a reddening law according to ~\citet{1999PASP..111...63F} with $R_V = 3.1$, and standard transmission for the Landolt $\ubvri$, the SDSS $r'i'$, and 2MASS \jhk\ filters. These extinction corrections are based on star spectra and we do not correct the extinction for the SN SED. Based on~\citet{2007ApJ...659..122J}, who studied this effect for SN Ia, we estimate the correction on $R_V$ to be $\la 4\%$.
No host reddening or cosmological corrections are applied. 
While with spectra and NIR data the host reddening can be constrained (Bianco et al., in preparation), here we use the observed color. 
This approach is sensible to aid photometric differentiation of subtypes.

\begin{figure}[th!]
\begin{center}
\includegraphics[width=0.73\columnwidth]{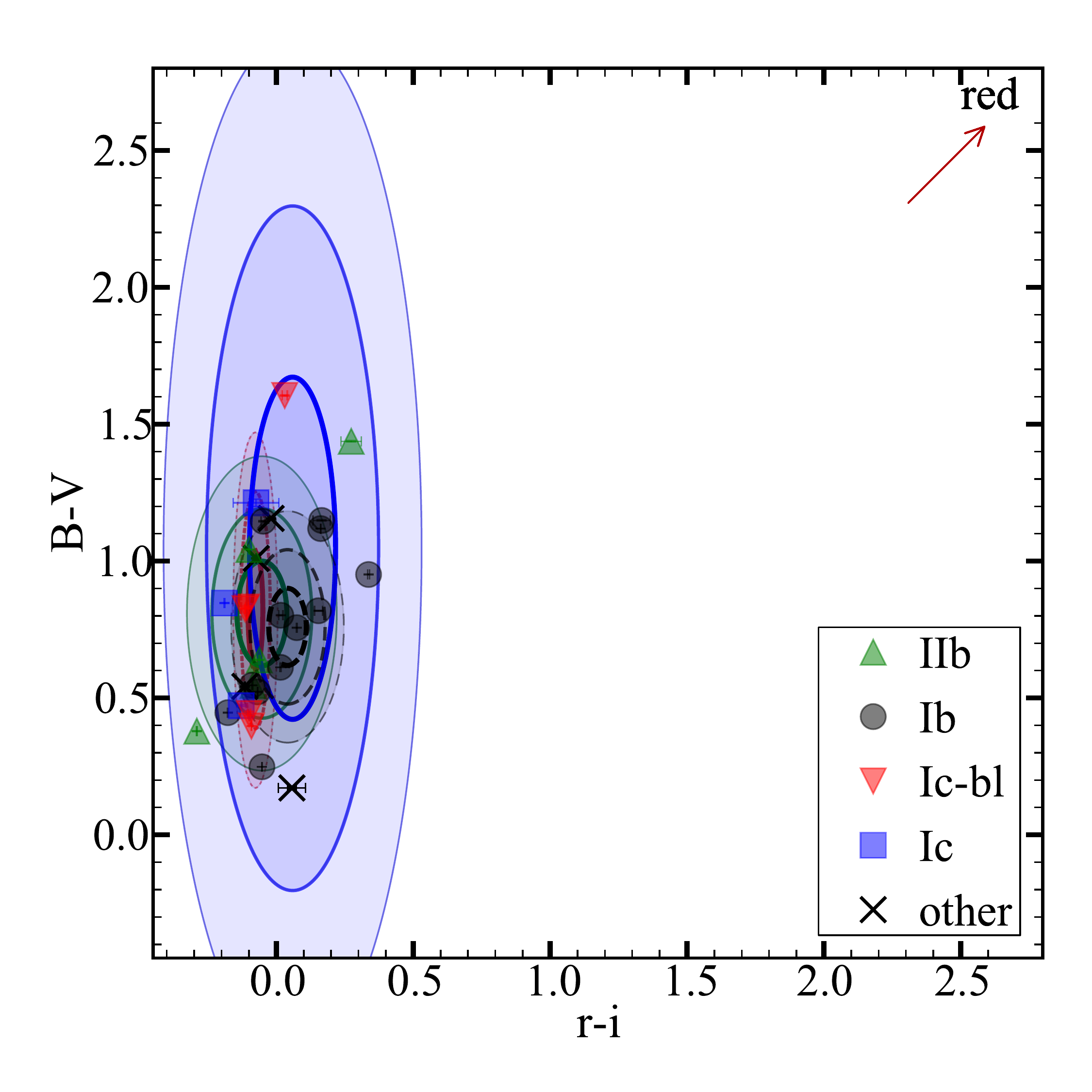}
\includegraphics[width=0.66\columnwidth]{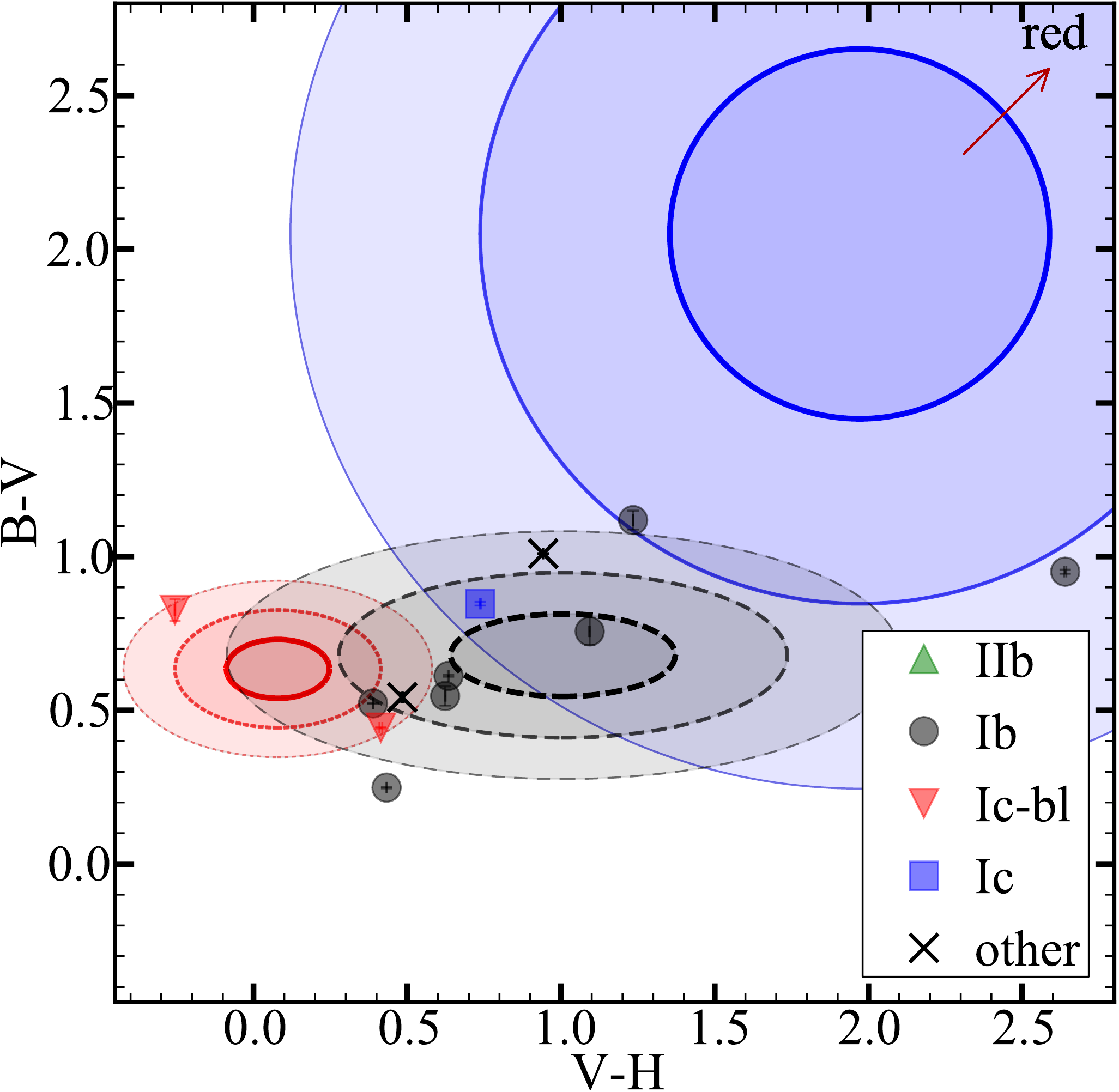}
\includegraphics[width=0.73\columnwidth]{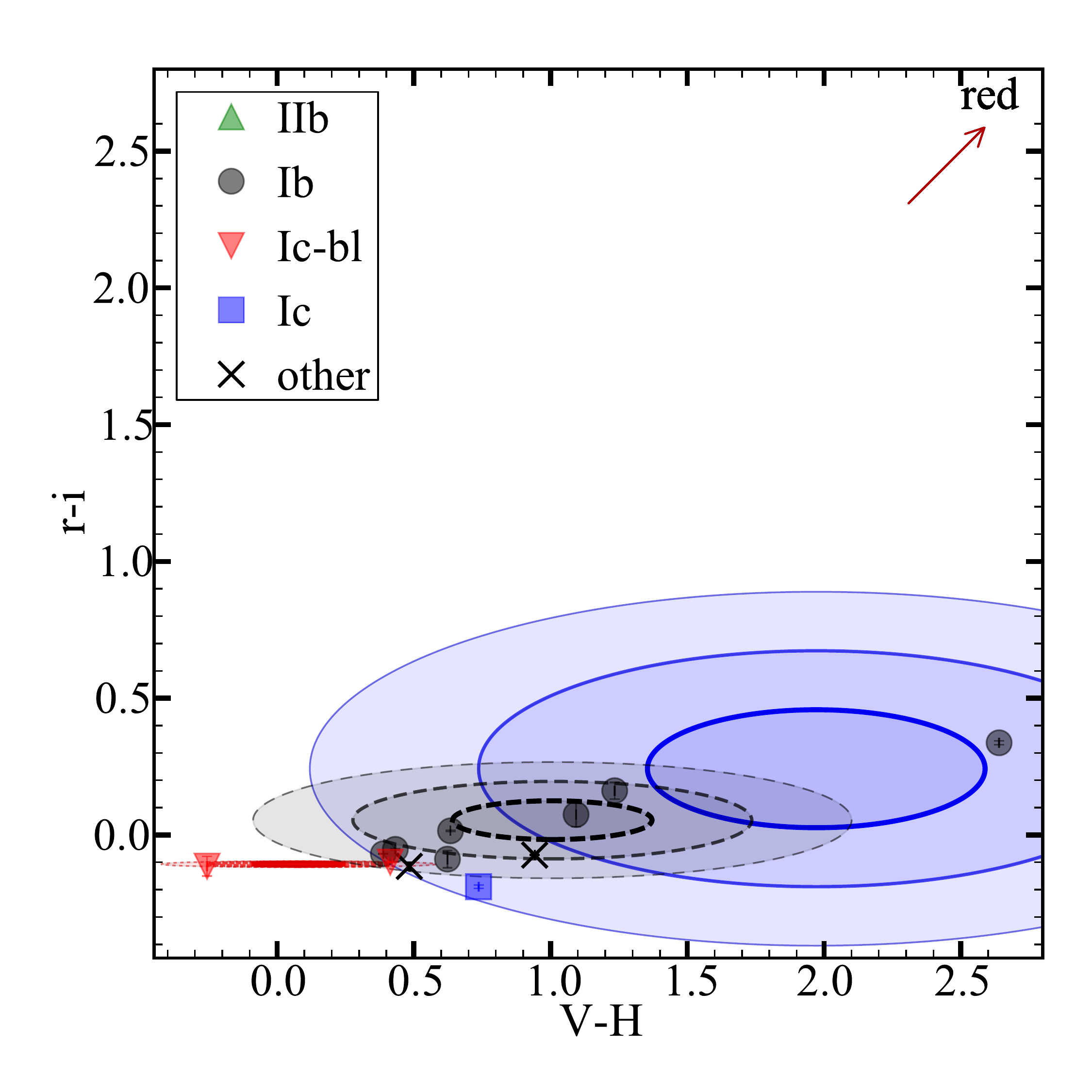}
\caption{Color-color distributions at \maxep\ for the SN presented in this paper: 
(Top) $B-V$~vs.~$r'-i'$; (Middle) $B-V$~vs.~$V-H$; (Bottom) $r'-i'$~vs.~$V-H$. 
Different SN types are indicated by different symbols: 
IIb -- green triangles; Ib -- gray circles; Ic -- blue squares; Ic-bl -- red upside-down triangles;  other subtypes are indicated by exes. 
Ellipses (same color mapping by subtype) are centered on the mean of each subtype distribution with their semi-major and semi-minor axes given by the 1-, 2-, and 3-$\sigma$ projected single-color distributions.  The large standard deviation of the Ic sample (shown in blue) is dominated by the contributions of the very red SN~2005kl which is outside the plotted range at $B-V=3.25 \pm 0.08$~mag, $r-i=0.6734 \pm 4\times10^{-4}$~mag, and $V-H=3.21 \pm 0.05$~mag.  All colors are the closest available to \maxep\ and only included if the epoch in both bands is within 5 days of  \maxep\ epoch. The arrow simply indicates the direction of redder colors (i.e. a reddeining vector with A=1).}
\label{fig:colorcolor}
\end{center}
\end{figure}

\figref{fig:colorcolor}  shows color-color plots for our sample of stripped SN in $B-V$~vs.~$r'-i'$, $B-V$~vs.~$V-H$, and $r'-i'$~vs.~$V-H$ space. The errorbars in the plot are generated by adding photometric errors in each band  in quadrature (disregarding correlation). All objects with a solid determination of \maxep, either in the literature or derived from our data (\secref{sec:litdata}), are included in these plots when photometry data are available within 8 days of \maxep\ for all four bands used in each plot. In $B-V$~vs.~$r'-i'$ this amounts to 47 objects: 8 SN IIb, 15 Ib, 13 Ic, 6 Ic-bl. Identified as ``other'' are objects for which we could not determine a subtype (SN~2007iq, which is of uncertain classification Ic or Ic-bl, as only late spectra are available), or that appear spectroscopically atypical (the Ca-rich transient SN~2007ke, the narrow line SN~2005la, and the SN Ib-pec 2007uy and SN~2009er). The spectroscopic characteristics of these objects are discussed in detail in M14 and the photometric properties of SN~2007ke, SN~2005la are discussed  in \secref{specSN_sec}. 
The ellipses in the plot represent the mean (center) and standard deviation ($\sigma$) of each subtype distribution.

In $B-V$~vs.~$r'-i'$ all distributions are well consistent with each other to the 1$\sigma$ level: the subtypes cluster in overlapping distributions and appear indistinguishable in this color-color space (\figref{fig:colorcolor}, top panel). When NIR colors are used the subtypes seem to separate in color-color space, although the number of objects available for this analysis is smaller. When including NIR colors ($V-H$) the number of objects that can be used for this plot drops significantly, as only 15 objects in our sample have a determination of \maxep\ \emph{and} optical and NIR photometry within 8 days of it. These objects include: 8 SN Ib, 2 SN Ic, and 2 Ic-bl, and the peculiar objects SN~2007uy and SN~2009er (in these plots under the label ``other''). 
Subtypes separate to $\ga 1\sigma$ level, particularly in $r'-i'$~vs.~$V-H$. The $V-H$ distribution in fact shows the broadest variation, while the $r'-i'$ is the narrowest. Amongst the SN IIb only SN~2008ax has NIR coverage, but unfortunately there is no optical coverage from FLWO for this SN, thus no SN IIb were included in the  $B-V$~vs.~$V-H$ or $r'-i'$~vs.~$V-H$ plot. 
The mean of the distribution of SN Ib and Ic is still consistent within 2$\sigma$, but SN Ib appear redder in $V-H$, and SN Ic-bl bluer, though this observation is based on only 2 SN Ic-bl and it needs to be verified in larger samples.

While these trends are based on small samples, they highlight the importance of NIR photometry, and we suggest populating such $r'-i'$~vs.~$V-H$ color-color plots in the future  to verify the SN-type-dependent color trends  observed here,  which ultimately  could be used to differentiate core-collapse SN subtypes photometrically.

\begin{figure}[ht!]
\begin{center}
\includegraphics[width=0.795\columnwidth]{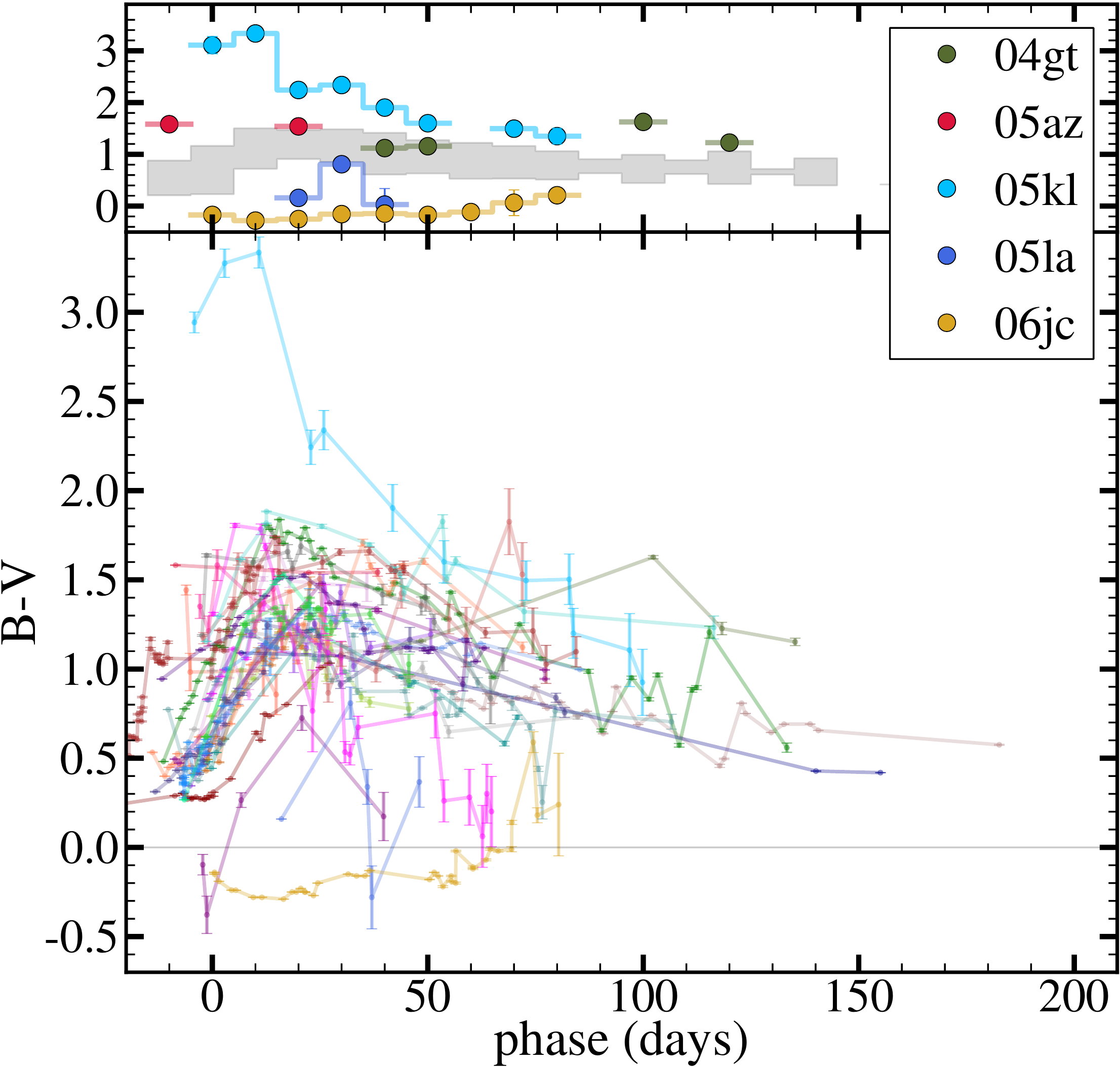}
\includegraphics[width=0.795\columnwidth]{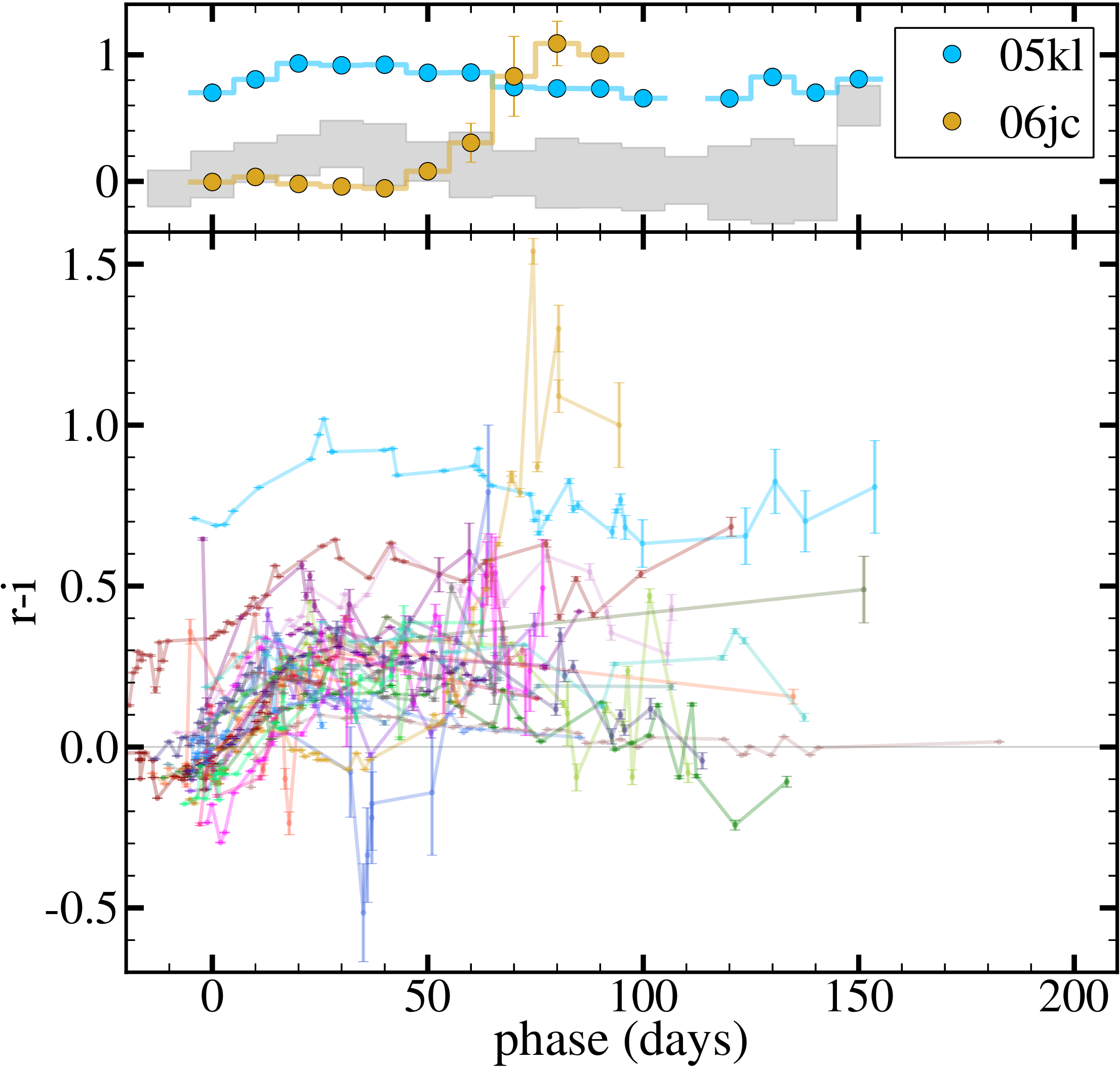}
\includegraphics[width=0.795\columnwidth]{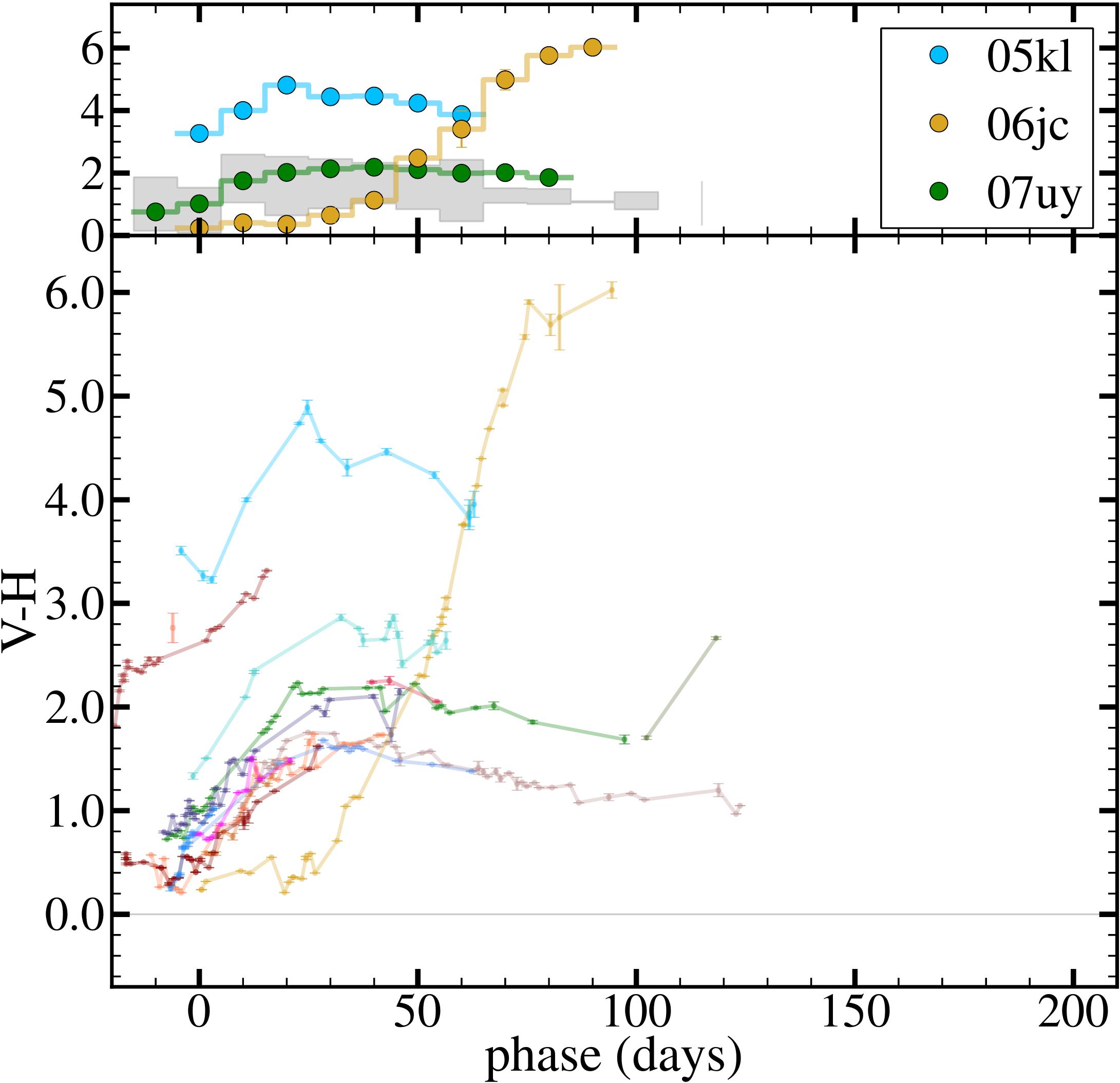}
\caption{Color evolution of the objects in our sample with determined  \maxep: 
(Top) $B-V$; (Middle) $r-i$; (Bottom) $V-H$. 
For each color plot the bottom panel shows the color evolution of all objects along with their photometric errors. 
The top panel shows the 1$\sigma$ weighted average color evolution range (gray region) for objects with defined colors at \maxep.  
Outliers are illustrated and named separately in these top panels with their color curves binned in 10-day intervals. 
}
\label{colallsn}
\end{center}
\end{figure}

We plot the color evolution of our objects in $B-V$, $r'-i'$, and $V-H$ in \figref{colallsn}. In each plot all photometric data between -20 and 210 days with respect to the epoch of peak V magnitude, \maxep, are plotted in the bottom panel for all SN with well determined \maxep. \emph{All} objects in our sample with known \maxep\ are on display here. The errorbars represent the photometric errors. The scatter in the bottom panels of this plot thus represents the diversity in the observed photometry of stripped SN.

\begin{figure*}[t!]
\begin{center}
\centerline{
\includegraphics[width=0.95\columnwidth]{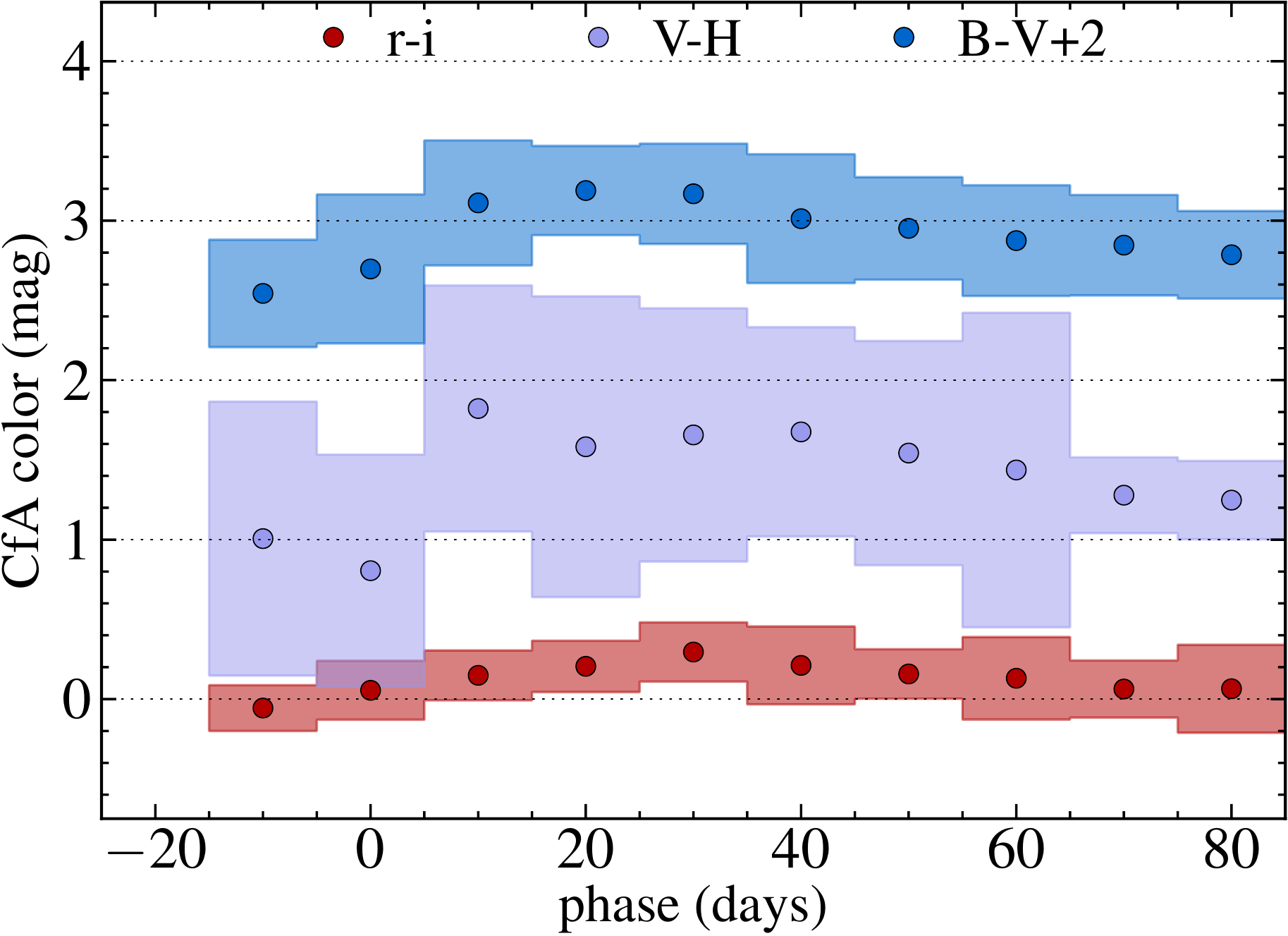}
\includegraphics[width=0.948\columnwidth]{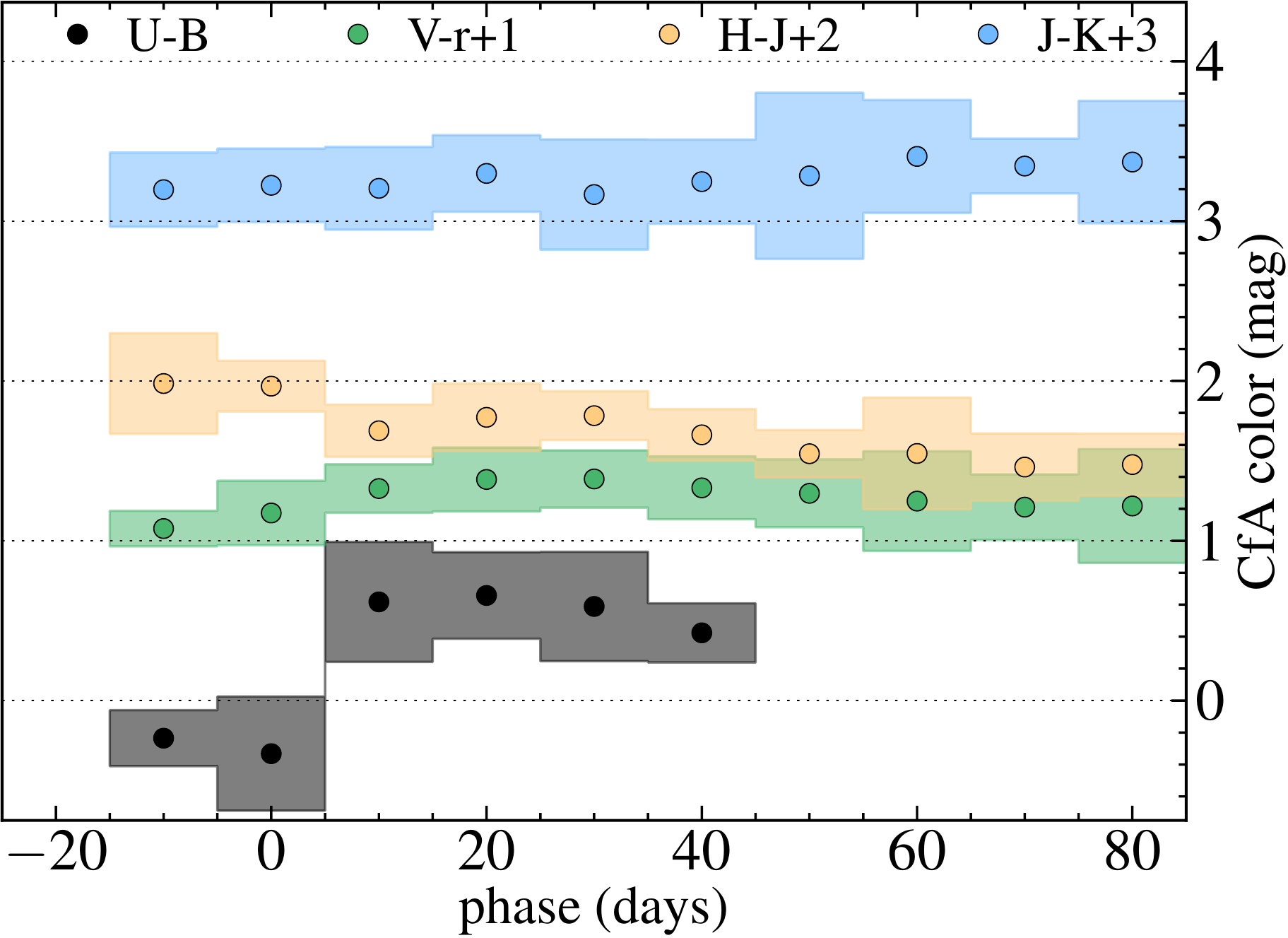}
}
\caption{Average color evolution of the stripped SN in our sample in several color bands. The weighted average, weighted by the photometric errors, is generated in each color band excluding SN~2006jc.
Colors are calculated from our photometry, then binned in 10 day intervals. 
The sample standard deviation is visualized as a shaded region.
The SN are corrected for galactic reddening, but not for host extinction. 
We have 8 and 5 data points for $V-H$ at phase 70 and 80 days respectively, and  6 for $U-B$ at both 30, 40 days phase. All other bins include 10 or more data points. Epochs later than 40 days are omitted for $U-B$ as $\la 3$ data points are available.  These weighted averages for $B-V$, $r-i$, and $V-H$ are also shown in the top panels of figure \figref{colallsn}.}\label{fig:meancol}
\end{center}
\end{figure*}

The top panel shows the mean color evolution, binned in 10 day intervals, and its standard deviation as a gray area. This is the weighted average of the photometry for all objects calculated over 10-day bins, weighted by the photometric errors.  The standard deviation in the average is calculated as the second moment of the distribution of photometric measurements, disregarding the photometric errors. The weighted average colors for a more complete set of color spaces is shown in \figref{fig:meancol}. 
Outliers are plotted in color in \figref{colallsn} in each top panel,
 with the same 10 day binning, and errorbars representing the standard deviation within the bin. All objects with a binned color data-point with a 2$\sigma$ lower (upper) limit above (below) the average by more than 2 standard deviations (standard deviations of the average in this case) are considered outliers, are plotted in this panel, and identified in the legend. Note that SN~2006jc (\secref{sec:06jc}) is an obvious outlier in each of these plots (gray circles) with early blue and late red colors. SN~2006jc is removed from the calculation of the mean color evolution, as it is known to be spectroscopically peculiar and  its late-time color-evolution is driven by non-intrinsic SN processes, such as dust formation (see \secref{sec:06jc} and references therein). 
Other outliers are discussed in~\secref{specSN_sec}.

Finally, we present the average color evolution across our sample, and its standard deviation, in \figref{fig:meancol} for $B-V$, $r'-i'$, and $V-H$ (left), and $U-B$, $V-r'$, $H-J$, $J-K$ (right). For $B-V$, $r'-i'$, and $V-H$ these averages also appear as shaded regions in \figref{colallsn}. The average, weighted by the photometric errors, is generated in each color band as described in the previous paragraph, again excluding SN~2006jc.

We notice that:
\begin{itemize}
\item
the largest  color variation is observed in $V-H$: in time, (\figref{fig:meancol}), as well as
amongst different SN types (as already noted in \secref{colorcolor} and \figref{fig:colorcolor}, and indicated by the large standard deviation in this plot). Two causes may contribute to this effect: the small size of the sample that has photometry near \maxep\ in both $V$ and $H$, which is only 20 objects, and host galaxy reddening
effects. High reddening would have the most impact on the bluest and the least on
the reddest band, and would result in a large effect in $V-H$, since we are bridging an interval of over 1100nm in wavelength. However reddening would affect the spread in color equally at all epochs (the standard deviation of the mean color, which is in fact large, typically $\sim1.5$~mag). The color \emph{evolution}
 over time is due to intrinsic changes in the SN SED as the SN evolves. The mean $V-H$ color spans a dramatic 1.6 magnitudes between $10$ days before and 50 days after peak; 
\item
 the least variation  is in $r'-i'$:  only 0.5 mag total as seen in \figref{fig:meancol}
both for the standard deviation, and for the change in mean color over time. Narrow standard deviations are also observed in  $H-J$ and $V-r'$, while $J-K_s$ shows remarkably little color evolution; 
\item 
the $r'-i'$ colors at peak are intriguingly similar for all objects. This was also noticeable in \figref{fig:colorcolor}, especially for SN Ic-bl, although only 6 and 2 SN Ic-bl are plotted, respectively, in the top and two bottom panels, due to the availability of photometry in all four bands needed for each plot. However, we can measure the $r'-i'$ color within 10 days of \maxep\ for 7 SN Ic-bl. We find a mean $r'-i'$ peak color for SN Ic-bl of \mbox{$<r'-i'>_\mathrm{peak}~=~-0.025~\pm~ 0.01$ mag}. The standard deviation for the distribution of other types is at least twice as large.
 
\end{itemize}

These color evolutions and the relation between colors are worth a more thorough investigation, since they will be valuable in typing and classification in synoptic surveys where the data volume renders spectroscopic identification infeasible, and followup resources will be scarce compared to the number of discoveries. A more complete analysis of the colors of stripped SN, and their correlations with types will be presented in Bianco et al. (in preparation) including SN data from the literature.

\section{Discussion of specific SN}\label{specSN_sec}

\subsection{SN~2005bf}
Some CfA optical and NIR photometric measurements of SN~2005bf were published in~\citet{jaz_Hicken_Challis_et_al__2005}, but the data presented here used template subtraction and more comparison stars for the reduction, and they supersede the measurements in~\citet{jaz_Hicken_Challis_et_al__2005}.
SN~2005bf was an unusual SN Ib with
unique photometric and spectroscopic properties \citep{2008MNRAS.389..141M,jaz_Hicken_Challis_et_al__2005,Lee_Hamuy_Gonzalez_et_al__2006,2007ApJ...666.1069M,2007MNRAS.381..201M}, interpreted as a strongly aspherical explosion of a Wolf-Rayet WN star, perhaps with a unipolar jet \citep{jaz_Hicken_Challis_et_al__2005,Lee_Hamuy_Gonzalez_et_al__2006,2007MNRAS.381..201M}, and possibly powered by a magnetar at late times \citep{2007ApJ...666.1069M}.

\begin{figure}[tb]
\begin{center}
\centerline{
\includegraphics[width=0.52\columnwidth]{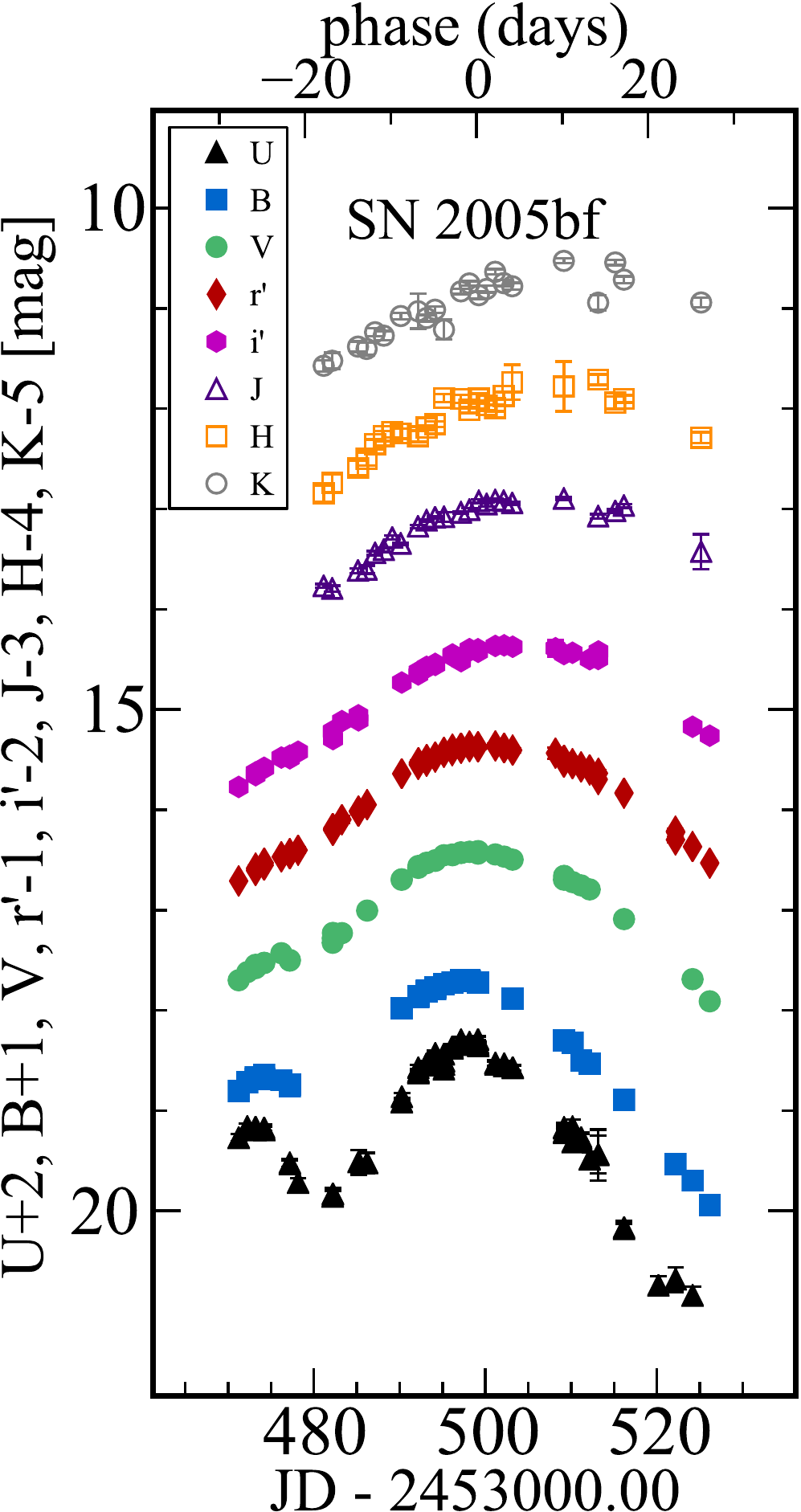}
\includegraphics[width=0.477\columnwidth]{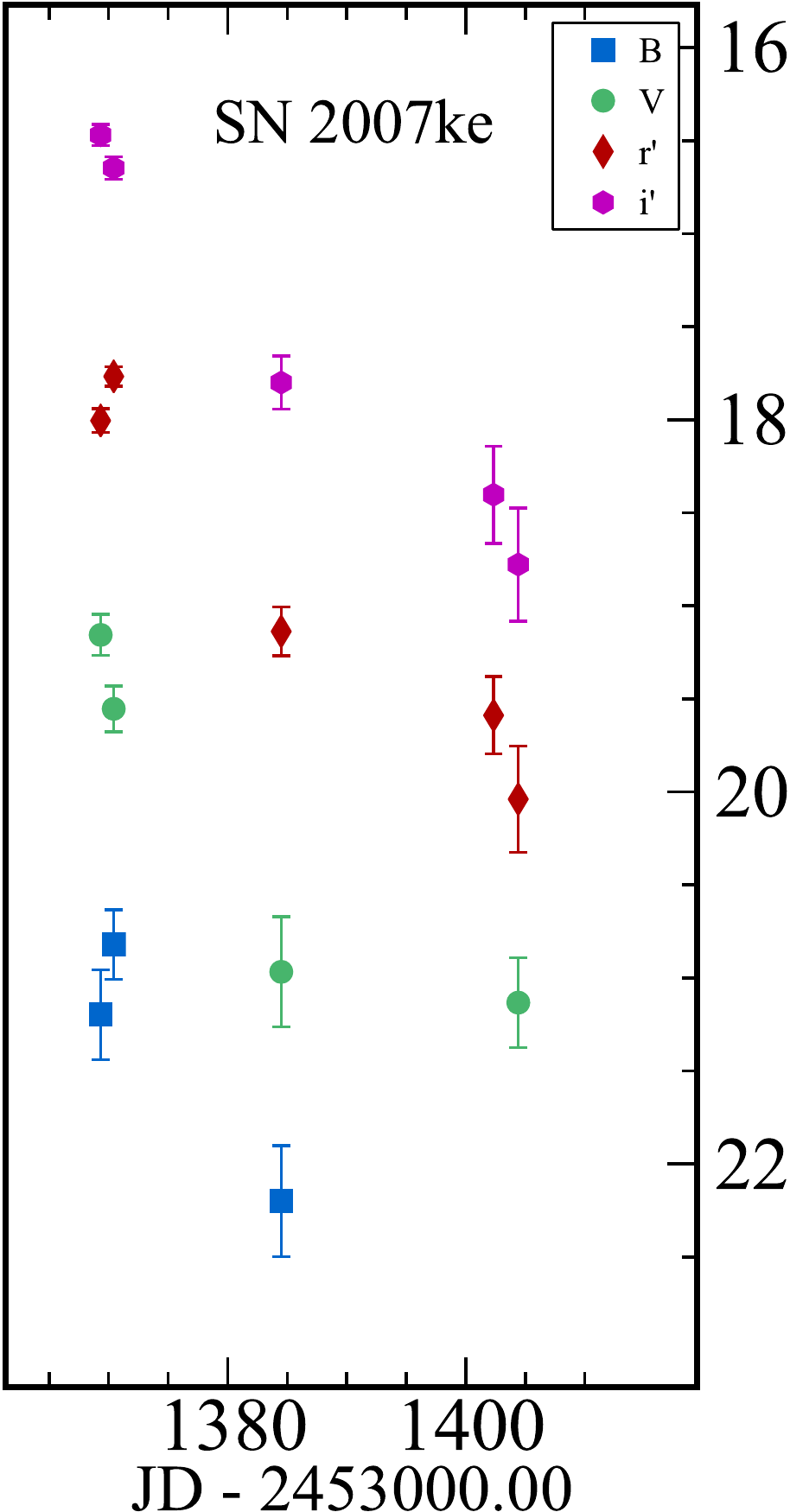}
}
\caption{Optical and NIR photometry of SN~2005bf (Left), and SN~2007ke (Right), symbols as described in \figref{fig:optphot}.}
\label{fig:05bf07kephot}
\end{center}
\end{figure}

Our light curve has excellent multi-band coverage of the region around peak brightness, both before and after peak: the first CfA optical epochs were collected on JD~2553471.742 ($U$), while NIR coverage began 10 days later. The CfA optical and NIR light curve of SN~2005bf is shown in \figref{fig:05bf07kephot}, left panel. An early peak is clearly identifiable in the bluer bands $U$ and $B$, and (less clearly) in $V$ on $\mathrm{JD}~2453476.75$. A later, more prominent peak, visible in all bands occurs in $V$ on $\mathrm{JD}$~2453498.96 (\maxep). Our photometric coverage continues through JD~2553526 in optical wavelengths and JD~2553525 in NIR. An in-depth phenomenological discussion of these peaks was presented in \citet{jaz_Hicken_Challis_et_al__2005}, and \citet{Lee_Hamuy_Gonzalez_et_al__2006}. We notice  that the first peak is too late after explosion to be consistent with a standard shock breakout (as seen in SN~2008D -- \citealt{Ofek_Cucchiara_Rau_et_al__2008,Kirshner_Kocevski_et_al__2009}), while if the second peak is considered, the rise time for this SN is unusually long (over 30 days!). The second maximum and long rise time has been attributed to the highly aspherical distribution of a large amount of~\synNi~ synthesized in the explosion \citep{jaz_Hicken_Challis_et_al__2005,Lee_Hamuy_Gonzalez_et_al__2006}.
 M14 discusses how  the choice of the epoch for the peak affects the spectroscopic analysis of SN~2005bf. 

\subsection{SN~2007ke}\label{sec:07ke}
SN~2007ke is a known Ca-rich transient with 7 photometric data points in KAIT $R$ published in \citet{Kasliwal13}. Our data does not cover the rise of the light curve, thus we estimated the $V$-band maximum using the \citet{Kasliwal13} data according to the prescription described in \secref{sec:sample_stat}, and applying a shift corresponding to the median difference in time between $R$ and $V$ reported in \tabref{tab:stats1}. Adding the error to the fit (0.32 days) and the standard deviation in the time offset (1.4 days, \tabref{tab:stats1}) in quadrature, we obtain \maxep$=2454371.2 \pm 2.1$. While our photometry is sparse, and our coverage begins near peak, we have coverage in four bands $BVr'i'$, with four epochs in $B$ and $V$, including a non detection at JD=2454402.8, and five in $r'i'$. This allows us to probe its color evolution, and we notice that SN~2007ke shows  scatter (\figref{fig:05bf07kephot}, right panel). Additionally, it was noted \citep{Kasliwal13} that SN~2007ke is well removed from its host galaxy 
 (compare its separation of $\sim 0.5$~arcmin from the center of the host, \tabref{tab:hostgal_table}, with  NGC~1129's radius of 1.6 arcmin: the 
 extinction corrected apparent semi major axis of the 25 $B~\mathrm{mag/arcsec^2}$ isophote
  reported by HyperLEDA, \citealt{1991A&A...243..319P}).
 Additionally, SN~2007ke is the only SN in our sample to arise in an elliptical host galaxy (see \tabref{tab:hostgal_table}), confirming the environments of Ca-rich transients are unusual, compared to stripped SN host environments, as pointed out in \citep{Kasliwal13}.

\subsection{Unusual color evolution}

\figref{colallsn} shows the color evolution of our SN in $B-V$, $r'-i'$, and $V-H$. The top panel in each of the three plots shows the mean color evolution, as described in \figref{fig:meancol}, and over plotted are the color curves for outliers: objects whose color, binned in 10 day intervals, is at least 2$\sigma$ away from the 2$\sigma$ limit of the color average in one or more bins.
Three obvious outliers in these plots are discussed in this Section: SN~2006jc, SN~2005kl (both outliers in all three color spaces), and SN~2005la (outlier in $B-V$, and for which we have no NIR coverage). Additionally SN~2008D is a $>2\sigma$ outlier in both $r'-i'$ and $V-H$.
Although SN~2004gt, SN~2005az, and SN~2007uy, 
also appear in the top plots in \figref{colallsn} as outlier in one color space,
their classification as outlier is weak, due only to the distance of one early (SN~2005az) or one  late 10-day bin (SN~2004gt, SN~2007uy) from the sample average, and are not discussed individually.

\subsubsection{SN~2008D}
SN 2008D is a well studied SN~Ib 
\citep{Ofek_Cucchiara_Rau_et_al__2008, 2008Sci...321.1185M,Kirshner_Kocevski_et_al__2009,2009ApJ...692L..84M}, discovered in the X-Ray while SWIFT monitored the host galaxy to observe the evolution of SN~2007uy \citep{Ofek_Cucchiara_Rau_et_al__2008}, thus yielding very stringent optical and NIR pre-explosion limits only hours before explosion \citep{Kirshner_Kocevski_et_al__2009}.
 Our data on SN~2008D were already published in \citet{Kirshner_Kocevski_et_al__2009}, and it is presented again here for completeness. SN~2008D does not appear as an outlier in ~\figref{colallsn}, but it is a $> 2\sigma$ outlier in both $r'-i'$ and $V-H$ colors. Its light curve is redder than the mean over the entire evolution, and in fact SN~2008D is known to suffer significant host reddening ($A_v\sim1.5-2.5$~mag - \citealt{Ofek_Cucchiara_Rau_et_al__2008}), in addition to the early (prior to $-10$ days to \maxep) blue excess attributed to cooling of the shock-heated stellar envelope~\citep{Kirshner_Kocevski_et_al__2009}. 

\subsubsection{SN~2006jc}\label{sec:06jc}
\begin{figure}[tb]
\centerline{
\includegraphics[width=0.52\columnwidth]{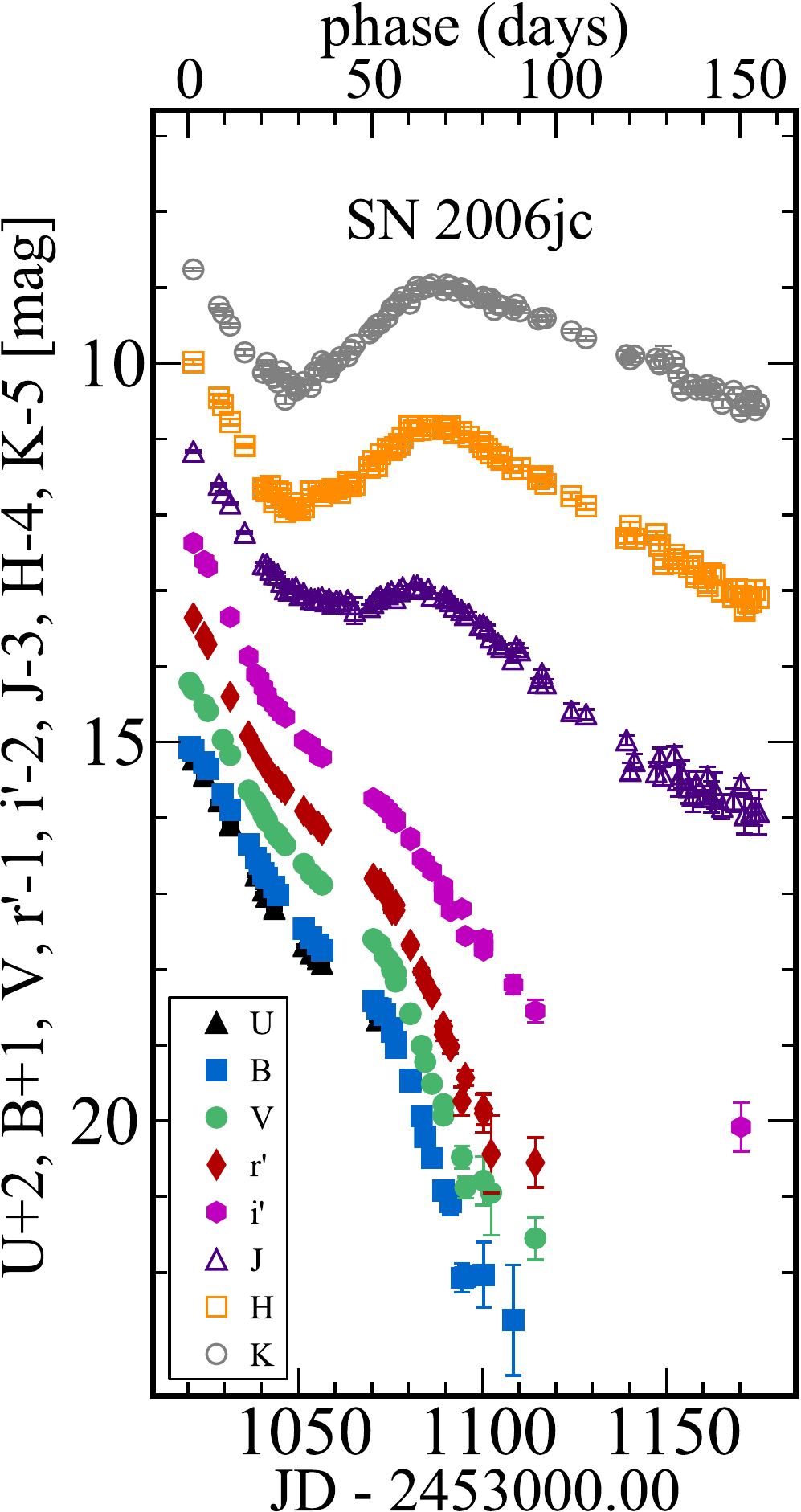}
\includegraphics[width=0.477\columnwidth]{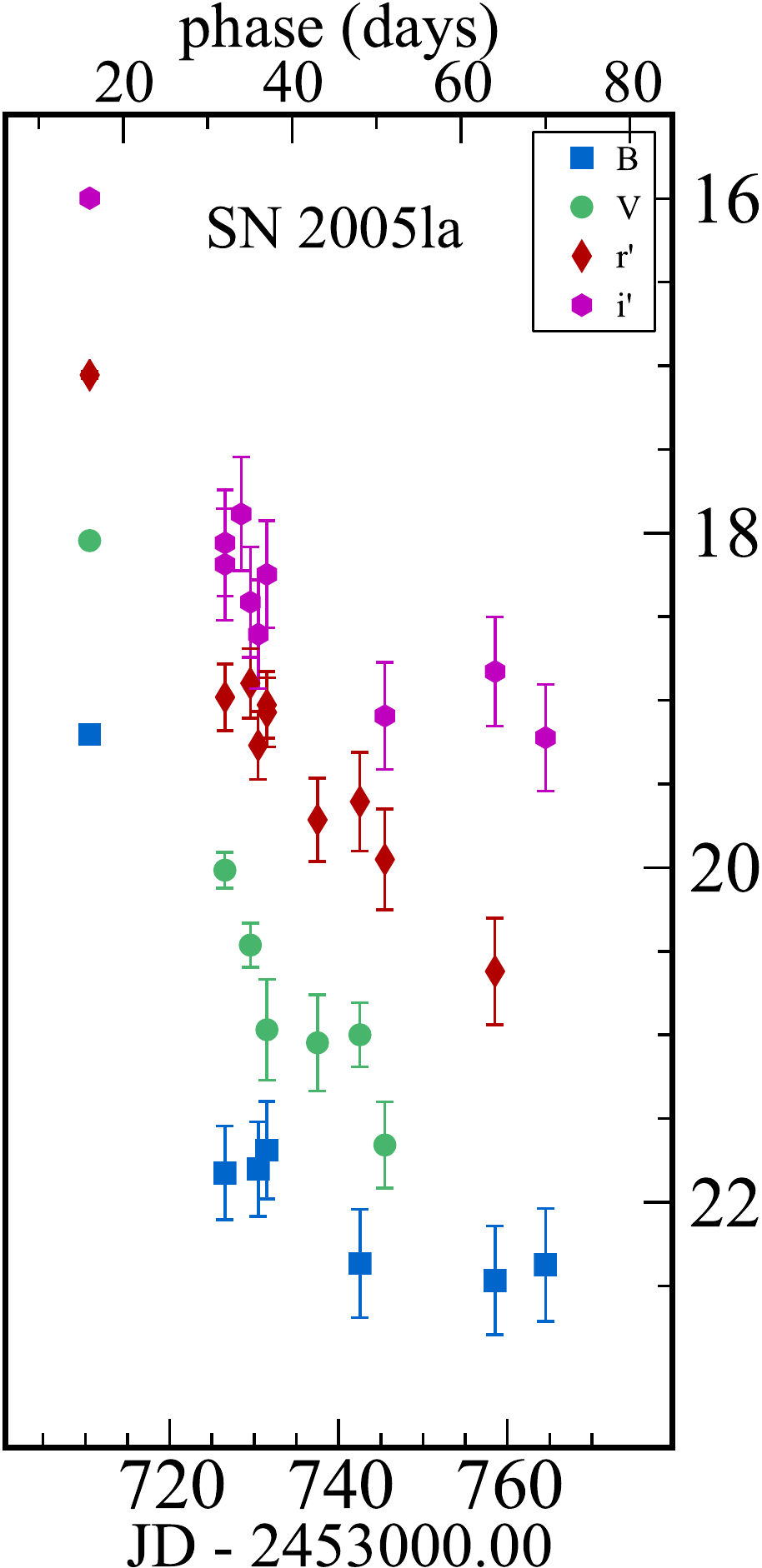}
}
\caption{Optical and NIR photometry of SN~2006jc (Left) and SN~2005la (Right), symbols  as describes in \figref{fig:optphot}.}
\label{fig:06jcphot}
\end{figure}

\begin{figure}[tb]
\begin{center}
\includegraphics[width=0.95\columnwidth]{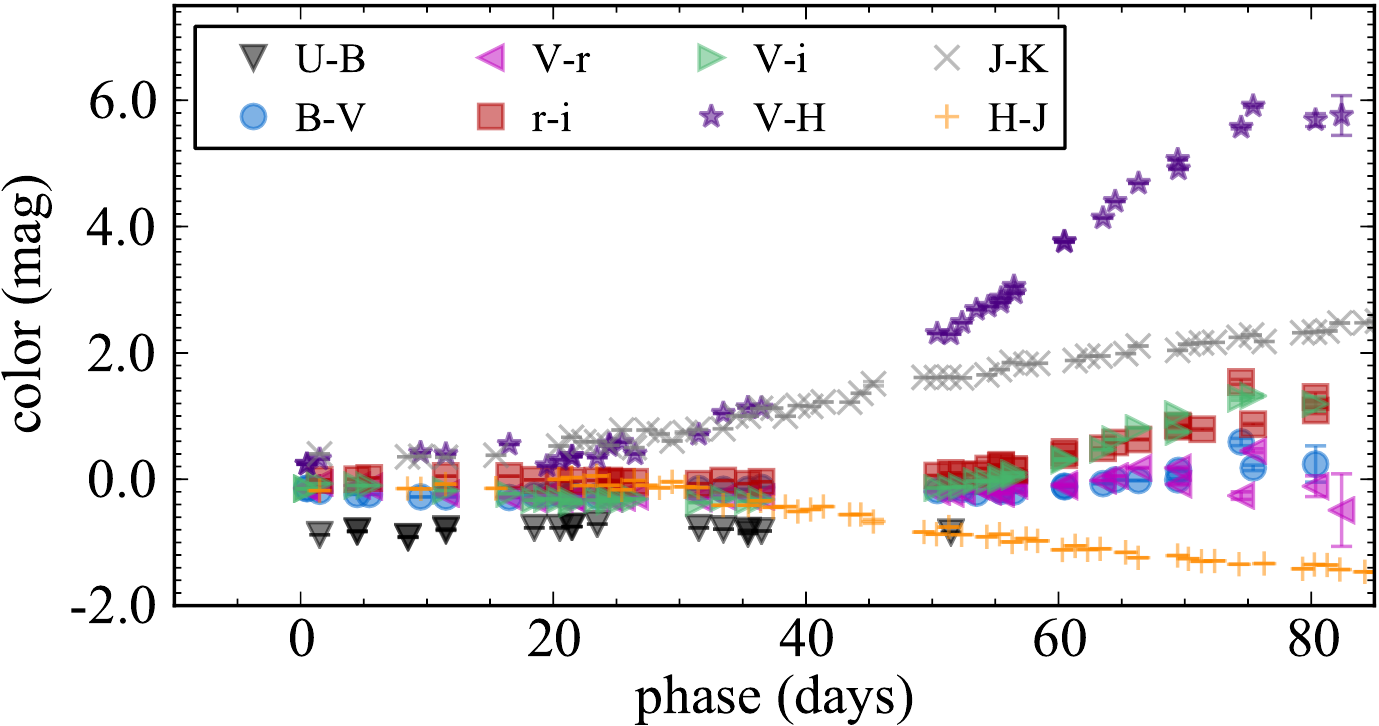}
\caption{Color time series for SN~2006jc.  The error bars are the quadrature sum of the photometric errors for the constituent bands.}
\label{fig:06jccol}
\end{center}
\end{figure}

SN~2006jc is classified as a SN Ib-n. A re-brightening in the NIR light curve of SN~2006jc was noticed first by \citet{2006ATel..961....1A}. Our NIR data has exquisite sampling of the NIR re-brightening, which begins near JD 24541050, or roughly 40 days after peak, adopting the peak determination of \citet{2008MNRAS.389..113P}, and continues with regular sampling through $\sim$160 days after peak. We also present regular photometry in optical bands that continues through 100 days after peak. Optical and NIR photometry of SN~2006jc can be found in the literature \citep{2008MNRAS.389..141M,2008MNRAS.389..113P}. We plot our light curve in \figref{fig:06jcphot} (left panel), and color time series in \figref{fig:06jccol}. The unusually blue early color of SN~2006jc at early times, and its later NIR re-brightening  have been explained by complex interaction with circumstellar material. The early blue color is due to interaction with He-rich material ejected by the progenitor in prior outbursts \citep{2007Natur.447..829P, 2007ApJ...657L.105F,2008ApJ...680..568S}, and the reddening is due to production of dust triggered in the ejecta at later epochs \citep{2008MNRAS.389..141M,2008MNRAS.389..113P}.

\subsubsection{SN~2005la}\label{sec:05la}
SN~2005la (\figref{fig:06jcphot}, right panel) is spectroscopically peculiar, showing narrow He and H lines in emission, indicative of interaction with circumstellar material \citep{2008MNRAS.389..131P}, and is considered a transitional object between SN~2006jc-like events, and SN IIn. SN~2005la appears as a blue outlier in $B-V$, and it is bluer than the mean in $r'-i'$, evolving redward at late times (phase $\ga 50$~days). SN~2005la is also included in the D11 sample.
We note that SN~2006jc and SN~2005la, both SN interacting with a
He-rich, and for SN~2005la also H-rich, circumstellar medium have
bluer $B-V$ colors than the rest of the normal stripped SN. 
Figure 3 in \citet{2008MNRAS.389..131P}  shows the
$B-V$ evolution of SN Ib-n (also including SN~2000er and SN~2002ao): 
all of these SN have very blue colors ($-0.4 \lesssim B-V \lesssim 0.6$~mag for
0 to 100 days after maximum) when compared to non-interacting stripped SN, as 
in our \figref{colallsn}. The bluer colors of these SN Ib-n and SN~2005la are
due to both a bluer continuum 
(most likely due blending of FeII lines from fluorescence, \citealt{2007ApJ...657L.105F})
and to strong He lines in emission.

\subsubsection{SN~2005kl}\label{sec:05kl}
SN~2005kl (light blue in \figref{colallsn}) is consistently and significantly redder than the sample mean throughout its color evolution in each color space. However  it is not spectroscopically peculiar. 

\begin{figure}[tb]
\centerline{
\includegraphics[width=0.52\columnwidth]{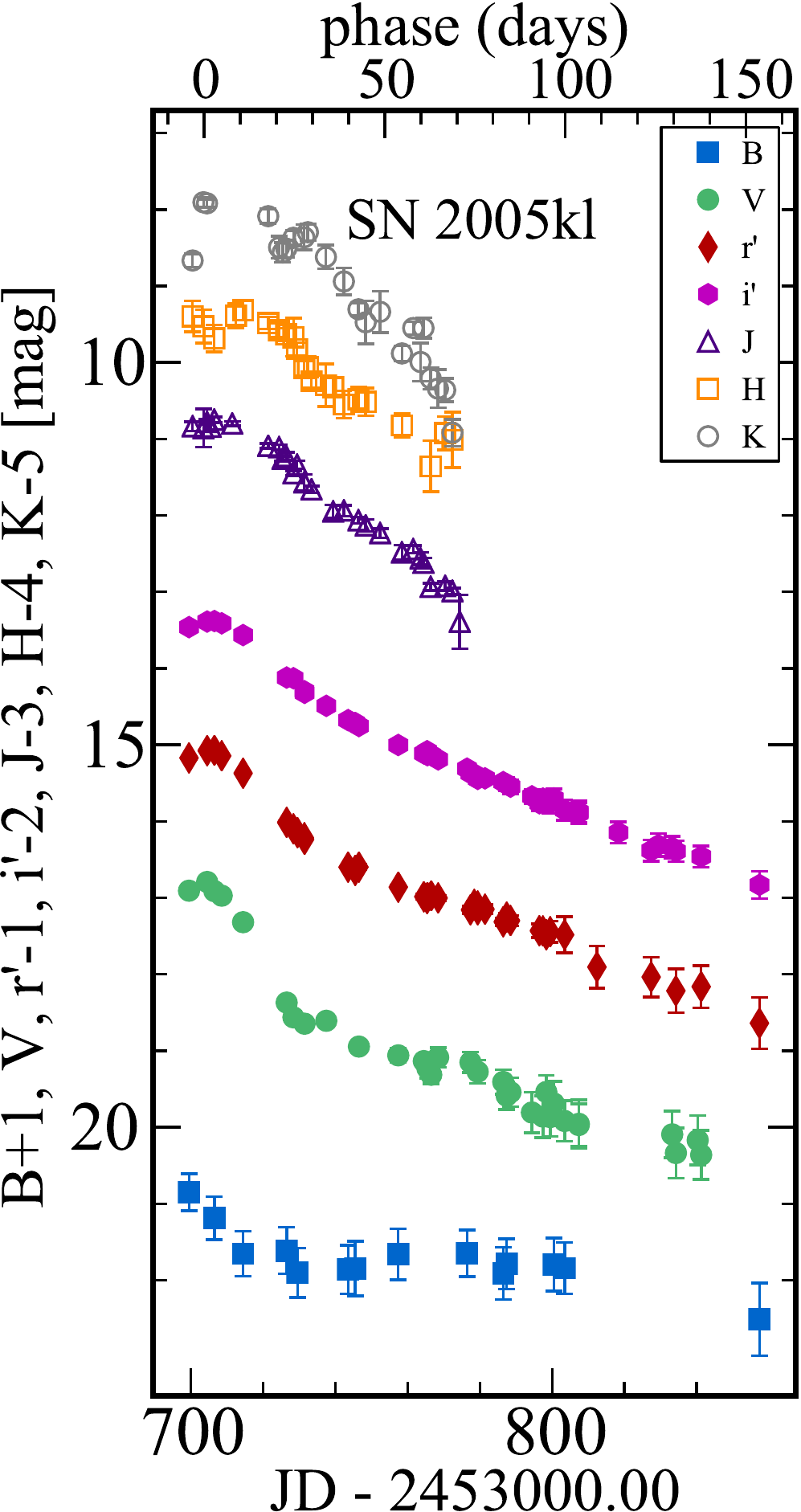}
\includegraphics[width=0.477\columnwidth]{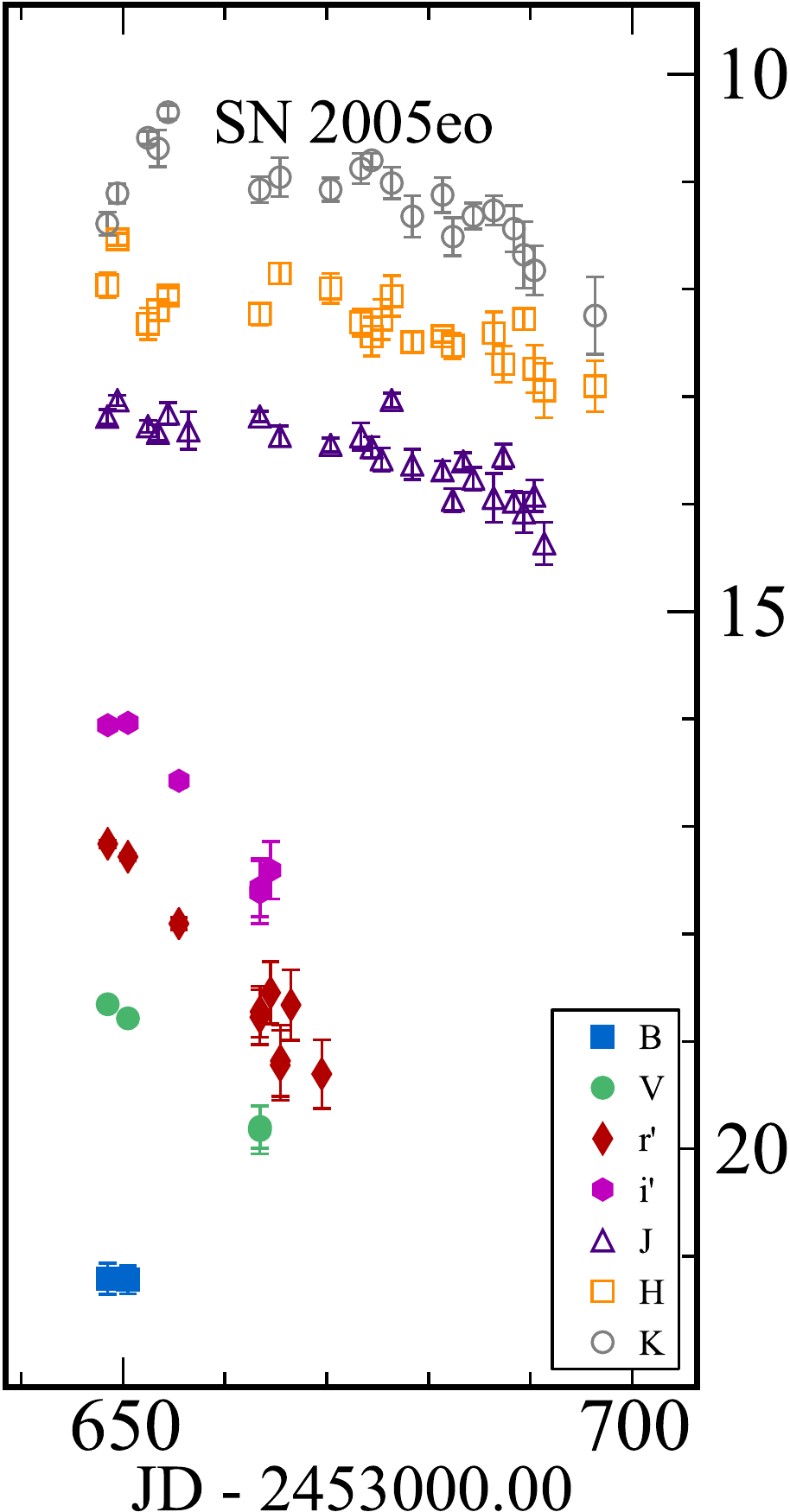}
}
\caption{SN Ic SN~2005kl (Left) and SN~2005eo (Right), which was reclassified as a late SN Ia in M14 (right). Both show uncharacteristically red colors. Note however the steeper slope of flux decay in SN~2005kl, consistent with a young SN evolution. }
\label{fig:sn05kl}
\end{figure}

The SN is in a bright and high-gradient galaxy, making image subtraction difficult, and resulting in the large errorbars, especially in $B$. From the spectra, published in M14 it is evident that the SN sits in an HII region. SN~2005kl is classified as a SN Ic in M14. The sparse spectral coverage cannot rule out the development of He I lines near peak, which would modify the classification to SN IIb.  We notice a supernova that shows a qualitatively similar color evolution, with suppressed $B$ flux, and red colors, is SN~2005eo. CfA light curves for both objects are shown in \figref{fig:sn05kl}. While SN~2005eo was initially classified as a SN Ic, and thus included in the D11 sample,
we now reclassify it as a late SN Ia (M14)\footnote{The light curve was fitted with SNANA \citep{2009PASP..121.1028K}, but with the little available data the photometric fit remains inconclusive, although SN Ic provided worse fits than SN Ia's. Although SN~2005eo would not be a standard SN Ia, as its $J$-band flux fall is slow, the poor light curve fit to SN Ic templates reinforces our trust in the new spectroscopic classification.}. With its reclassification the earliest optical photometry in D11 is actually catching the second $R$-band peak of the late SN Ia. Both SN~2005kl and SN~2005eo are found in early type spiral galaxies. Neither object suffers from significant galactic reddening (see \tabref{tab:hostgal_table}). 
While SN~20005kl may show similarly red colors as SN~2005eo, a SN Ia, we are certain that SN~2005kl is not a SN Ia since our CfA late-time, nebular spectra of SN~2005kl show strong emission lines of [OI] and [CaII], characteristic of stripped SN (M14).

\section{Conclusions}
This paper presents a densely-sampled, homogeneous suite of photometric measurements of stripped SN at optical and NIR wavelengths: $U$ (before 2009 January) and $u'$ (after 2009 January); $\!BV$; $RI$ (before 2004 September) and  $r'i'$ (after 2004 September); and $JHK_s$ bands, including \nsnopt\ objects covered in optical bands, and \nsnir\ in NIR. These data were collected between 2001 and 2009 at FLWO.

Our photometry provides additional data for \ninlit\ supernovae already discussed in the literature, and the first published measurements for \nnoninlit\ new supernovae. Among the objects previously published, we are publishing photometry for \nphotnoninlit\ objects studied in the literature at other wavelengths (radio or UV), or with methods different than photometry (e.g., spectroscopy, or host and progenitor studies, see~\tabref{tab:discoverytable}), but for which photometric measurements had not yet been published. Stripped SN spectroscopy from our group is presented in M14, complementing this data set with coverage for \nsnspec\ of our total \nsntot\ objects. 

This is the largest stripped SN data set to date, containing multi-color photometry in bands from optical $U$ to NIR $K_s$. Our sample includes \nsntot\ stripped SN, \nsnboth\ with both optical and NIR measurements, \nsnopt\ with optical measurements, and \nsnir\ with NIR measurements. This doubles the current supply of stripped SN objects in the literature. All SN have multi-band data, allowing the determination of multiple colors. Our photometry (described in \secref{sec:optphot_sec} and  \ref{sec:irphot_sec}) is produced from template-subtracted images, for all but 6 objects in optical bands and 5 in NIR, since those objects are well removed from their host galaxy.
We compare our photometry with literature data (\secref{sec:litdata}) and find agreement within the errors for most published SN.  However, we find D11 photometry to be brighter than ours by as much as one magnitude at peak, and even more at late times. We attribute this discrepancy to host galaxy contamination in the D11 data, since the
D11 photometric data are not based on template-subtracted images.

A solid determination of the epoch of maximum brightness in $V$-band (\maxep) is now possible for 36 objects using our data in combination with existing literature data (\secref{sec:sample_stat}). 

With these data we investigated the color behavior of stripped SN (\secref{colorcolor}), with the caveat that no corrections for host galaxy extinction have been applied to our data.  This approach captures the observed
color behavior of stripped SN (also adopted for SN rates, in \citealt{2011MNRAS.412.1441L} for example), and simulates the parameter space of current and future optical
and NIR surveys such as RATIR~\citep{2012SPIE.8446E..10B} and LSST \citep{LSST}, assuming the reddening of the SN
in our sample experienced is representative.

We present an intriguing separation of different stripped SN subtypes in the $r'-i'$~vs~$V-H$ color space, with SN Ic-bl appearing bluer than both SN Ib and Ic, based on the 19 stripped SN for which \maxep\ optical+NIR colors can be determined, but cautioning the reader that in our dataset only two SN Ic and two SN Ic-bl has sufficient photometric coverage to be included in a $r'-i'$~vs~$V-H$ color-color plot evaluated near peak (\figref{fig:colorcolor}).
We also observe a very narrow distribution in $r'-i'$ color for SN Ic-bl, with a standard deviation of only 0.01~mag around a mean  \mbox{$<r'-i'>_\mathrm{peak}~=~-0.025$~mag}.  This standard deviation is at least two times smaller than for any other subtype. 
As the data set grows, these color-color plots hold the promise to separate supernova types by photometry alone, especially with the advent of new NIR surveys
(e.g., RATIR -- \citealt{2012SPIE.8446E..10B,2012SPIE.8453E..1OF}, and SASIR -- \citealt{2009arXiv0905.1965B}).

In addition we identify a number of individual SN with peculiar color
behavior - some of which were known to be peculiar from spectra (e.g.,
SN~2005la and SN~2006jc, both SN whose ejecta are interacting with He-rich
and for SN~2005la, also H-rich circumstellar material), while others are
spectroscopically normal (SN~2005kl).

Finally, spectra for over 80\% of the objects we presented here are presented in a companion paper (M14).           
The availability of spectra with ample coverage at several epochs for most of our objects, offers an opportunity for an accurate statistical assessment of the 
photometric diversity among stripped SN types, which is necessary for classification in upcoming large synoptic surveys, as well as for future progenitor studies.

\acknowledgements

We are immensely grateful for the efforts of the service observers at
 the 1.2m FLWO, who obtained the majority of data
presented here. In addition, we thank the staff of the F. L. Whipple
 Observatories for their extensive support and assistance. 

The authors would like to thank Saurabh Jha, Tom Matheson, Alex Filippenko, Ryan Foley,
Nathan Smith, John Raymond, Rob Fesen, Chris Stubbs,
Avishay Gal-Yam, Claes Fransson, Alicia Soderberg, and Eli Dwek for
illuminating discussions. We thank Brandon Patel and Saurabh Jha for running SNANA fits. 

FBB is supported by a James Arthur fellowship at the Center for
Cosmology and Particle Physics at NYU.
Supernova research at Harvard University has been supported in part by
the National Science Foundation grant AST06-06772 and R.P.K. in part
by the NSF grant AST09-07903 and AST12-11196, 
and in part by the Kavli Institute for Theoretical Physics NSF grant PHY99-07949.
Observations reported here were obtained 
at the F.L Whipple Observatory, which is operated by the
Smithsonian Astrophysical Observatory. PAIRITEL is
operated by the Smithsonian Astrophysical Observatory (SAO) and was
made possible by a grant from the Harvard University Milton Fund, the
camera loan from the University of Virginia, and the continued support
of the SAO and UC Berkeley. The data analysis in this paper has made
use of the Hydra computer cluster run by the Computation Facility at
the Harvard-Smithsonian Center for Astrophysics.

This research has made use of NASA's Astrophysics Data
System Bibliographic Services (ADS), the HyperLEDA database and the
NASA/IPAC Extragalactic Database (NED) which is operated by the Jet
Propulsion Laboratory, California Institute of Technology, under
contract with the National Aeronautics and Space Administration.  This
publication makes use of data products from the Two Micron All Sky
Survey, which is a joint project of the University of Massachusetts
and the Infrared Processing and Analysis Center/California Institute
of Technology, funded by NASA and NSF.

\tabletypesize{\tiny}
\begin{turnpage}
\begin{deluxetable*}{lccllclll}
\singlespace
\setlength{\tabcolsep}{0.0001in} 
\tablewidth{514.88pt}
\tablecaption{Discovery and Classification Data for SN Sample}
\tablehead{\colhead{SN Name}  & 
\colhead{CfA spectra\tablenotemark{a}}&
\colhead{CfA NIR\tablenotemark{b}}&
\colhead{RA}&
\colhead{Dec}& 
\colhead{SN Type\tablenotemark{c}}  &  
\colhead{Discovery Date}  &  
\colhead{Discovery Reference}  &  
\colhead{Spectroscopic ID\tablenotemark{c}} } 
\startdata
{SN~2001ej}                     &M14&                           &7:23:43 &+33:26:38.0 &		Ib		&	2001-09-17	&	IAUC 7719	& IAUC 7721 \\
{SN~2001gd}\tablenotemark{3} 	&M14&                           &13:13:23& +36:38:17.7&		IIb		&	2001-11-24	&	IAUC 7761	& IAUC 7765\\
{SN~2002ap}\tablenotemark{1}	&M14&                           &1:36:23 &+15:45:13.2 &		Ic-bl		&	2002-01-29	&	IAUC 7810	& IAUC 7811/7825 \\
{SN~2003jd}\tablenotemark{1}    &M14&                           &21:03:38& -4:53:45   &		Ic-bl		&	2003-10-25	&	IAUC 8232	& IAUC 8234\\
{SN~2004ao}\tablenotemark{1} 	&M14&	                        &17:28:09& +07:24:55.5&		Ib		&	2004-03-07	&	IAUC 8299	& IAUC 8430\\
{SN~2004aw}\tablenotemark{1} 	&M14&                           &11:57:50& +25:15:55.1&		Ic		&	2004-03-19	&	IAUC 8310 	& IAUC 8331        \\
{SN~2004fe}			&M14&    	                &0:30:11 &+02:05:23.5 &		Ic		&	2004-10-30	&	IAUC 8425	& IAUC 8426\\
{SN~2004gk}\tablenotemark{1,2} &M13&Y                           &12:25:33& +12:15:39.9&		Ic		&	2004-11-25	&	IAUC 8446	& IAUC 8446\\
\enddata
\tablenotetext{1}{Indicates SN whose early-time behavior have been studied in the literature. $^\mathrm{2}$ Included in the D11 sample. }
\tablenotetext{3}{Radio studies, no published optical lightcurve.$^4$Progenitor studies, no published lightcurve.$^5$Host studies, no published lightcurve.$^6$Only spectra published.$^7$Only UV data published.}    
\tablenotetext{a}{These objects have spectra presented in M14}
\tablenotetext{b}{Object for which NIR data is available within the CfA survey. $^\mathrm{*}$Indicates \emph{only} NIR data is available within the CfA survey for this object.}
\tablenotetext{c}{SN reclassified with SNID using updated templates in M14 have the new classification reported.} 
\tablenotetext{d}{SN 2005la is spectroscopically peculiar showing narrow He and H lines in emission (Pastorello 2008).}
\tablenotetext{e}{SN~2006jc is spectroscopically peculiar and  showed narrow He lines in emission. See text for details. }
\tablenotetext{f}{The type classification of SN 2009er remains ambiguous. SNID classifies SN 2009er differentlly at different epochs. We also note the presence of high velocity He I lines.}
\label{tab:discoverytable}
\end{deluxetable*}

\clearpage
\begin{deluxetable*}{lllrrcrcrrr}
\singlespace
\tablecaption{SN Host Galaxy Basic Data}
\tablehead{\colhead{SN Name} & 
\colhead{SN Type\tablenotemark{a}} & 
\colhead{Galaxy\tablenotemark{b}} & 
\colhead{$cz_{\rm{helio}}$} & 
\colhead{$M_{\rm{B}}$\tablenotemark{c}} & 
\colhead{$m_{\rm{B}}$\tablenotemark{c}} & 
\colhead{$m-M$}\tablenotemark{c}&
\colhead{Morphology}\tablenotemark{c}&
\multicolumn{2}{c}{SN Offsets\tablenotemark{d}} &
\colhead{$E_{B-V}$\tablenotemark{e}}\\
\colhead{} &
\colhead{} &
\colhead{} &
\colhead{[\kms]} &
\colhead{[mag]} &
\colhead{[mag]} &
\colhead{[mag]} &
\colhead{} &
\colhead{''} &
\colhead{''} &
\colhead{} 
} 
\startdata
SN~2001ej & Ib 	     &	 UGC	3829   		&    4031  &        -            &  13.80 (0.2)             &   33.97         &   Sb                    &4W  & 7N                            & 0.0460\\ 
SN~2001gd & IIb      &	 NGC	5033   		&    1024  &    {\bf -20.95}    & {\bf 10.70 (0.13)}        &  {\bf 30.92}    & {\bf Sc}                   	& 52W & 161N                         & 0.0098\\
SN~2002ap & Ic-bl    &	 M74           		&      659 &    {\bf -20.71}    & {\bf  9.70 (0.26)}        &  {\bf 29.67}    & {\bf Sc}                        &258W & 108S                         & 0.0620\\
SN~2003jd & Ic-bl    & MCG-01-59-21 \  		&     5635 &    {\bf -19.81}    & {\bf 15.21 (0.50)}        &  34.49          & {\bf SABm}                      & 8E  & 8S                           & 0.0784\\
SN~2004ao &       Ib &   UGC     10862 		&     1691 &     {\bf -18.53}   & {\bf 14.41 (0.59)}        & 31.95           & {\bf SBc}               	&6.3E  & 23.8S                       &  0.0893\\
SN~2004aw &      Ic  &           NGC 3997       &    4770  &     {\bf -20.78}   & {\bf 14.04 (0.13)}        & 34.24           & {\bf SBb}               	           &       28E    &     20S  &  0.0180\\     
SN~2004fe &      Ic  & NGC 132                  &  5363    &      {\bf -21.13}   & {\bf 13.74 (0.33)}       & 34.21           & {\bf SABb}                         &9E & 12S                         & 0.0210\\
SN~2004gk &       Ic &    IC      3311 		&   $-$122 &    {\bf -18.00}    & {\bf 14.71 (0.18)}        &  {\bf 31.48}    & {\bf Scd}                      &     2E               &     3N       & 0.0247\\
\enddata
\tablenotetext{a}{The SN Type "Ic-bl" stands for broad-lined SN~Ic. SN~2006jc is a SN Ib-n ("n" for narrow) which showed narrow He lines in emission. See text for details. }
\tablenotetext{b}{``Anonimous'' stands for non-catalogues host galaxy.}
\tablenotetext{c}{From HYPERLEDA catalog (bold) or NED catalog.}
\tablenotetext{d}{From \url{http://www.cbat.eps.harvard.edu/lists/Supernovae.html}.}
\tablenotetext{e}{Galactic extintion calculated according to \citet{2011ApJ...737..103S}}
\tablenotetext{f}{Abbreviated galaxy name for ``SDSS~J074726.40+265532.4'' .}
\tablenotetext{g}{Abbreviated galaxy name for ``SDSS~J153930.49+242614.8'' .}  
\label{tab:hostgal_table}
\end{deluxetable*}

\begin{deluxetable*}{ccccccccccccc}
\tablecolumns{13}
\tablewidth{0pc}
\tabletypesize{\scriptsize}
\tablecaption{SN \protect{2005{\MakeLowercase{hg}} Comparison Photometry}}
\tablehead{ \colhead{Star} & \colhead{RA(J2000)} &  \colhead{DEC(J2000)} & \colhead{$V$} & \colhead{$N_V$} & \colhead{$U-B$} & \colhead{$N_U$} & \colhead{$B-V$} & \colhead{$N_B$} & \colhead{$V-r'$} & \colhead{$N_{r'}$} & \colhead{$V-i'$} & \colhead{$N_{i'}$} }
\startdata
01 & 01:56:08.386 & +46:51:51.91 &  16.217$\pm$.013 &  3 &    0.113$\pm$.052 &  3 &   0.729$\pm$.020 &  3 &   0.189$\pm$.009 &  3 &   0.407$\pm$.011 &  3\\
02 & 01:56:03.007 & +46:52:47.44 &  16.338$\pm$.013 &  3 &    0.382$\pm$.043 &  3 &   0.740$\pm$.016 &  3 &   0.204$\pm$.011 &  3 &   0.395$\pm$.011 &  3\\
03 & 01:56:02.641 & +46:52:31.82 &  16.032$\pm$.012 &  3 &    0.091$\pm$.042 &  3 &   0.593$\pm$.017 &  3 &   0.150$\pm$.010 &  3 &   0.314$\pm$.011 &  3\\
04 & 01:56:02.558 & +46:43:25.08 &  15.460$\pm$.012 &  3 &    0.170$\pm$.042 &  3 &   0.725$\pm$.015 &  3 &   0.192$\pm$.009 &  3 &   0.381$\pm$.009 &  3\\
05 & 01:55:59.139 & +46:51:51.01 &  15.467$\pm$.012 &  3 &    0.102$\pm$.047 &  3 &   0.620$\pm$.017 &  3 &   0.159$\pm$.011 &  3 &   0.330$\pm$.010 &  3\\
06 & 01:55:59.060 & +46:46:57.56 &  15.492$\pm$.013 &  3 &    0.250$\pm$.056 &  3 &   0.759$\pm$.017 &  3 &   0.211$\pm$.009 &  3 &   0.431$\pm$.009 &  3\\
07 & 01:55:58.585 & +46:49:29.73 &  15.898$\pm$.013 &  3 &    0.105$\pm$.050 &  3 &   0.632$\pm$.016 &  3 &   0.166$\pm$.011 &  3 &   0.344$\pm$.009 &  3\\
08 & 01:55:57.677 & +46:44:26.44 &  14.789$\pm$.013 &  3 &    0.095$\pm$.043 &  3 &   0.603$\pm$.016 &  3 &   0.135$\pm$.010 &  3 &   0.264$\pm$.009 &  3\\
\enddata
\label{tab:compstar_2005hg}
\end{deluxetable*}

\begin{deluxetable*}{ccccccccccccccc}
\tablecolumns{15}
\singlespace
\setlength{\tabcolsep}{0.0001in}
\tablewidth{514.88pt}
\tablewidth{0pc}
\tabletypesize{\scriptsize}
\tablecaption{\protect{{SN~2005hg}} Optical Photometry}
\tablehead{\colhead{MJD}&
\colhead{$U$}&
\colhead{d$U$}&
\colhead{MJD}&
\colhead{$B$}&
\colhead{d$B$}&
\colhead{MJD}&
\colhead{$V$}&
\colhead{d$V$}&
\colhead{MJD}&
\colhead{$r'$}&
\colhead{d$r'$}&
\colhead{MJD}&
\colhead{$i'$}&
\colhead{d$i'$}}
\startdata
53671.188 & 19.08 & 0.16 &  53671.176 & 18.64 & 0.07 &  53671.184 & 18.86 & 0.03 &  53671.180 & 18.46 & 0.03 &  53671.184 & 18.49 & 0.04\\  
53671.191 & 18.84 & 0.11 &  53671.180 & 18.55 & 0.04 &  53674.402 & 18.43 & 0.03 &  53674.398 & 17.98 & 0.03 &  53674.398 & 17.97 & 0.03\\  
53674.402 & 18.43 & 0.09 &  53674.395 & 18.06 & 0.02 &  53676.309 & 18.32 & 0.04 &  53676.305 & 17.71 & 0.02 &  53676.309 & 17.74 & 0.02\\  
53674.406 & 18.52 & 0.09 &  53676.305 & 17.89 & 0.03 &  53677.348 & 18.20 & 0.04 &  53677.336 & 17.71 & 0.01 &  53677.336 & 17.68 & 0.02\\  
53676.316 & 18.36 & 0.13 &  53677.340 & 17.81 & 0.02 &  53678.359 & 18.23 & 0.02 &  53678.348 & 17.63 & 0.02 &  53678.348 & 17.60 & 0.02\\  
53677.340 & 18.22 & 0.06 &  53678.348 & 17.73 & 0.02 &  53679.219 & 18.22 & 0.03 &  53679.215 & 17.56 & 0.02 &  53679.219 & 17.56 & 0.02\\  
53678.352 & 18.24 & 0.07 &  53679.215 & 17.70 & 0.02 &  53680.359 & 18.16 & 0.02 &  53680.348 & 17.53 & 0.02 &  53680.348 & 17.49 & 0.02\\  
53678.355 & 18.17 & 0.06 &  53680.352 & 17.65 & 0.02 &  53681.289 & 18.14 & 0.02 &  53681.277 & 17.48 & 0.02 &  53681.277 & 17.47 & 0.02\\
\enddata
\label{tab:snoptphot}
\end{deluxetable*}

\begin{deluxetable*}{ccccccccc}
\tablecolumns{9}
\singlespace
\setlength{\tabcolsep}{0.0001in}
\tablewidth{514.88pt}
\tablewidth{0pc}
\tabletypesize{\scriptsize}
\tablecaption{\protect{{SN~2005\MakeLowercase{hg}} NIR Photometry}}
\tablehead{\colhead{MJD}&
\colhead{$H$}&
\colhead{d$H$}&
\colhead{MJD}&
\colhead{$J$}&
\colhead{d$J$}&
\colhead{MJD}&
\colhead{$K_s$}&
\colhead{d$K_s$}}
\startdata
53676.109 & 17.10 & 0.09 &  53676.109 & 17.02 & 0.05 &  53676.109 & 17.34 & 0.25\\  
53677.129 & 17.03 & 0.09 &  53677.129 & 16.98 & 0.04 &  53677.129 & 17.19 & 0.10\\  
53678.141 & 16.78 & 0.07 &  53678.141 & 16.77 & 0.03 &  53679.148 & 16.90 & 0.09\\  
53679.148 & 16.89 & 0.05 &  53679.148 & 16.93 & 0.05 &  53681.352 & 16.81 & 0.09\\  
53681.352 & 16.77 & 0.07 &  53681.352 & 16.85 & 0.04 &  53682.230 & 16.58 & 0.08\\  
53682.160 & 16.58 & 0.05 &  53682.160 & 16.79 & 0.03 &  53683.359 & 16.53 & 0.17\\  
53683.340 & 16.65 & 0.05 &  53683.340 & 16.72 & 0.03 &  53684.238 & 16.75 & 0.07\\  
53684.238 & 16.70 & 0.05 &  53684.238 & 16.69 & 0.03 &  53686.301 & 16.47 & 0.06\\
\enddata
\label{tab:snnirphot}
\end{deluxetable*}

\begin{deluxetable*}{lccccccccc}
\tabletypesize{\scriptsize}
\singlespace
\setlength{\tabcolsep}{0.0001in} 
\tablewidth{514.88pt}
\tablecaption{SN characteristics}
\tablehead{\colhead{SN Name} &
\colhead{$U$}& 
\colhead{$U$}& 
\colhead{$U$}& 
\colhead{$B$}& 
\colhead{$B$}& 
\colhead{$B$}& 
\colhead{$V$}& 
\colhead{$V$}& 
\colhead{$V$}\\
\colhead{}& 
\colhead{$\mathrm{JD}$ ($d\mathrm{JD}$)}&
\colhead{$\mathrm{max}$ ($d\mathrm{max}$)}&
\colhead{\delm ($d$\delm)}&
\colhead{$\mathrm{JD}$ ($d\mathrm{JD}$)}&
\colhead{$\mathrm{max}$ ($d\mathrm{max}$)}&
\colhead{\delm ($d$\delm)}&
\colhead{$\mathrm{JD}$ ($d\mathrm{JD}$)}&
\colhead{$\mathrm{max}$ ($d\mathrm{max}$)}&
\colhead{\delm ($d$\delm)}
}
\startdata
SN~2003jd\tablenotemark{a} &  - - & - - & - - &   52942.50 (0.39) & 16.33 (0.05) & - - & 52942.78 (0.47) & 15.93 (0.02) & - - \\
SN~2004aw\tablenotemark{b} &  - - & - - & - - &   53085.55 (0.32) & 17.91 (0.03) & 0.85 (0.26) & 53090.95 (0.85) & 17.23 (0.03) & - - \\
SN~2004fe &  - - & - - & - - &    - - & - - & - - &   53318.80 (0.46) & 17.09 (0.02) & - - \\
SN~2004gq & 53353.86 (2.61) & 15.67 (0.09) & - - & 53353.68 (1.87) & 16.12 (0.02) & - - &  - - & - - & - - \\
SN~2004gt &  - - & - - & - - &   53388.52 (1.41) & 18.07 (0.05) & - - &  - - & - - & - - \\
SN~2005az &  - - & - - & - - &   - - & - - & - - & 53473.36 (0.75) & 16.40 (0.05) & 0.33 (0.03) \\
SN~2005bf\tablenotemark{c} & 53473.20 (0.05) & 17.16 (0.03) & - - & 53474.22 (0.18) & 17.66 (0.02)  & - - & 53476.75 (0.42) & 17.46 (0.03)  &- - \\
SN~2005bf\tablenotemark{c} & 53498.54 (0.09) & 16.42 (0.02) & 1.13 (0.08) & 53497.73 (0.14) & 16.71 (0.02) & 1.12 (0.06) & 53498.46 (0.33) & 16.43 (0.02) & 0.86 (0.02) \\
\enddata
\tablenotetext{a}{Includes data near peak from Valenti et al. (2008).}
\tablenotetext{b}{Includes data near peak from Taubenberger et al (2006).}
\tablenotetext{c}{Two peaks are visible for SN 2005bf in $V$-band and both are reported here.}
\tablenotetext{d}{Includes data near peak from Drout et al. (2011).}
\tablenotetext{e}{Includes KAIT data near peak from Sahu et al. (2009).}
\tablenotetext{f}{Includes KAIT data near peak from Modjaz et al. (2009), and Soderberg et al. (2008).}
\tablenotetext{g}{Includes data near peak from Valenti et al. (2011).}
\label{tab:maxsUBV}
\end{deluxetable*}

\begin{deluxetable*}{lcccccc}
\tabletypesize{\scriptsize}
\singlespace
\setlength{\tabcolsep}{0.0001in} 
\tablewidth{514.88pt}
\tablecaption{SN characteristics}
\tablehead{\colhead{SN Name} &
\colhead{$r` (R)$}& 
\colhead{$r` (R)$}& 
\colhead{$r` (R)$}& 
\colhead{$i` (I)$}& 
\colhead{$i` (I)$}& 
\colhead{$i` (I)$}\\
\colhead{}&
\colhead{$\mathrm{JD}$ ($d\mathrm{JD}$)}&
\colhead{$\mathrm{max}$ ($d\mathrm{max}$)}&
\colhead{\delm ($d$\delm)}&
\colhead{$\mathrm{JD}$ ($d\mathrm{JD}$)}&
\colhead{$\mathrm{max}$ ($d\mathrm{max}$)}&
\colhead{\delm ($d$\delm)}
}
\startdata
SN 2003jd\tablenotemark{a} &  - - & - - & - - &   52945.05 (2.54) & 15.61 (0.03) & - - \\
SN 2004aw\tablenotemark{a} & 53096.04 (0.02) & 16.82 (0.03) & 0.39 (0.01) & 53095.29 (0.80) & 16.79 (0.02) & 0.57 (0.11) \\
SN 2004fe &   53321.52 (0.68) & 17.01 (0.04) & - - & 53320.09 (0.75) & 17.09 (0.03) & - - \\
SN 2004gq &    - - & - - & - - &   53364.31 (0.24) & 15.50 (0.04) & 0.25 (0.02) \\
SN 2005az &   53476.10 (0.54) & 16.06 (0.03) & - - & 53479.84 (0.76) & 15.97 (0.04) & 0.45 (0.05) \\
SN 2005bf &   53499.93 (0.38) & 16.36 (0.02) & 0.65 (0.05) & 53503.58 (0.30) & 16.36 (0.02) & - - \\
SN 2005eo &    - - & - - & - - &   53648.97 (0.63) & 18.02 (0.04) & 2.57 (1.00) \\
SN 2005hg &   53686.44 (0.40) & 17.45 (0.04) & 0.64 (0.20) & 53688.69 (0.63) & 17.33 (0.03) & 0.60 (0.14) \\
\enddata
\tablenotetext{a}{Photometry collected with Johnson R and I filters}
\label{tab:maxsRI}
\end{deluxetable*}

\begin{deluxetable*}{lccccccccc}
\tabletypesize{\scriptsize}
\singlespace
\setlength{\tabcolsep}{0.0001in} 
\tablewidth{514.88pt}
\tablecaption{SN characteristics}
\tablehead{\colhead{SN Name} &
\colhead{$J$}& 
\colhead{$J$}& 
\colhead{$J$}& 
\colhead{$H$}& 
\colhead{$H$}& 
\colhead{$H$}& 
\colhead{$K_s$}& 
\colhead{$K_s$}& 
\colhead{$K_s$}\\
\colhead{}&
\colhead{$\mathrm{JD}$ ($d\mathrm{JD}$)}&
\colhead{$\mathrm{max}$ ($d\mathrm{max}$)}&
\colhead{\delm ($d$\delm)}&
\colhead{$\mathrm{JD}$ ($d\mathrm{JD}$)}&
\colhead{$\mathrm{max}$ ($d\mathrm{max}$)}&
\colhead{\delm ($d$\delm)}&
\colhead{$\mathrm{JD}$ ($d\mathrm{JD}$)}&
\colhead{$\mathrm{max}$ ($d\mathrm{max}$)}&
\colhead{\delm ($d$\delm)}
}
\startdata
SN 2004gk&   53458.33 (2.49) & 16.39 (0.12) & - -  &  - -            &  - -         & - -         &  - - & - - & - - \\
SN 2004gt&    - -            & - -          & - -          &  - -            & - -          & - -         &   53472.00 (0.45) & 15.63 (0.28) & - - \\
SN 2005bf& 53505.32 (0.44)   & 15.90 (0.01) & 0.39 (0.02)  & 53511.45 (2.03) & 15.81 (0.07) & 0.18 (0.11) & 53512.49 (1.69) & 15.57 (0.03) & 0.26 (0.04) \\
SN 2005eo&   - - & - - & - - &  - -            & - -          & - -         & 53659.95 (1.62) & 15.57 (0.09) & 0.40 (0.13) \\
SN 2005hg& 53692.40 (0.61)   & 16.54 (0.03) & 0.53 (0.07)  & 53693.73 (0.44) & 16.55 (0.06) & 0.29 (0.03) & 53695.95 (0.70) & 16.37 (0.08) & 0.44 (0.08) \\
SN 2005kl& 53706.57 (1.20)   & 13.78 (0.04) & 0.28 (0.04)  & 53713.94 (1.70) & 13.34 (0.12) & 0.50 (0.09) & 53706.71 (3.34) & 12.97 (0.09) & 0.22 (0.05) \\
SN 2005mf& 53740.91 (0.41)   & 17.41 (0.03) & 0.61 (0.17)  & 53742.27 (0.27) & 17.27 (0.05) & 0.97 (0.25) & 53743.06 (0.38) & 16.93 (0.09) & 0.80 (0.12) \\
SN 2006aj& 53793.79 (2.09)   & 16.95 (0.12) & - -          & 53793.85 (1.62) & 16.70 (0.13) & - -         & 53793.11 (0.76) & 16.39 (0.11) & - - \\
\enddata
\label{tab:maxsHJK}
\end{deluxetable*}

\end{turnpage}
\end{document}